\documentclass{nature}
\usepackage{graphicx}
\usepackage{textcomp}
\usepackage{soul,color}
\usepackage{amsfonts}
\usepackage{upgreek}

\makeatletter
\let\saved@includegraphics\includegraphics
\AtBeginDocument{\let\includegraphics\saved@includegraphics}
\renewenvironment*{figure}{\@float{figure}}{\end@float}
\makeatother

\title{EGaIn tube memristors offer reliable switching on a biological time scale}

\author{Yuriy V. Pershin$^{1}$, Liya Patel$^2$, Bapi Berra$^2$, Doug Aaron$^2$, \& Stephen A. Sarles$^2$}

\begin{document}

\maketitle

\begin{affiliations}
 \item Department of Physics and Astronomy, University of South Carolina, Columbia, SC 29208, USA
 \item Department of Mechanical and Aerospace Engineering, University of Tennessee, Knoxville, TN 37996, USA

\end{affiliations}

\section*{Abstract}
Memristive devices have been considered promising candidates
for nature-inspired computing and in-memory information processing.
However, experimental devices developed to date typically show significant variability and function at different time scales than biological neurons and synapses. This study presents a new kind of memristive device comprised of liquid-metal eutectic gallium indium (EGaIn) contained within a mm-scale tube that operates via a bulk, voltage-dependent switching mechanism and exhibits distinct unipolar resistive switching characteristics that occur on a biological time scale (tens of milliseconds). The switching mechanism involves voltage-controlled growth and dissolution of an oxide layer on the surface of the liquid metal in contact with an aqueous electrolyte. Through comprehensive measurements on many devices, we observed remarkably consistent cycle-to-cycle behavior and uniformity in the voltage-controlled memristance. We present our findings, which also include an experimental demonstration of logic gates utilizing EGaIn tube memristors. Furthermore, we observe both accelerated and decelerated switching behaviors and identify signatures indicative of a fractional dynamic response.

\section*{Introduction}

A primary challenge hindering the large-scale commercialization of memristive~\cite{chua76a} technologies is their inconsistent switching behaviors, commonly termed device-to-device and cycle-to-cycle variability. In the most technologically advanced resistive random access memories (ReRAMs), such as electrochemical metallization (ECM)~\cite{valov2011electrochemical} and valence change mechanism (VCM)~\cite{yang2012metal} devices, the switching typically occurs due to the growth or dissociation of a conductive filament. The conductivity of a filament, which is a nanoscale structure made up of cations in ECM devices and anions in VCM devices, is greatly influenced by the arrangement of atoms that cannot be precisely controlled. Thus, the variation in response in such devices can be attributed to the stochastic nature of filament formation. Currently, it seems very difficult or even impossible to eliminate this in the existing solid-state technology.

In the present work, we utilize the advantages of the liquid phase material system, which consists of two EGaIn:sodium hydroxide solution interfaces contained within a narrow tube, as an alternative to a solid phase material architecture. In fact, liquid-state memristors have recently attracted increasing attention~\cite{kim2023liquid,Fan24a}. But what makes liquid memristors different? Unlike in the solid phase, atoms and molecules within a liquid are mobile. Consequently, large ensembles of atoms/molecules of the same type undergo on-average uniform interactions with other species in the surrounding liquid, resulting in a homogeneous collective response. As a result, an important distinction between our EGaIn liquid metal device and conventional memristors lies in the bulk switching mechanism, which provides an extra level of cycle-to-cycle and device-to-device averaging. Although the structure of our devices is very different from that of biological synapses, they  both operate on nearly the same time scale.~\footnote{Although it may be an ambiguous explanation, it is possibly because they both rely on a liquid.} Therefore, our EGaIn memristive devices may be relevant to several important applications such as neuroprosthetics~\cite{mikhaylov2020neurohybrid} and brain-computer interfaces.

In addition to various nanofluidic and hydrogel-based ionic memristors\cite{emmerich2024nanofluidic,xu2024nanofluidic}, liquid metals (LMs), including eutectic gallium indium (EGaIn) and gallium indium tin (Galinstan), have attracted interest for building memristors, generally via one of three approaches. In the first, LMs were used as flexible metal contacts on one side of a dielectric thin film (e.g., polymer~\cite{Zaheer2020}, metal organic framework~\cite{Yi2019}, inorganic oxide~\cite{Zaheer2022}, or self-assembled monolayer~\cite{Cao2024}) on a solid metal substrate. In this metal-insulator-metal (MIM) configuration, applied voltage drives Ga ions from the LM through the dielectric to form conductive nanofilaments, resulting in bipolar nonvolatile memristance. An advantage of using LM as one of the metal contacts is its native flexibility, enabling devices to operate under significant mechanical deformations~\cite{Yi2019,Cao2024}. A second approach harvested the solid gallium oxide layer from the surface of the LM to form the central dielectric material in an MIM sandwich. To do so, Xu, et al. squeeze printed the $2.5-3.5$~nm thick Ga$_2$O$_3$
LM coating onto an n+-Si solid surface~\cite{Xu2023}. Upon the addition of a Pt top contact, the MIM device displayed bipolar memristive switching, attributed to oxygen vacancy migration through Ga$_2$O$_3$. The third approach, which is most similar to our work presented herein, utilizes one or more LM volumes interfacing an aqueous electrolyte. Koo, et al. demonstrated that bipolar memristance could be achieved with two EGaIn volumes separated by a bi-layer arrangement of acidic (PAA, pH $\sim 3$) and basic (PEI, pH $\sim 10$) aqueous hydrogels~\cite{Koo2011}. Bipolar memristance was attributed to the asymmetric gel arrangement that enabled nonvolatile oxide growth at the PAA interface (stable switching from the low resistance state, LRS, to the high resistance state, HRS) under positive voltages ($3-5$~V) and volatile oxide growth at the PEI interface under negative voltages ($\sim -1$~V). Recent works~\cite{Ivanov2021,Yuan2023} have also shown that liquid metal EGaIn interfacing aqueous electrolytes at controlled pH levels enables programmable, voltage-dependent changes in device impedance, offering pathways for building physical neural networks for signal and flexible random access memory for information storage. 

Herein, we demonstrate that a symmetric, all-liquid configuration of EGaIn/electrolyte/EGaIn exhibits volatile memristance through voltage-controlled electrochemical oxidation and reduction that occurs alternatively at the two EGaIn-electrolyte interfaces. The use of a fully liquid system ensures atomically smooth interfaces (on average), while the electrochemical mechanism of resistive switching occurs uniformly, in a spatially distributed manner across the interface, which provides repeatable switching characteristics on biological time scales (milliseconds), low cycle-to-cycle performance variation, and device longevity (weeks). Reliable switching voltages and memory states enable consistent changes in the device conductance state, which are leveraged to demonstrate logic gate functionality.


\section*{Results and Discussion}

\begin{figure}[]
\centering
(a) \includegraphics[width=0.2\textwidth]{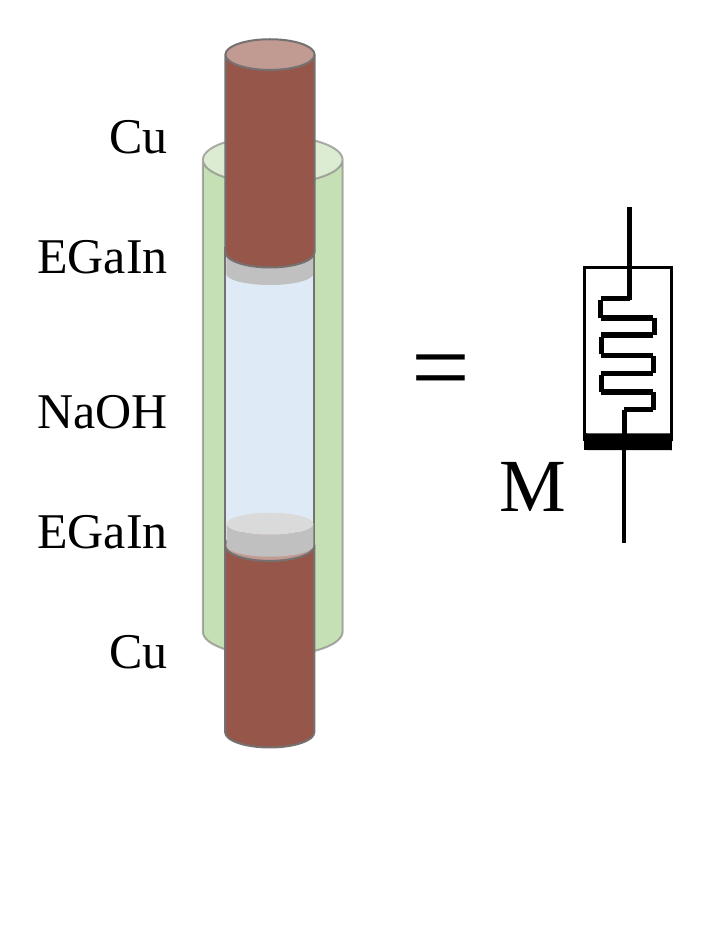} \;
(b)\includegraphics[width=0.2\textwidth]{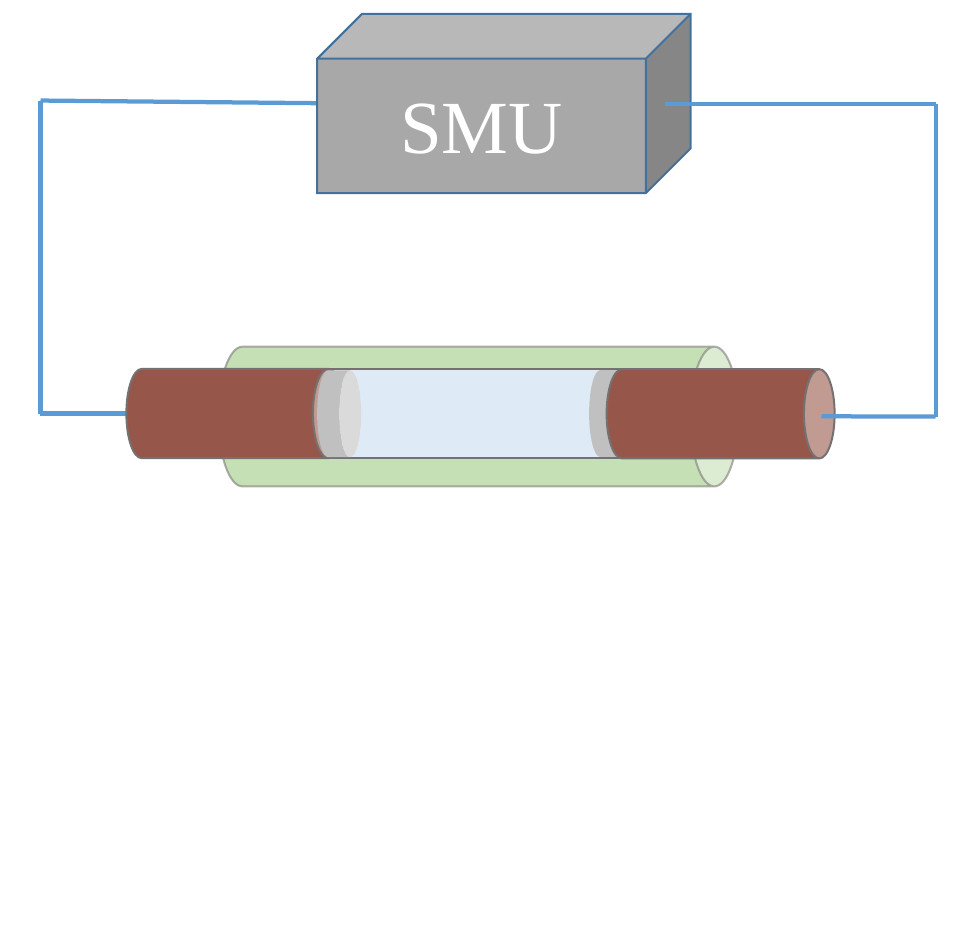} 
(c) \includegraphics[width=0.45\textwidth]{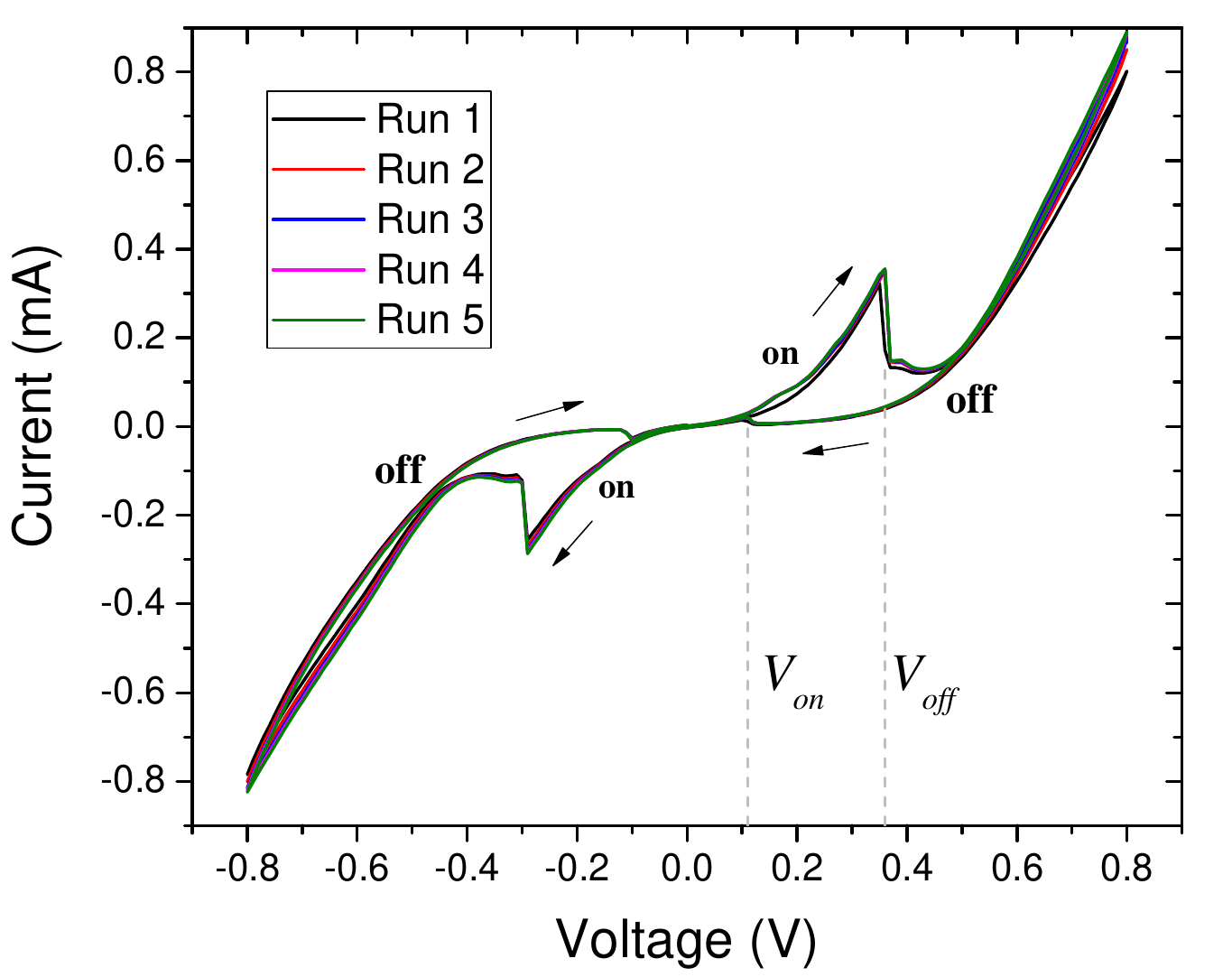}\\
(d) \includegraphics[width=0.45\textwidth]{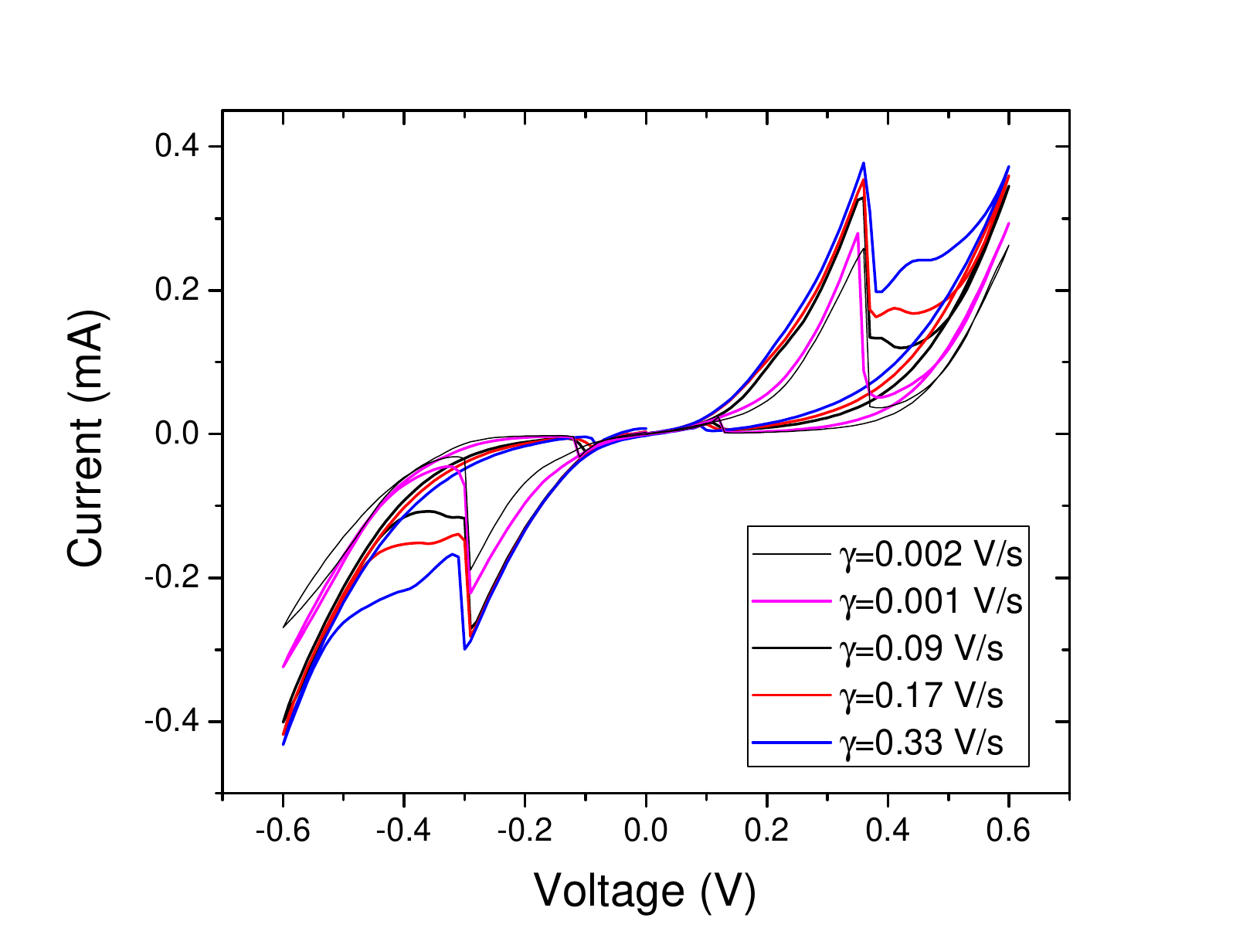}
(e) \includegraphics[width=0.45\textwidth]{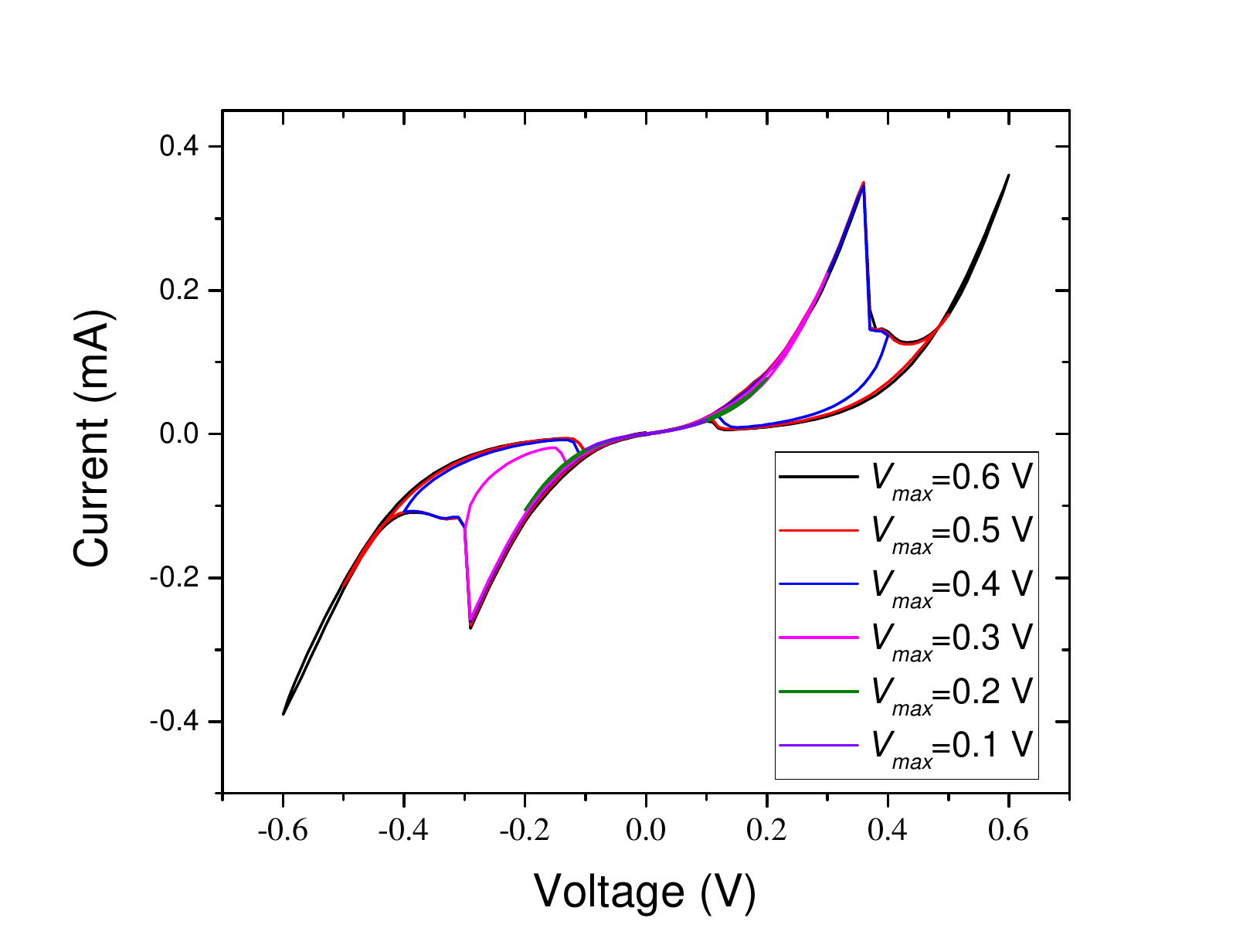}
(f) \includegraphics[width=0.45\textwidth]{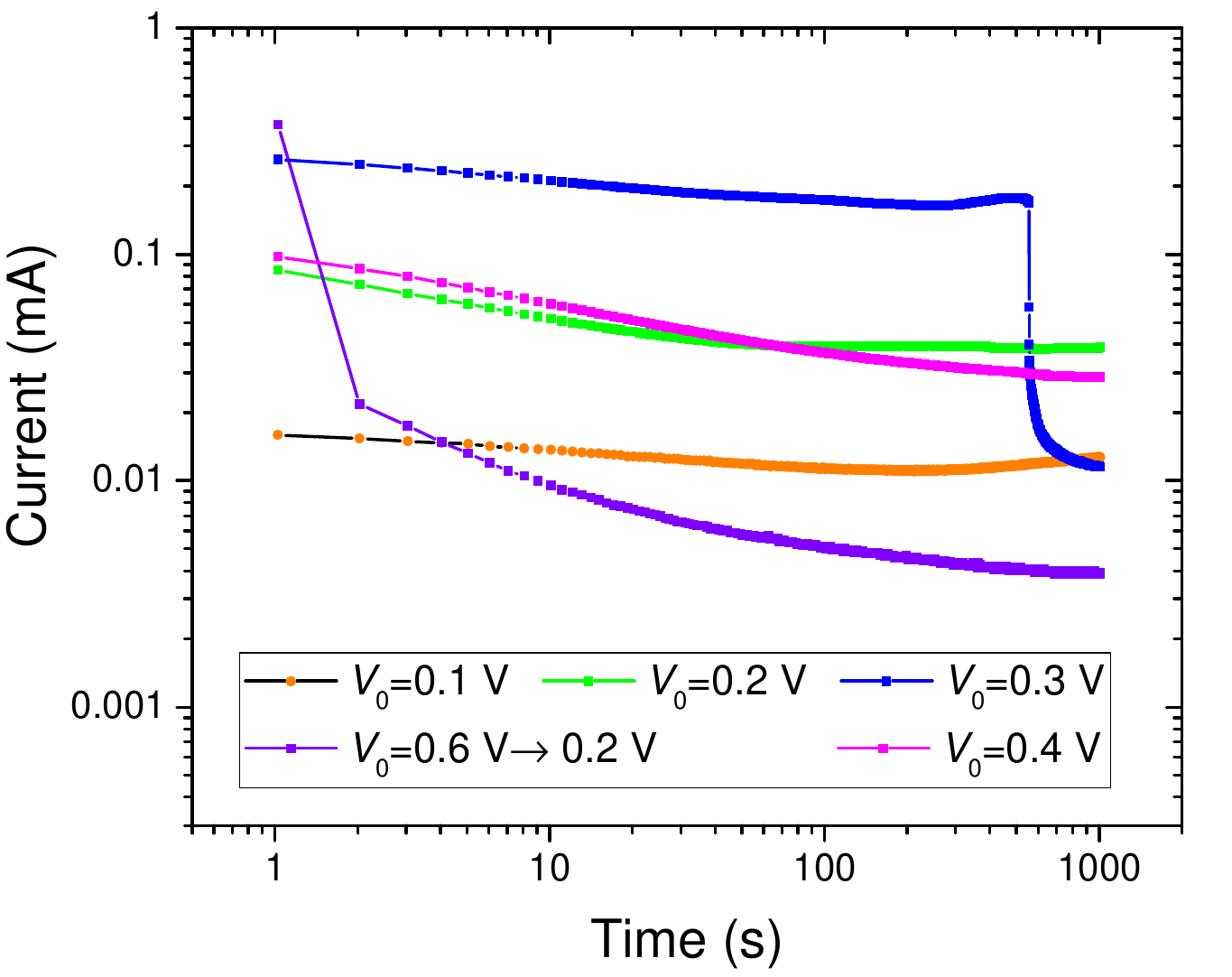}
\caption{Schematics of a tube memristor and its response recorded using a source measure unit (SMU).
(a) EGaIn tube memristor is composed of two external copper electrodes inserted into a plastic tube. The electrodes' surfaces facing each other are covered with EGaIn and are separated by 3M NaOH solution. (b) Measurement setup. (c)-(e) Representative examples of current-voltage curves. The recording in (c) was obtained using the voltage ramp, $\gamma$, of $0.09$~V/s. (f) Temporal current response to a constant voltage level, $V_0$.  ``$V_0=0.6$ V $\rightarrow 0.2$ V'' indicates the measurement occurred at $V=0.2$ V after switching the device to its {\bf off} state by applying $V=0.6$ V. \label{fig:1}} 
\end{figure}

{\bf $I-V$ curves and switching mechanism.} The schematic diagram of an EGaIn tube memristor is shown in Fig.~\ref{fig:1}(a) (for the fabrication details, see the {\bf Methods} section). Within these devices, all switching operations take place entirely in the liquid segment, which consists of two EGaIn electrodes separated by a NaOH aqueous solution. The $I-V$ characteristics recorded using a source-measure unit (Fig.~\ref{fig:1}(b)) are consistently reproducible and free of noise, see Figs.~\ref{fig:1}(c)-(e). These $I-V$ curves are classified as binary unipolar. The switching is characterized by two voltage thresholds, $V_{off}\approx 0.3-0.5$~V and $V_{on}\approx 0.1-0.2$~V, with a hysteresis interval between $V_{on}$ and $V_{off}$. Although the shape of $I-V$ curves is somewhat similar to that of solid-state diffusive memristors~\cite{wang2017memristors,yoon2018artificial}, we emphasize an important difference: the direction around the loops is clockwise in our devices. This indicates that our devices start in a lower resistance state (LRS = {\bf on}) and switch to a higher resistance state (HRS = {\bf off}) when the voltage eclipses $V_{off}$. In the reverse direction, a return to the lower resistance state occurs when voltage reaches $V_{on}$.

These pinched, hysteretic $I-V$ features, including the values of $V_{off}$ and $V_{on}$, are maintained across a range of sweep rates of the applied voltage (Fig.~\ref{fig:1}(d)). At higher sweep rates, the $I-V$ traces no longer pinch at zero due to the presence of significant capacitive charging currents. Varying the amplitude of the sweep confirms that the hysteresis requires the applied voltage to surpass $V_{off}$ (Fig.~\ref{fig:1}(e)).

Fig.~\ref{fig:1}(f) shows the measured current through an EGaIn memristive device to different applied dc voltages. An important aspect of these responses is the power-law relaxation, suggesting a potential fractional-order effect~\cite{podlubny1998fractional}, similar to the behavior described by the Curie–von Schweidler law in capacitors~\cite{curie1888recherches,von1907studien}. Moreover, impedance measurements discussed below confirm the presence of pseudo-capacitance at intermediate frequencies ($\sim10-1000$ Hz). Nevertheless, in this work, we primarily concentrate on memristive phenomena, postponing the exploration of capacitive effects for subsequent investigations. Beyond their transient dynamics, these curves provide insight into the stability of the {\bf on} and {\bf off} states at specific potentials. For example, the $V_0=0.3$~V curve in Fig.~\ref{fig:1}(f)  (shown in blue) exhibits an abrupt transition at $t\approx 600$~s from the {\bf on} to the {\bf off} state, indicating that the {\bf on} state (at $V_0=0.3$~V) is metastable, while the {\bf off} state, recorded at $V=0.4$~V after increasing from $V=0.3$~V and at $V=0.2$~V upon returning from $V=0.6$~V, is stable. A metastable {\bf on} state will be associated with local minima in the Landau free energy in our model of EGaIn memristors (in the {\bf Memristive model} section below).


\begin{figure}[]
\centering
(a) \includegraphics[width=0.45\textwidth]{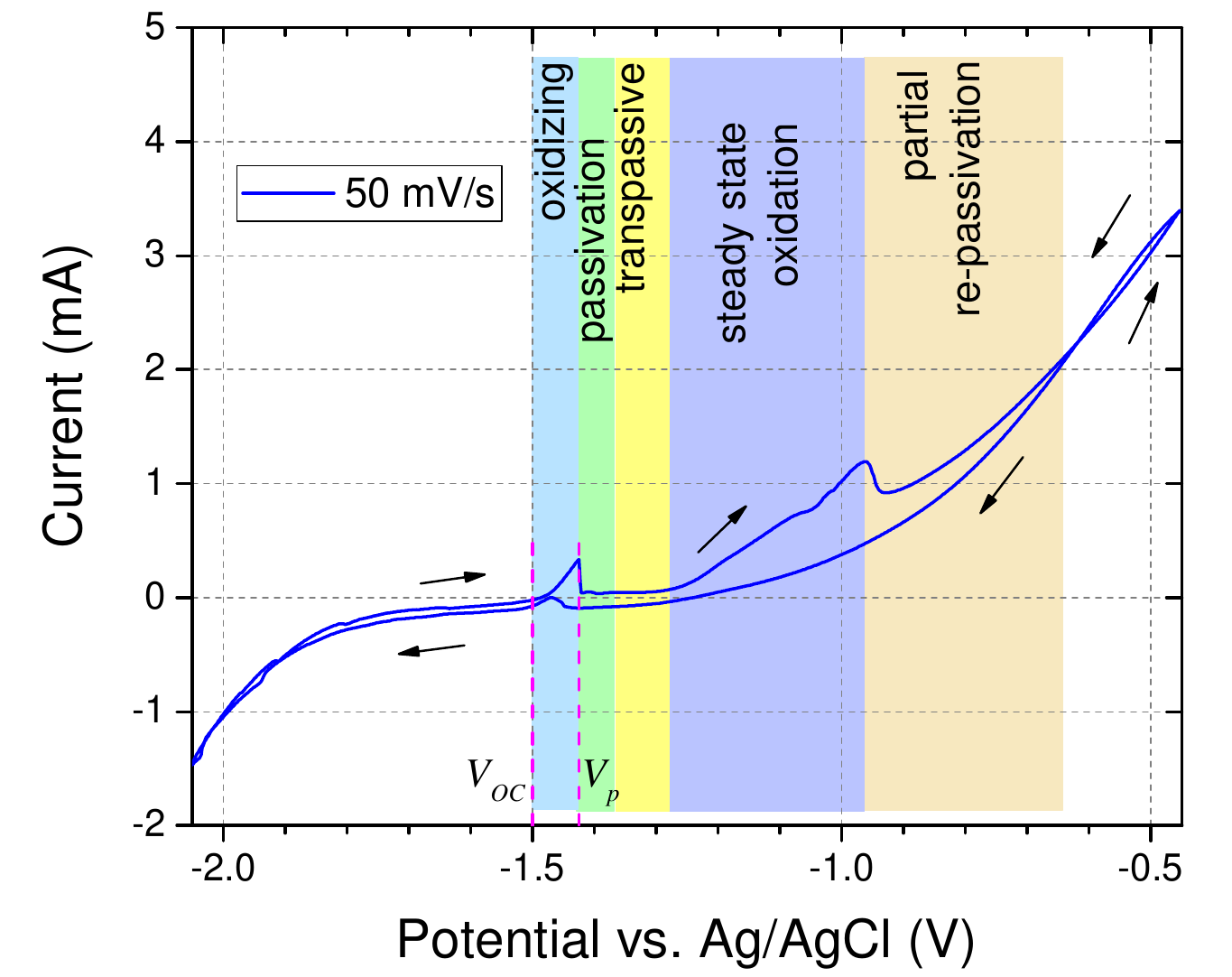} 
(b) \includegraphics[width=0.45\textwidth]{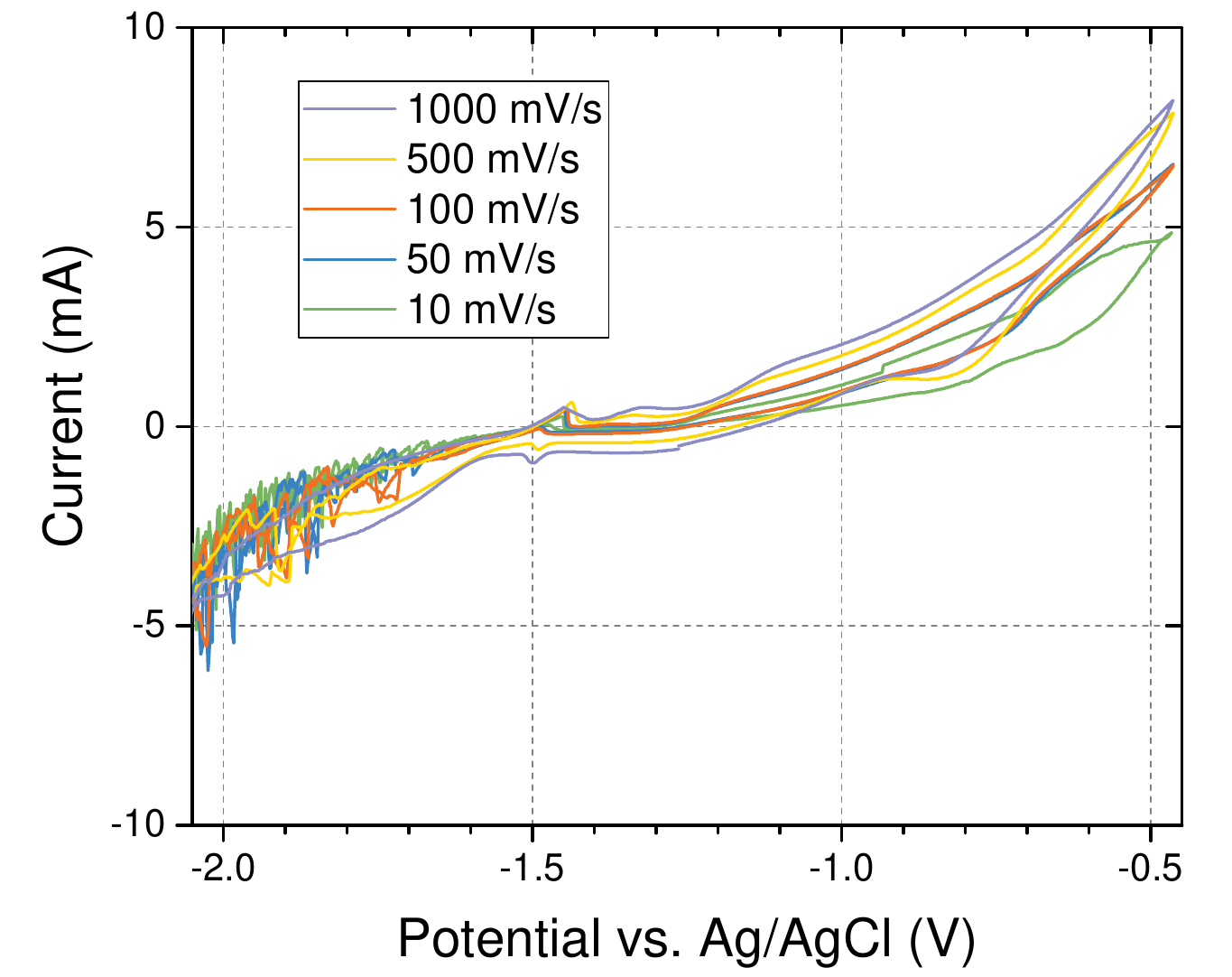}
\caption{half-cell EGaIn/NaOH current-voltage relationships.
(a) $I-V$ response measured in $1$ M NaOH using a scan rate of $50$ mV/s. The open circuit voltage ($V_{OC}$) and passivating potential ($V_P$) are denoted by the dashed lines the five different oxidation regions observed at positive potentials are shown. 
(b) Representative examples of half-cell $I-V$ curves in $3$ M NaOH at varying scan rates.
\label{fig:2andy}}
\end{figure}

Prior works describe the reversible electrochemical oxidation and reduction of EGaIn in basic solutions~\cite{khan2014giant,rashid2024role,hillaire2024interfacial}. Therefore, we hypothesized the symmetric, voltage-dependent resistive switching exhibited by an EGaIn tube memristor stems from the combination of opposite electrochemical activities of the two separate EGaIn/NaOH interfaces. Thus, we performed half-cell measurements on single EGaIn/NaOH interfaces to better understand the individual responses of a single interface at both positive and negative applied potentials. Fig.~\ref{fig:2andy}(a) shows a representative $I-V$ relationship obtained using a 3-electrode setup for a single EGaIn volume ($\sim20$~$\upmu$L) surrounded by $1$ M NaOH electrolyte. Fig.~\ref{fig:S2andy}(a) compares this response to the hysteretic $I-V$ response of an EGaIn half-cell containing 3M NaOH concentration. Fig.~\ref{fig:2andy}(b) and Fig.~\ref{fig:S2andy}(b) show how the voltage scan rate affects the $I-V$ relationship for $1$ and $3$ M NaOH, respectively. Details of this measurement method are provided in the {\bf Methods} section. 

These half-cell $I-V$ recordings agree well with previous measurements  made on EGaIn in NaOH at similar conditions~\cite{hillaire2024interfacial,rashid2024role} and they share some of the same features to the whole-cell memristor presented in Fig.~\ref{fig:1}.
For a single EGaIn/NaOH interface, the open circuit potential, $V_{OC}$, (measured with respect to a Ag/AgCl reference electrode saturated in $3$ M KCl), where quasi-static anodic current (i.e., where $\textnormal{d}V/\textnormal{d}t>0$) is $\approx 0$, occurs near a potential of $-1.5$~V (Fig.~\ref{fig:2andy}(a)). For $\textnormal{d}V/\textnormal{d}t>0$ and $V_{OC} < V < -1.41$~V, the measured current increases roughly linearly with voltage. In this \textit{oxidizing} region~\cite{hillaire2024interfacial,rashid2024role}, the increase in current corresponds to the electrochemical oxidation of Ga by hydroxide (OH$^-$) groups in the solution, which produces water soluble gallate (Ga(OH)$_4$$^-$) ions. The anodic current exhibits a local oxidation peak at a potential of $-1.44$ to $-1.41$ V, termed the passivation potential, $V_{P}$~\cite{hillaire2024interfacial,rashid2024role}. The stability of the location of $V_{P}$ across different scan rates indicates quasi-reversibility in the reaction.  

When $V>V_P$ (i.e., the \textit{passivation} regime), the measured current falls sharply due to the formation of an insulating anodic film, likely composed of water-soluble gallium oxyhydroxide (GaOOH) or GaO$_3$H$_y$ ($y<3$) facilitated by the adsorption of gallate ions~\cite{hillaire2024interfacial,rashid2024role}. At potentials near  $-1.35$ to $-1.39$ V, the interface enters the \textit{transpassive} regime where electrostatic attraction of OH$^-$ anions leads to solubilization of the GaOOH to the soluble Ga(OH$_4$)$^-$ form. This corresponds to a slight rise in the measured current. At around $-1.25$ V, the current slope again becomes linear with voltage, signifying the onset of \textit{steady-state oxidation} in which Ga(OH)$_3$ is formed. Depending on the potential scan rate and the NaOH concentration (Fig.~\ref{fig:S2andy}), a local peak is observed at potentials between $-1$ and $-0.8$~V, marking the transition to a \textit{partial-repassivation} region.

During the reverse sweep ($\textnormal{d}V/\textnormal{d}t<0$) at positive overpotentials ($V>V_{OC}$), the current takes a different path: it remains lower than that measured for $\textnormal{d}V/\textnormal{d}t>0$ across the same potential range. The arrows in Fig.~\ref{fig:2andy}(a) differentiate the paths of the $I-V$ relationship during forward and reverse sweeps. This hysteresis is due to the retention of the Ga(OH)$_3$ oxide layer on the surface that keeps the interface in a state of higher resistance. As $V$ approaches $V_{OC}$ during the reverse sweep, the interface exhibits either a re-oxidation peak (at low scan rates $\le100$ mV/s or a reduction peak (for scan rates $>100$ mV/s) at a potential between $-1.42$ and $-1.5$ V (Fig.~\ref{fig:2andy}(a)). This is likely a result of the finite diffusivity of oxidized species leaving (or staying) at the interface, which causes either re-oxidation or reduction, respectively, to occur. 

At negative overpotentials (where $V<V_{OC}$; the unshaded region in Fig.~\ref{fig:2andy}(a)), the EGaIn interface remains in a reduced state where current increases quadratically with increasing overpotential. Moreover, we observed bubble formation (presumably H$_2$ gas~\cite{hillaire2024interfacial,rashid2024role})\textemdash especially for the $3$ M NaOH case\textemdash on the surface of the EGaIn at sufficiently negative overpotentials and slow sweep rates, which created significant noise in the measured current of the half-cell configuration (Fig.~\ref{fig:2andy}(b)). More broadly, measurements in $3$ M NaOH produced larger oxidation and reductive currents than those in $1$ M NaOH (Fig.~\ref{fig:S2andy}). Additionally, $V_P$ occurs at $-1.45$ V in 3 M NaOH, versus appearing at $-1.42$ V for $1$ M NaOH. In both concentrations, sweep rates above $\sim100$ mV/s induced significant capacitive charging currents.

These data reveal key similarities and differences between the $I-V$ relationships of a single EGaIn/NaOH interface half-cell (Fig.~\ref{fig:2andy}) and of a two-interface, whole-cell tube memristor (Fig.~\ref{fig:1}). Both systems local current peaks at voltages more positive than $V_{OC}$, followed by a sharp reduction in current magnitude, a flattened current regime, and then a successive rise with continued increases in voltage. The shape of the $I-V$ hysteresis for each system is similar too; both exhibit lower net currents during the reverse sweep at positive potentials (provided the previous maximum potential was above the passivation potential). A re-oxidation peak (or reduction peak at high sweep rates) observed during the reverse sweep at potentials between $V_{OC}$ and $V_P$ are present in both recordings too.

However, the symmetric tube-memristor configuration causes the whole-cell response to be symmetric about $V_{OC}\approx0$ V, whereas behaviors of the half-cell are reported as the nominal electrochemical potential versus Ag/AgCl. Moreover, the relative voltages where the various $I-V$ features occur are significantly higher in the whole-cell device. For example, the passivation peak occurs at at overpotential of $\Delta V\equiv V-V_{OC}\approx 0.05-0.10$~V for the half-cell, versus current showing a sharp decline at a switching voltage of $\Delta V\approx 0.3-0.5$~V, depending on the device and NaOH concentration. The former potential difference represents the overpotential across a single EGaIn/NaOH interface, while the latter reflects the difference in overpotentials across both half-cell interfaces in the whole-cell memristor.

These comparisons suggest that an EGaIn tube memristor exhibits symmetric and hysteretic $I-V$ relationships as a result of alternating electrochemical responses at the two EGaIn surfaces, which occur at opposite voltage polarities. For $V>0$, the EGaIn interface ({\bf EGaIn1}) attached to the \textit{source} electrode of the voltage supply undergoes electrochemical oxidation and reduction, according to the responses seen in Fig.~\ref{fig:2andy}. Meanwhile, the EGaIn interface ({\bf EGaIn2}) attached to the \textit{ground} electrode of the voltage supply, which feels a net negative voltage difference compared to the NaOH, at a positive supply potential is held in a non-oxidizing state (unshaded region of Fig.~\ref{fig:2andy}(a)). This behavior flips when $V<0$: {\bf EGaIn1} is now held at in a non-oxidizing state, while {\bf EGaIn2} exhibits electrochemical oxidation and reduction according to the amplitude of the applied voltage. Thus, $V_{off}$ defined for the whole-cell memristor corresponds directly to $V_P$ from the half-cell recordings. These occur at different nominal values voltages because these two interfaces are coupled via the conductive NaOH electrolyte, which causes the total applied voltage between the source and ground to be dynamically divided across {\bf EGaIn1} and {\bf EGaIn2}.

\begin{figure}[]
\centering
(a) \includegraphics[width=0.2\textwidth]{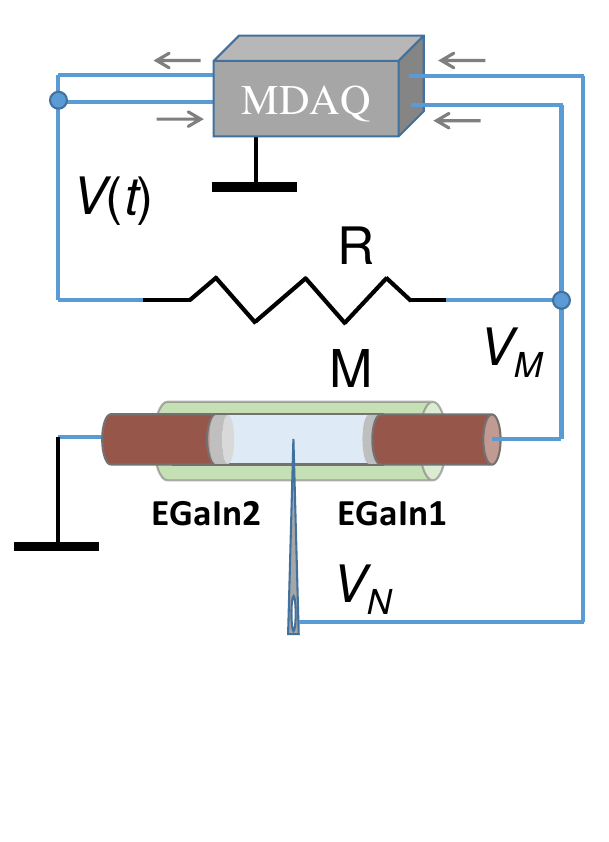}  \hspace{1cm}
(b) \includegraphics[width=0.45\textwidth]{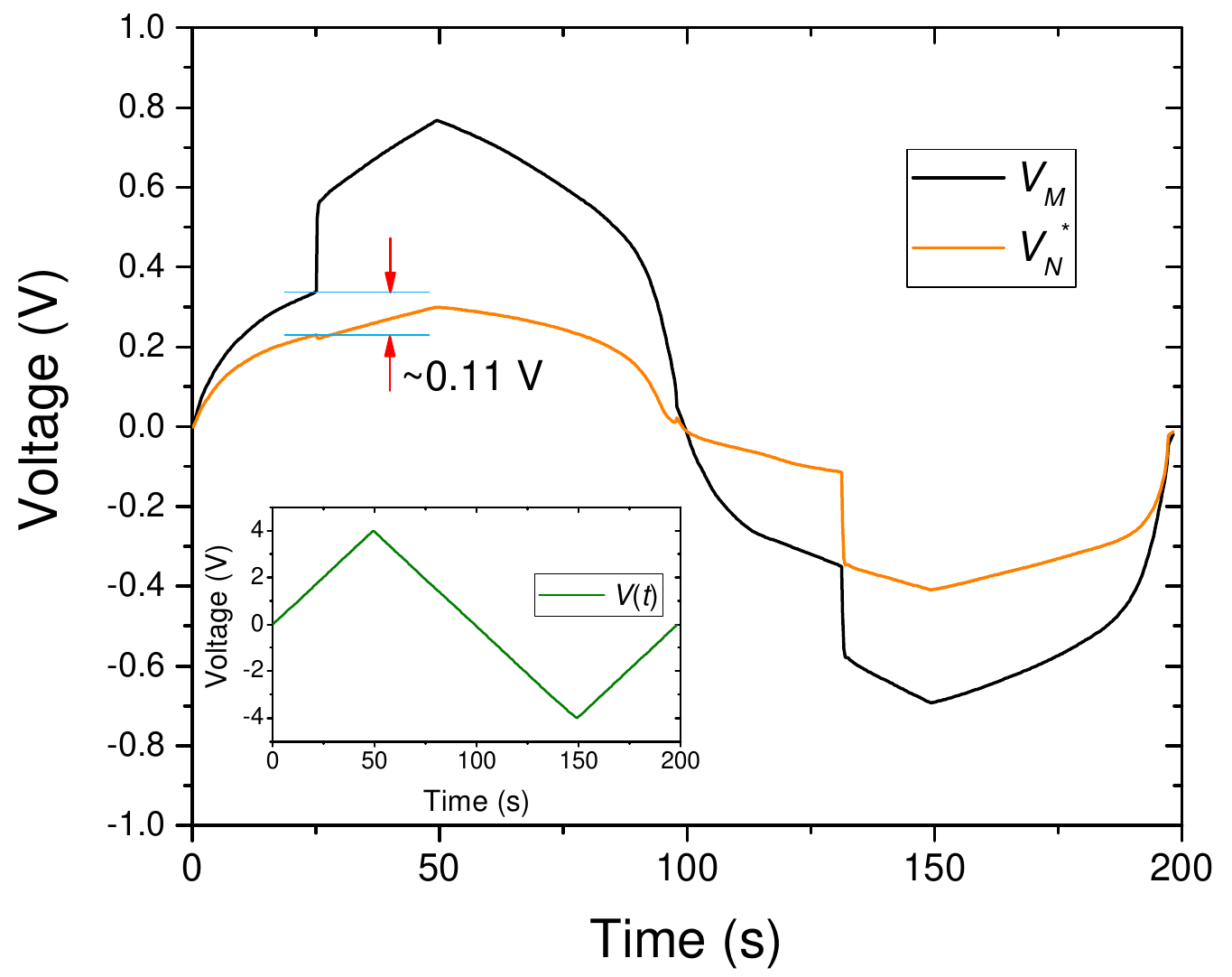} 
\caption{Three-terminal measurements of a tube memristor. (a) The measurement setup. To distinguish the separate contributions of the two electrodes to the signal, a sewing needle (acting as a pseudo-reference electrode) was inserted into the center of the tube. The measurements were performed using a multifunctional data acquisition unit (MDAQ) that was used to apply the driving signal $V(t)$ and read the response signals $V_M$ and $V_N$. The arrow's orientation shows if a signal is an input or an output. (b) Measured response to a triangular waveform (shown in the inset). Here, $V_N^*$ represents the needle voltage that has been adjusted by the open circuit needle voltage (approximately 1.17~V).
\label{fig:33}}
\end{figure}

This specific behavior was directly observed by inserting a pseudo-reference electrode into the NaOH solution between {\bf EGaIn1} and {\bf EGaIn2} (Fig.~\ref{fig:33}(a)). Here, a multi-channel data acquisition system was used to supply a $4$~V, $80$~mV/s triangular voltage waveform ($V(t)$) to a $10$~k$\Omega$ resistor wired in series with a tube-memristor. Both the voltage at {\bf EGaIn1} ($V_M$) and the potential at the middle electrode ($V_N$) were recorded (Fig. ~\ref{fig:33}(b)). During first quarter of the cycle ($V>0$ and $\textnormal{d}V/\textnormal{d}t>0$), $V_M$ and $V_N$ increase with increasing voltage. However, $V_M$ undergoes a sharp rise that occurs as it surpasses $\sim 0.33$~V. $V_N$ exhibits a much smaller decrease. The magnitudes and directions of these responses signify that {\bf EGAIn1}, which experiences a voltage roughly equal to $V_M-V_N$ undergoes an increase in effective resistance. Meanwhile, the small dip in $V_N$, is due to the fact that the resistance of {\bf EGaIn2} is now a smaller portion of the total resistance of the device. At negative potentials as $|V|$ increases, \textit{both} $V_M$ and $V_N$ exhibit sharp increases in their respective values when $|V_M|\approx0.33$ V. This shared response indicates that {\bf EGaIn2} undergoes a marked increase in resistance, causing both $V_M$ and $V_N$ to increase.

\subsection*{\bf Device Impedance}
\begin{figure}[]
\centering
(a) \includegraphics[width=0.35\textwidth]{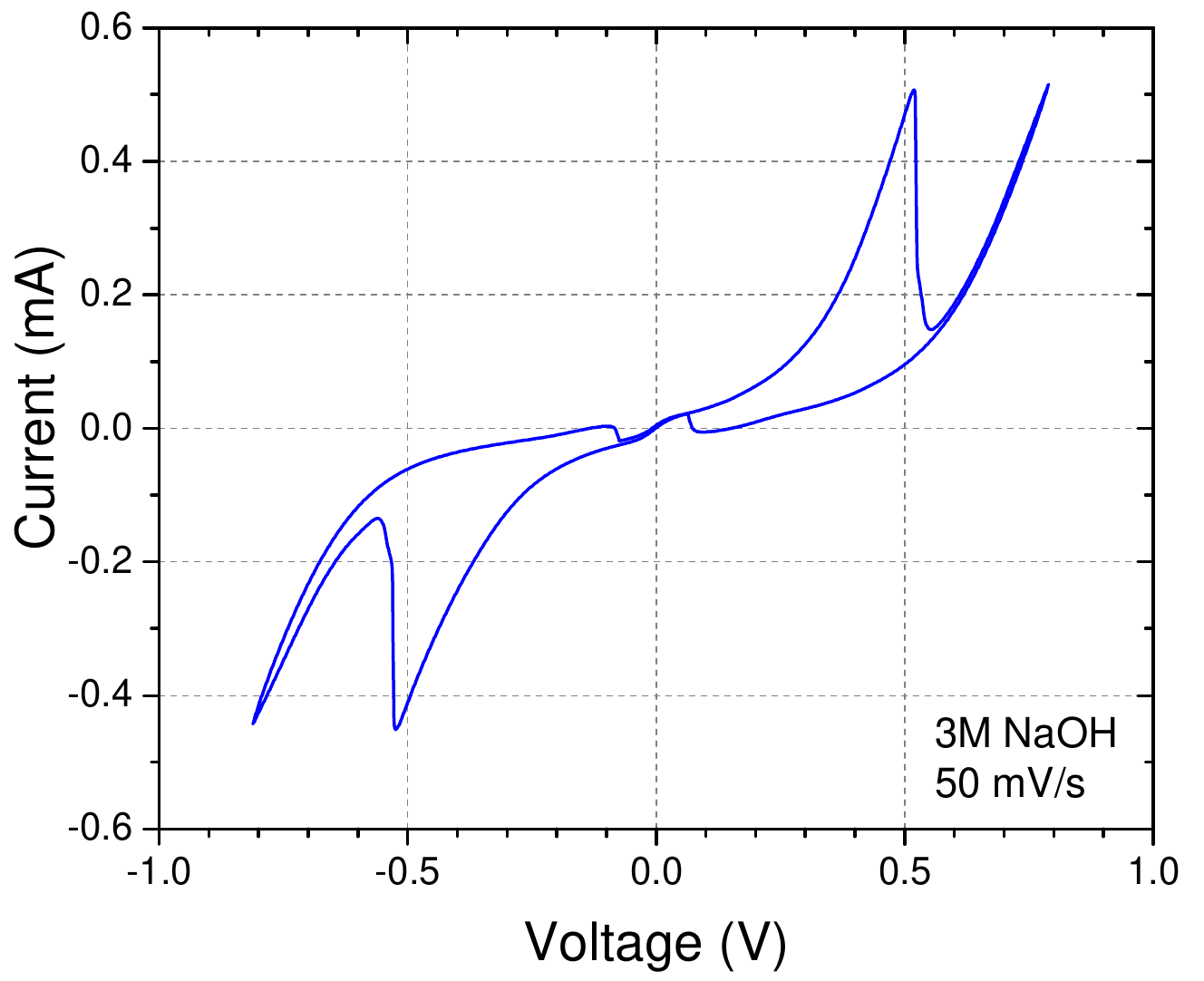}\\
(b) \includegraphics[width=0.45\textwidth]{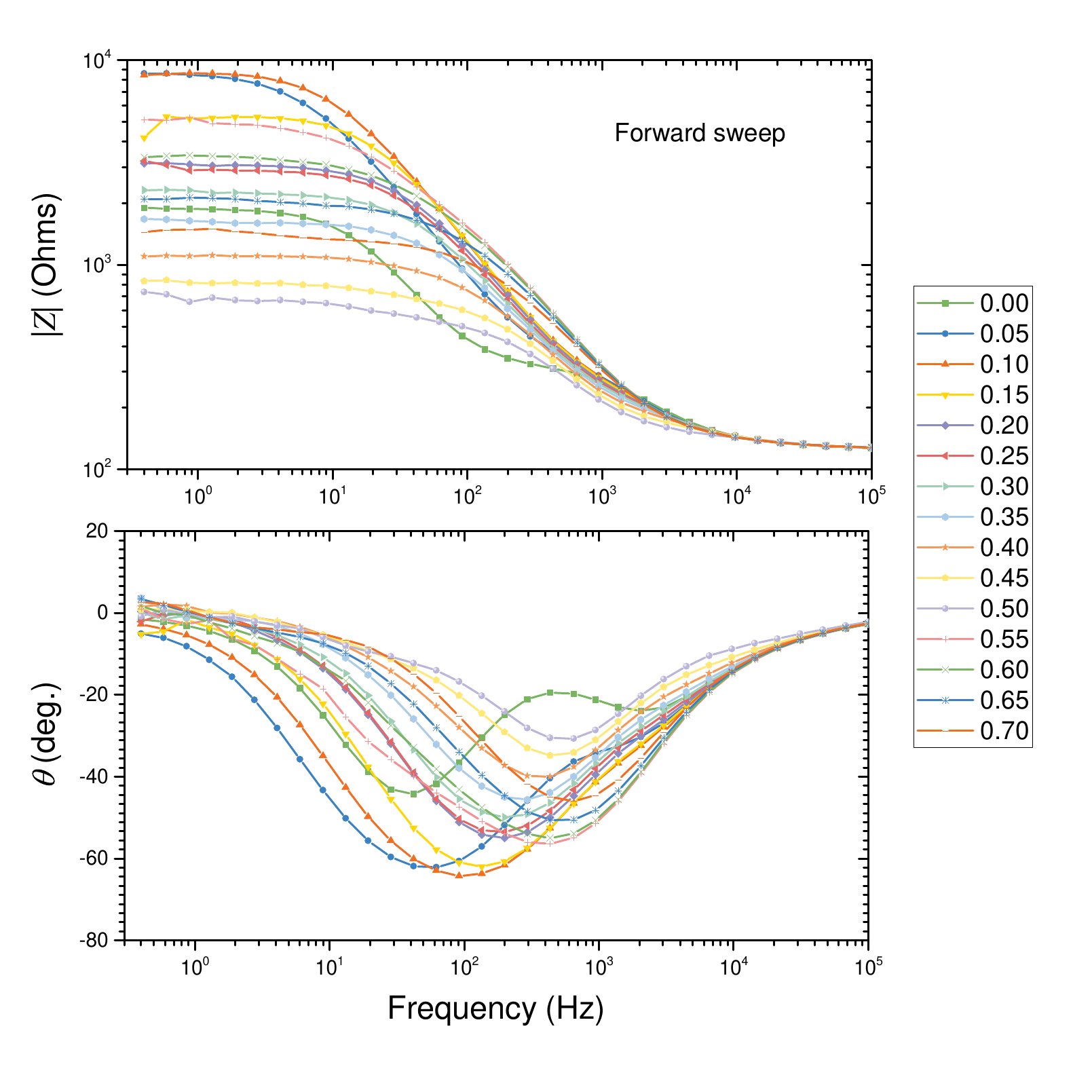} 
(c) \includegraphics[width=0.45\textwidth]{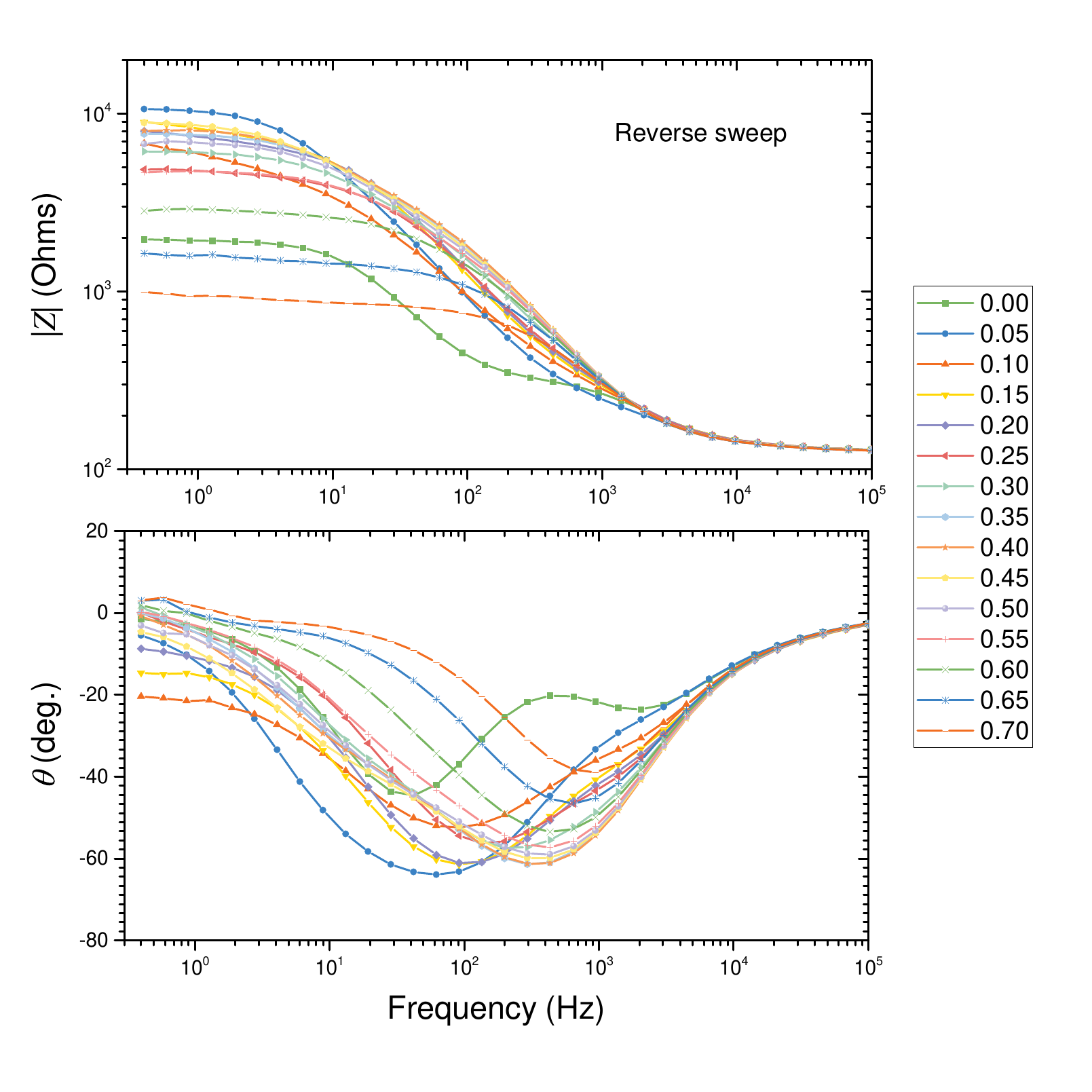}\\
\hspace{0.5in}(d) \includegraphics[width=0.3\textwidth]{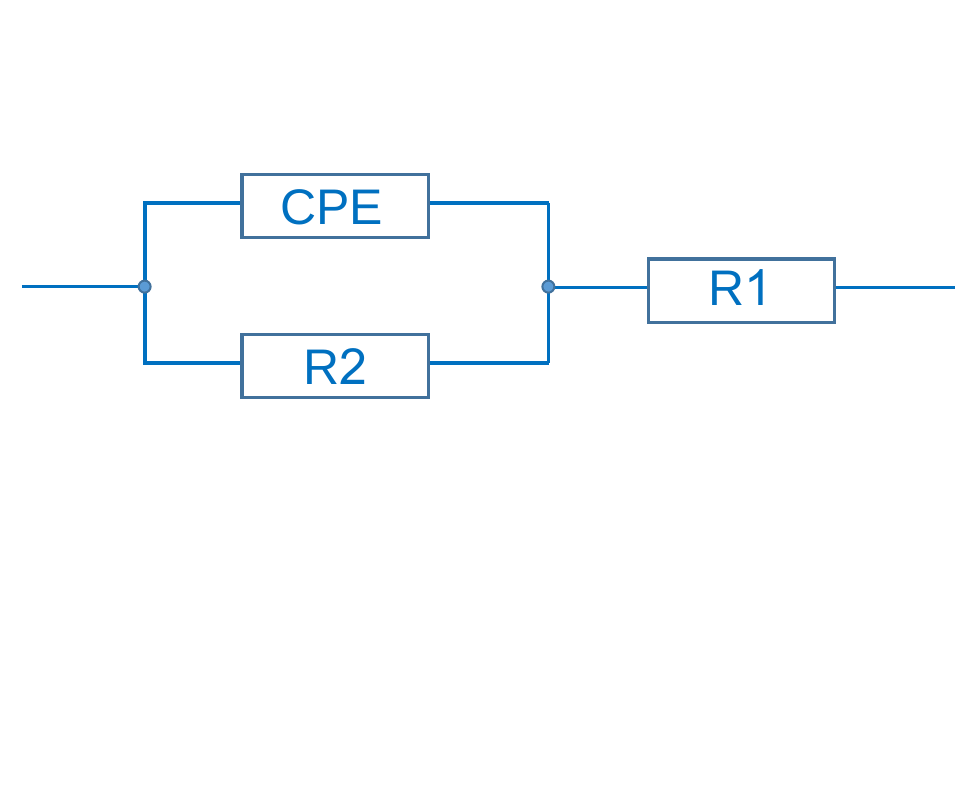} 
\hspace{0.5in}(e) \includegraphics[width=0.45\textwidth]{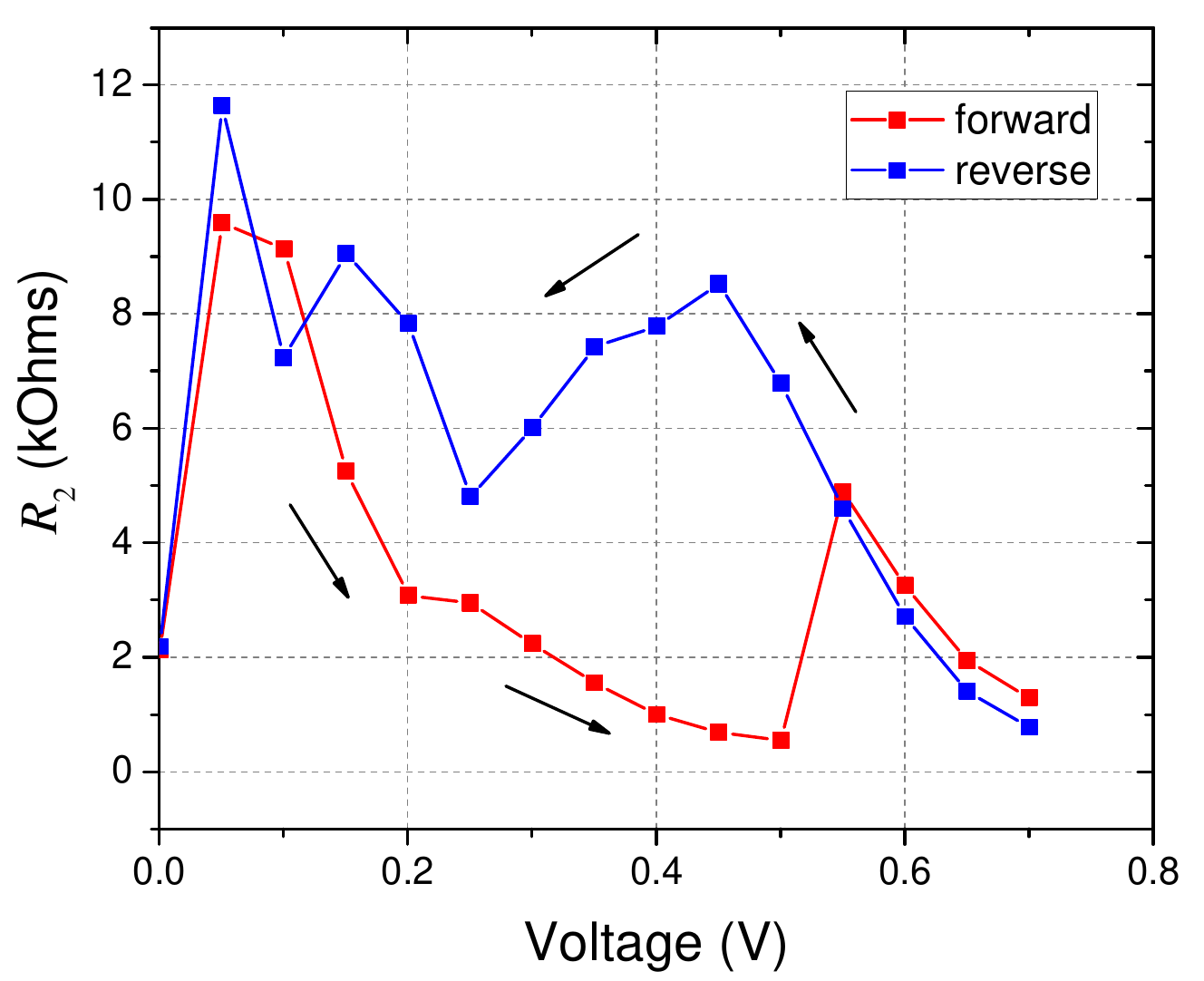}
\caption{Electrical impedance spectroscopy measurements of a whole-cell EGaIn tube memristor. (a) $I-V$ relationships from 2 consecutive cycles (sweep rate: $50$ mV/s) for a device containing 3 M NaOH. (b-c) Electrical impedance magnitude (top) and phase (bottom) versus frequency during sequential voltage increases (b) and decreases (c) between $0$ and $+0.7$ V. (d) Equivalent circuit used to fit the EIS data. (e) Low-frequency resistance ($R_2$) versus voltage during quasi-state forward and reverse sweeps.
\label{fig:4andy}}
\end{figure}

Electrical impedance spectroscopy (EIS) measurements were performed on whole-cell EGaIn tube memristors to quantify changes in device impedance at various dc biases. The corresponding $I-V$ curve of one representative device is shown in Fig.~\ref{fig:4andy}(a). Fig.~\ref{fig:4andy}(b)-(c) shows how the impedance magnitude (top) and phase (bottom) of this same device vary with frequency for both forward and reverse bias sweeps from between $0$ and $+0.7$~V. The impedance spectra reveal that the tube memristor is not simply a resistive device, but includes psuedo-capacitance and sometimes inductive components in its responses. Fig.~\ref{fig:andyS1} provides Nyquist representations of these same data, which are similar to those obtained on single EGaIn interfaces at positive potentials relative to $V_{OC}$ by Hillaire, et al.~\cite{hillaire2024interfacial}. However, our measurements combine changes in impedance at both EGaIn interfaces.

Similar to the $I-V$ relationships which show that dc current returns to zero as the voltage completes a full half-cycle, the EIS data show reversible changes in the impedance spectra. At dc potentials very close to $0$~V, EIS measurements reveal the presence of two different resistor-capacitor (RC) pairs (see the two separate dips in phase angle versus frequency). However, at non-zero dc voltages, the memristors exhibit only a single RC time constant and are well approximated by the equivalent circuit shown in Fig.~\ref{fig:4andy}(d). In this simplified circuit model, a constant phase element (CPE, or psuedocapacitance) and $R_2$ are used to describe the electrochemical state of the EGaIn/NaOH interface, while $R_1$ represents the ionic resistance of the NaOH solution. For this particular device, $R_1$ is approximately $130$~$\Omega$. More generally, $R_1$ is expected to increase with increasing tube length, decreasing tube diameter, or decreasing NaOH concentration. The larger valued $R_2$ parameter determines the low-frequency, or quasi-static, resistance of the device that affects the dc $I-V$ relationship.

A nonlinear least-squares fitting routine (see methods) was used to extract estimates for $R_2$ versus the dc bias (Fig.~\ref{fig:4andy}(e)). Matching the trend of the low-frequency impedance magnitude shown in Fig.~\ref{fig:4andy}(b)-(c), these data show that $R_2$ exhibits a hysteretic path with changes in voltage. Apart from the increase in $R_2$ observed between $V=0$ and $V=0.1$V, $R_2$ steadily-decreases from $\sim 10$~k$\Omega$ to $<1$~k$\Omega$ with increasing voltage until reaching $V=V_{off}\approx0.5$~V, where the memristor switches from {\bf on} to {\bf off}. At this location, $R_2$ sharply rises to approximately $5$~k$\Omega$. Further increases in $V$ cause $R_2$ to again reduce below $2$~k$\Omega$. Subsequently, decreasing the dc bias below $+0.7$~V causes the value of $R_2$ to increase steadily above $8$~k$\Omega$ by $V=0.4$~V, where it remains until falling back to the starting value of $2$~k$\Omega$ for $V<0.05$~V. This hysteretic path identifies the net resistance of the device as voltage changes and quantifies how much $R_2$ changes as each EGaIn interface undergoes electrochemical oxidation and reduction. For this device, a maximum $R_{off}/R_{on}$ resistance ratio of $8-10$ occurs at a voltage just below $V_{off}$. Given the symmetry of the $I-V$ curves, similar changes in device resistance are found at negative potentials too. Fitting the EIS spectra also revealed that the constant phase element changes hysteretically with voltage; details of this psuedo-capacitive switching and memory will be discussed in a separate publication.

\begin{figure}[]
\centering
 \hspace{1.5cm} \hspace{1cm}
(a) \includegraphics[width=0.22\textwidth]{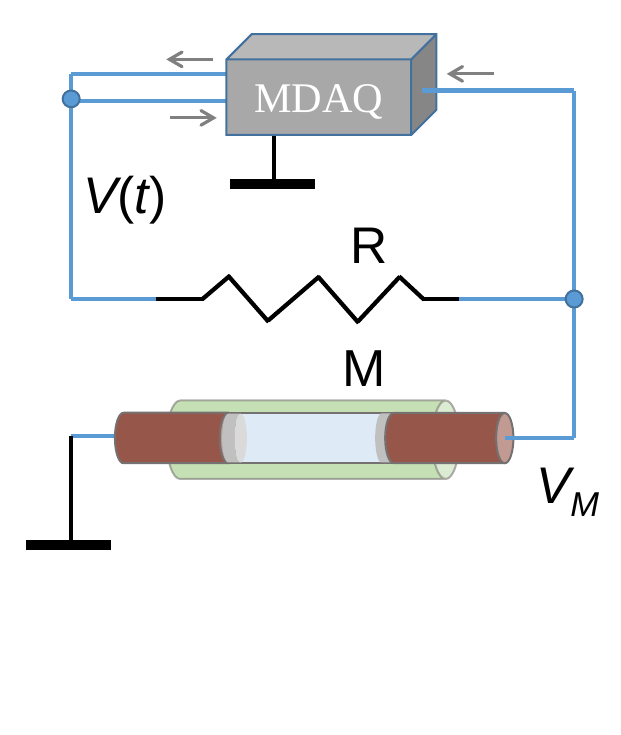}  \hspace{1cm}
(b) \includegraphics[width=0.45\textwidth]{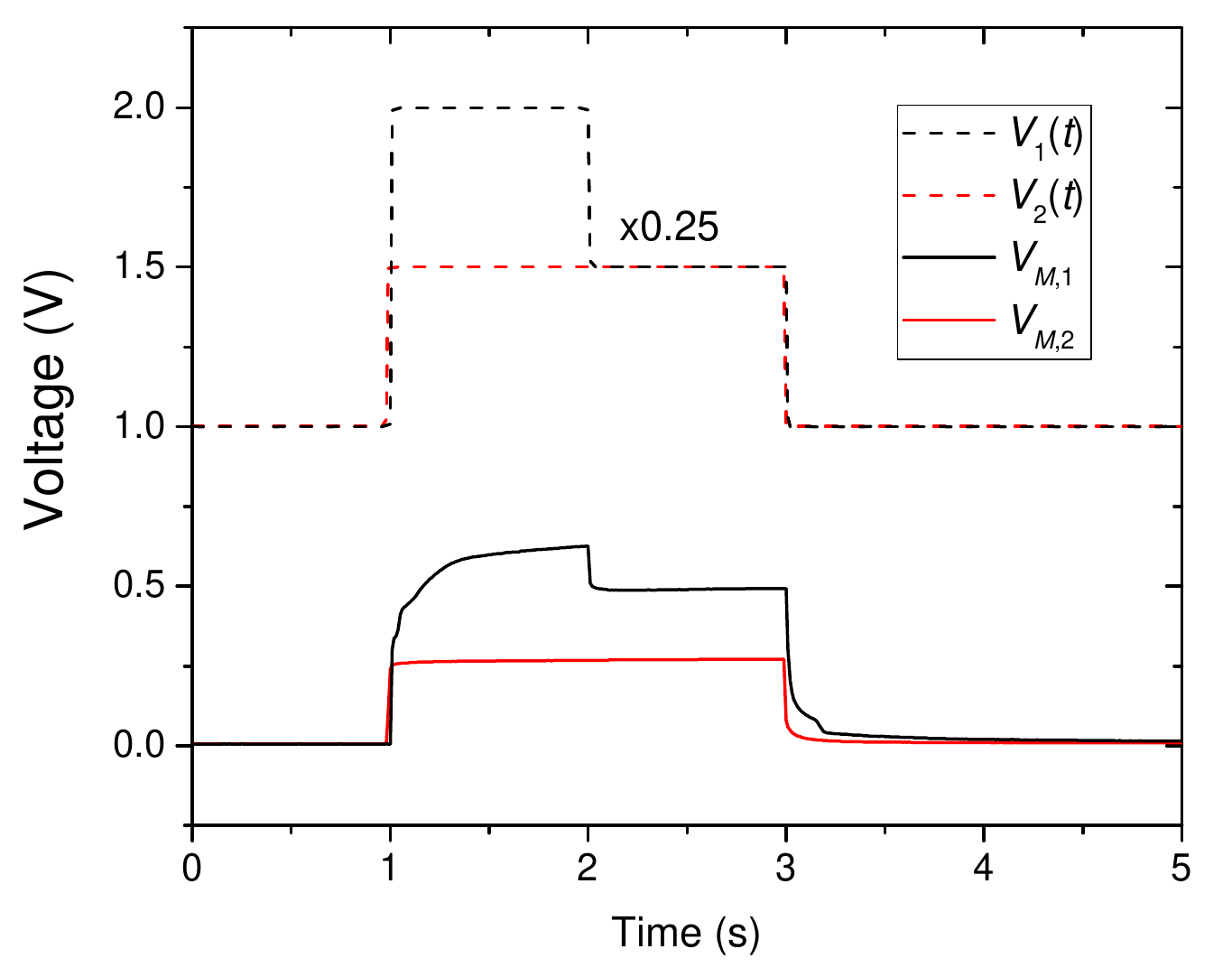} \\
(c) \includegraphics[width=0.45\textwidth]{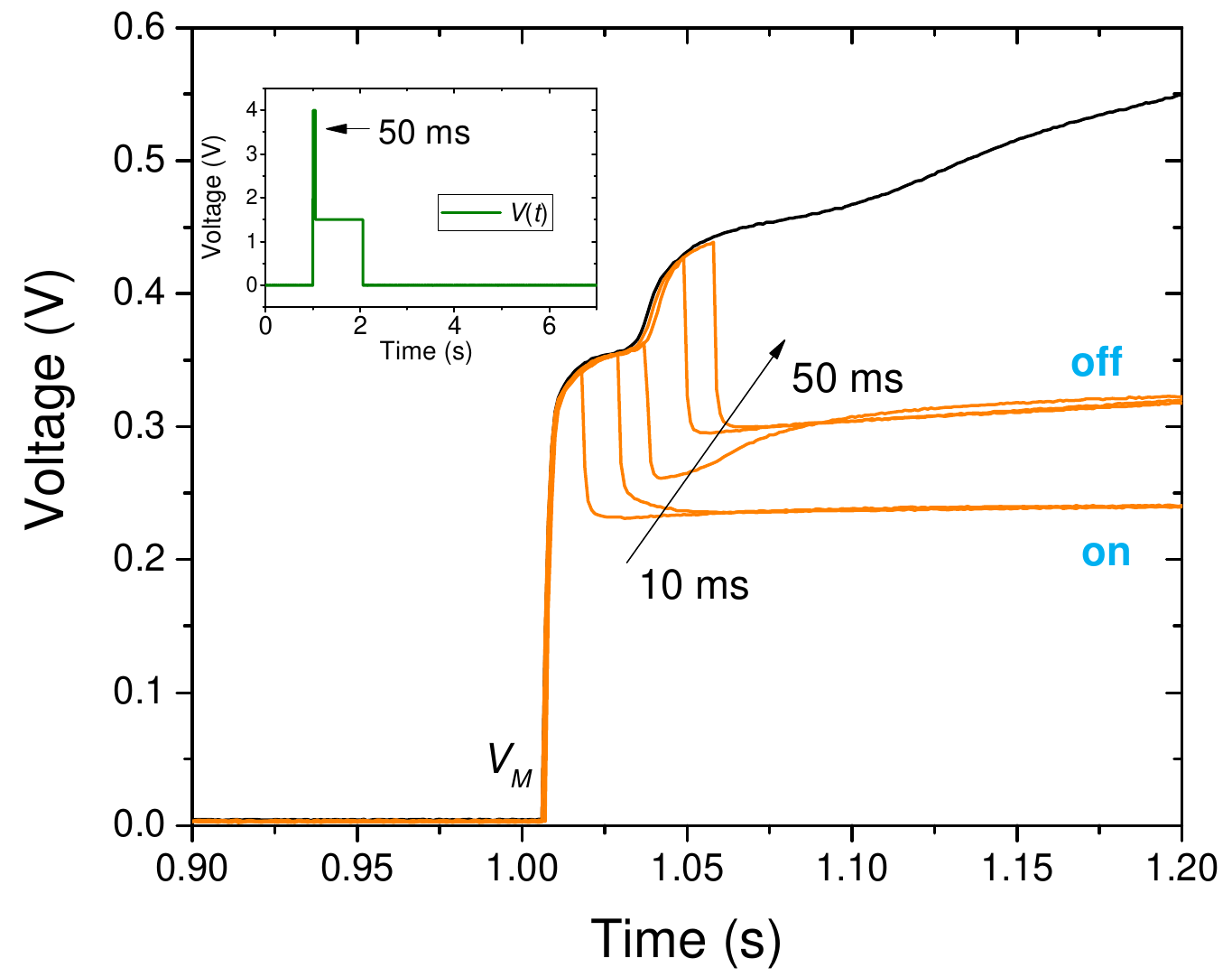}
(d) \includegraphics[width=0.45\textwidth]{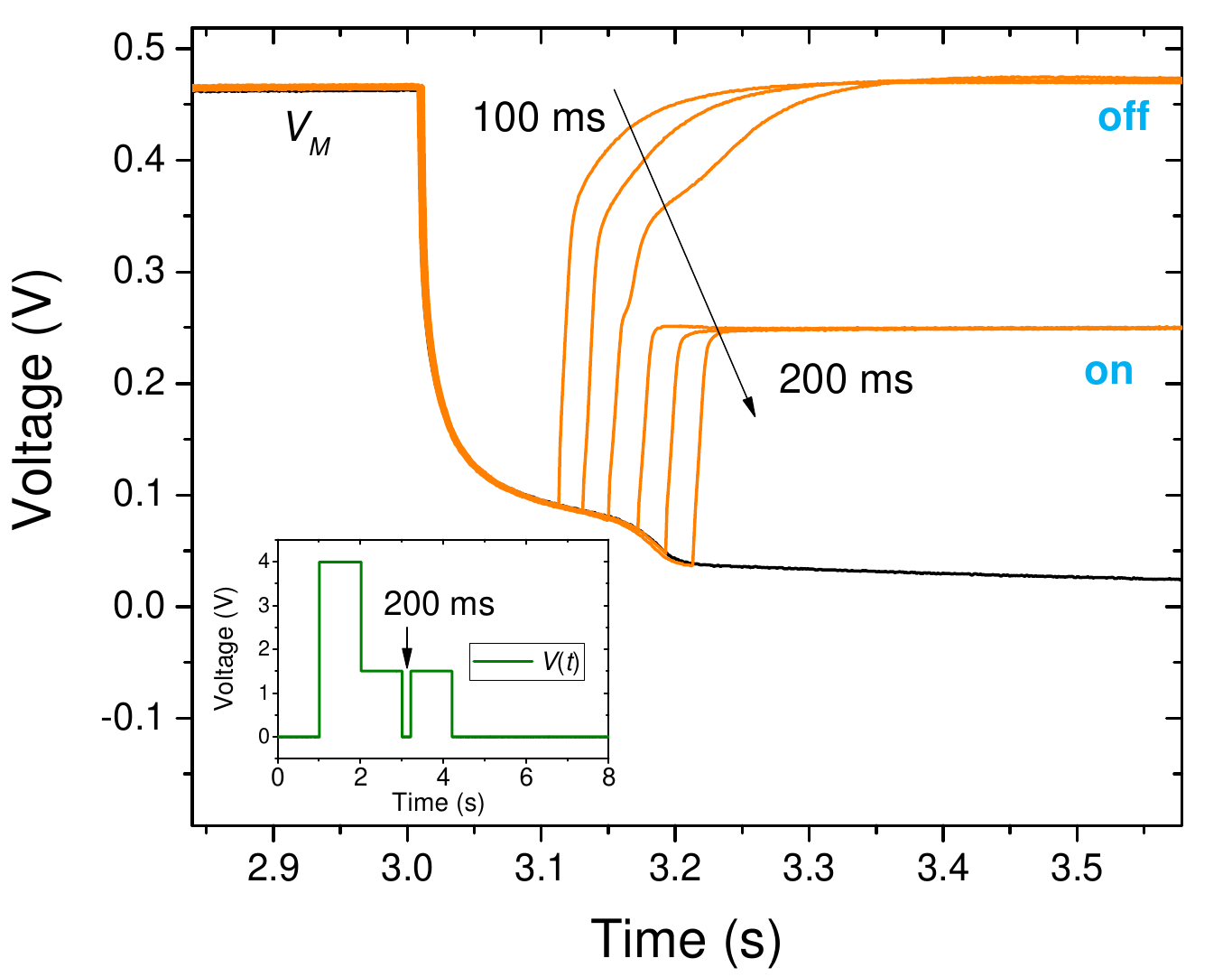} \\
(e) \includegraphics[width=0.45\textwidth]{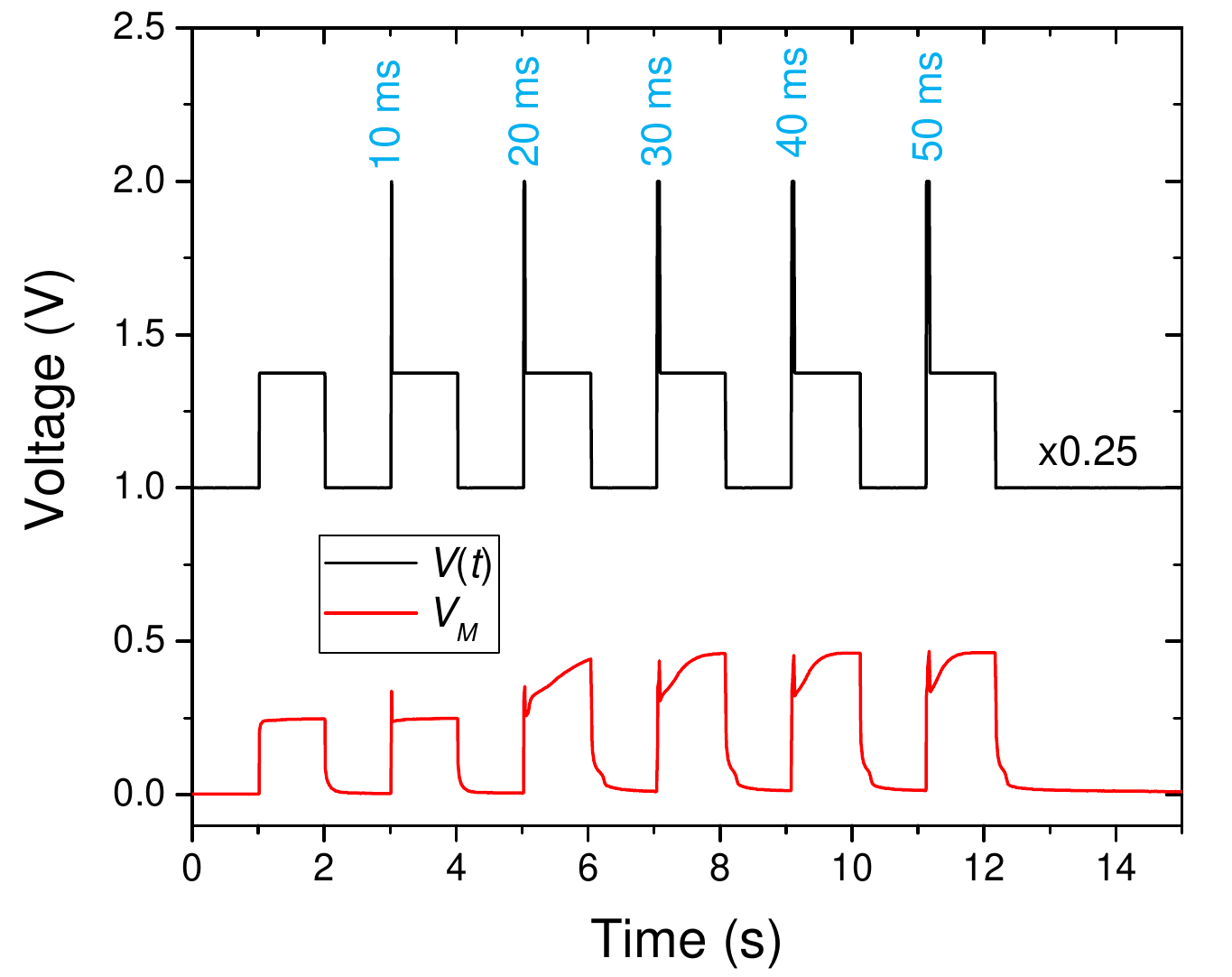}
(f) \includegraphics[width=0.45\textwidth]{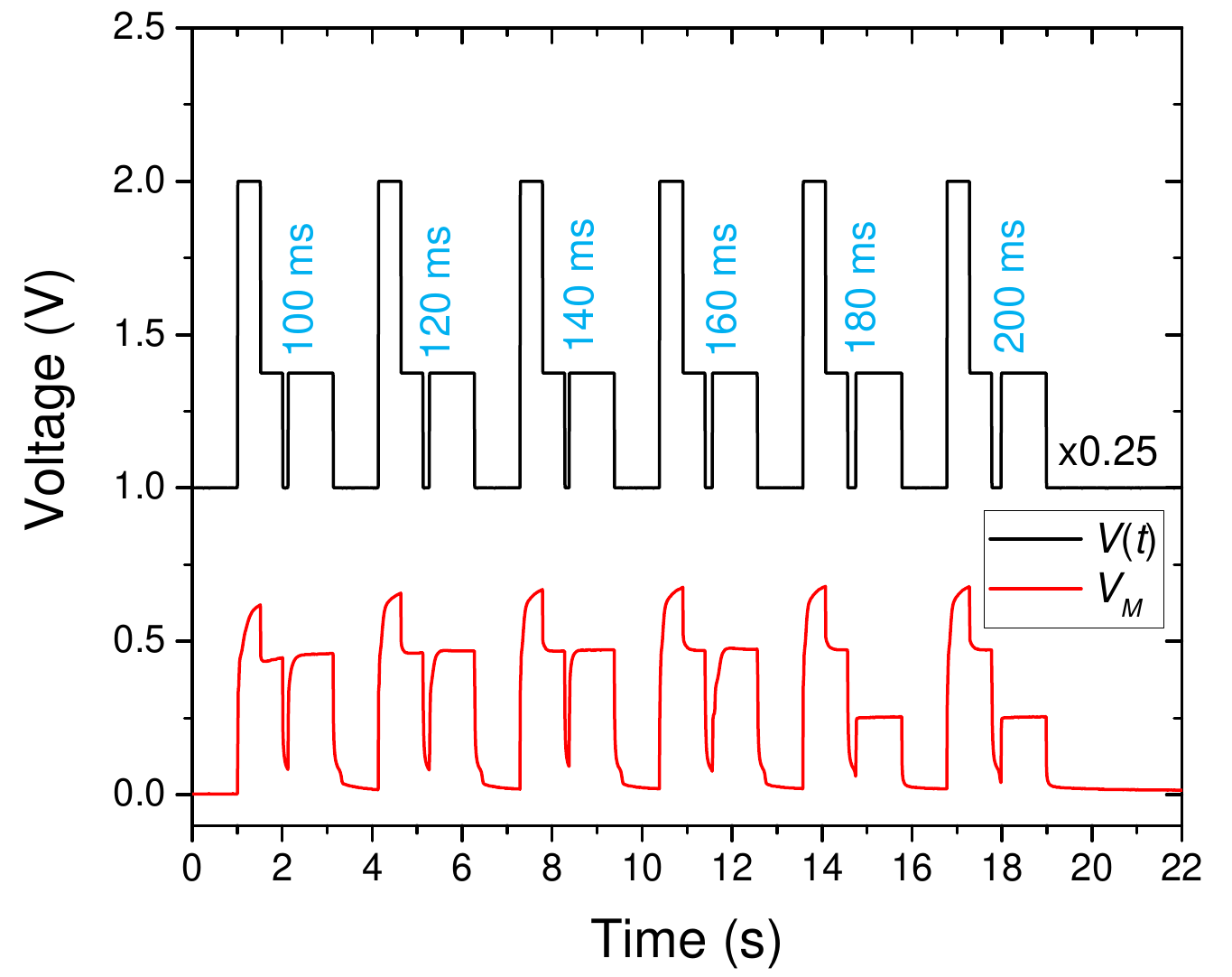}
\caption{ Pulse control of EGaIn memristive devices. In these measurements, we used $R=10$~k$\Omega$. The source voltage curves in (b), (e) and (f) were scaled and displaced by 1~V for clarity. \label{fig:3}}
\end{figure}

To understand the time constants associated with the switching process, square voltage pulses were used on a resistor-memristor circuit (refer to Fig.~\ref{fig:3}(a)). As depicted in Fig.~\ref{fig:3}(b), when the pulse amplitude is less than $V_{off}$, the voltage across the memristor closely mimics the form of the input voltage, with the exception of a minor decaying tail. When the applied voltage exceeds $V_{off}$, a shoulder-like step appears, somewhat similar in shape to 0.7 anomaly in quantum point contacts (but, definitely, of a different origin)~\cite{bauer2013microscopic}, followed by a region where the voltage rises gradually (refer to Fig.~\ref{fig:3}(c) for a detailed structure). Similarly, as the voltage drops to zero, there is a shoulder-like drop followed by a non-exponential extended tail (refer to Fig.~\ref{fig:3}(d) for a detailed structure). These rising (falling) shoulder-like steps correspond to the creation (dissolution) of the oxide film at the liquid metal surface.

To determine the reset time (the transition from {\bf on} to {\bf off}), we altered the duration of a high-amplitude pulse, followed by the hold voltage, as shown in the inset of Fig.~\ref{fig:3}(c). According to our measurements, the reset time is approximately $25$~ms, and can be reduced to below $20$~ms by using a higher amplitude voltage pulse (refer to Fig.~\ref{fig:3}(e)). As shown in Fig.~\ref{fig:3}(d), the set time is relatively longer, approximately $150$~ms. We emphasize that the measurements in Fig.~\ref{fig:3}(c) and (d) correlate the points for maximum curvature (in the ``shoulder'' regions) with the transitions between states {\bf on} and {\bf off}.

\subsection*{\bf Memristive model.}
Existing memristive models, including the VTEAM~\cite{Kvatinsky2013} and others, do not capture the binary nature of the $I-V$ characteristics of the EGaIn memristive devices. We propose a model inspired by concepts from the Landau phase transition theory~\cite{arovas2019lecture}. Consider a free energy 
\begin{equation}
    F(x,V_M)=\frac{1}{2}ax^2+\frac{1}{4}bx^4+\kappa (|V_M|-V_0)x, \label{eq:FE1}
\end{equation}
where $a<0$, $b>0$, $\kappa>0$, and $V_0>0$ are constants, $V_M$ is the applied device voltage, and $x$ is the order parameter. When $|V_M|=V_0$, the free energy in Eq.~(\ref{eq:FE1}) exhibits a $\mathbb{Z}_2$ symmetry, indicating that there exist two equivalent phases, corresponding to the {\bf on} and {\bf off} memristive states, within the hysteresis region. This symmetry breaks down when $|V_M|$ deviates from $V_0$, favoring one of the states. As the deviation continues to rise, a bifurcation is induced, resulting in only one energy minimum remaining ({\bf on} or {\bf off} state).

We model the dynamic switching behavior as a relaxation process, 
$\textnormal{d} x/ \textnormal{d} t\sim -\Gamma\; \partial F(x,V_M)/ \partial x$, where $\Gamma$ is the relaxation rate that may depend on the voltage. Substituting Eq.~(\ref{eq:FE1}), one obtains the equation of motion
\begin{equation}
    \frac{\textnormal{d} x}{\textnormal{d} t}=-\Gamma \left( -x+bx^3+\kappa(|V_M|-V_0)\right) \label{eq:FE2}
\end{equation}
where some of the constants were re-normalized.
Additionally, the memristive response is described by a generalized Ohm's law
\begin{equation}
    I=\left(G_{on}(V_M)f_{on}(x)+G_{off}(V_M)(1-f_{on}(x))\right)V_M,\label{eq:FE3}
\end{equation}
where $f_{on}(x)\in [0,1]$ is a function that describes the contribution of the  {\bf on}-state conductivity $G_{on}$ to the total conductivity at a particular value of the order parameter $x$.
 It is important to recognize that this model accommodates transient states beyond binary. Nevertheless, the system internally stabilizes into a binary state under zero or a finite constant voltage. 
 Overall, Eqs.~(\ref{eq:FE2}) and (\ref{eq:FE3}) describe a voltage-controlled first-order memristive system~\cite{chua76a}.

\subsection*{\bf In-memory computing.}
Neuromorphic and reservoir computing, access devices, and hardware security are among the most studied applications for volatile memristors~\cite{Wang20a,Kim22a}. In contrast, we present experimental evidence of in-memory computing using our EGaIn memristive devices. Their inherent bi-stable states make them quite suitable for storing Boolean data. Traditional memristive devices have been shown to be well suited for implementing material implication logic gates~\cite{Borghetti2010}. Moreover, volatile memristor emulators (with diffusive memristor hysteresis) were used to demonstrate material implication logic gates and their inverse~\cite{8480647}. 

\begin{figure}[]
\centering
(a)  \includegraphics[width=0.60\textwidth]{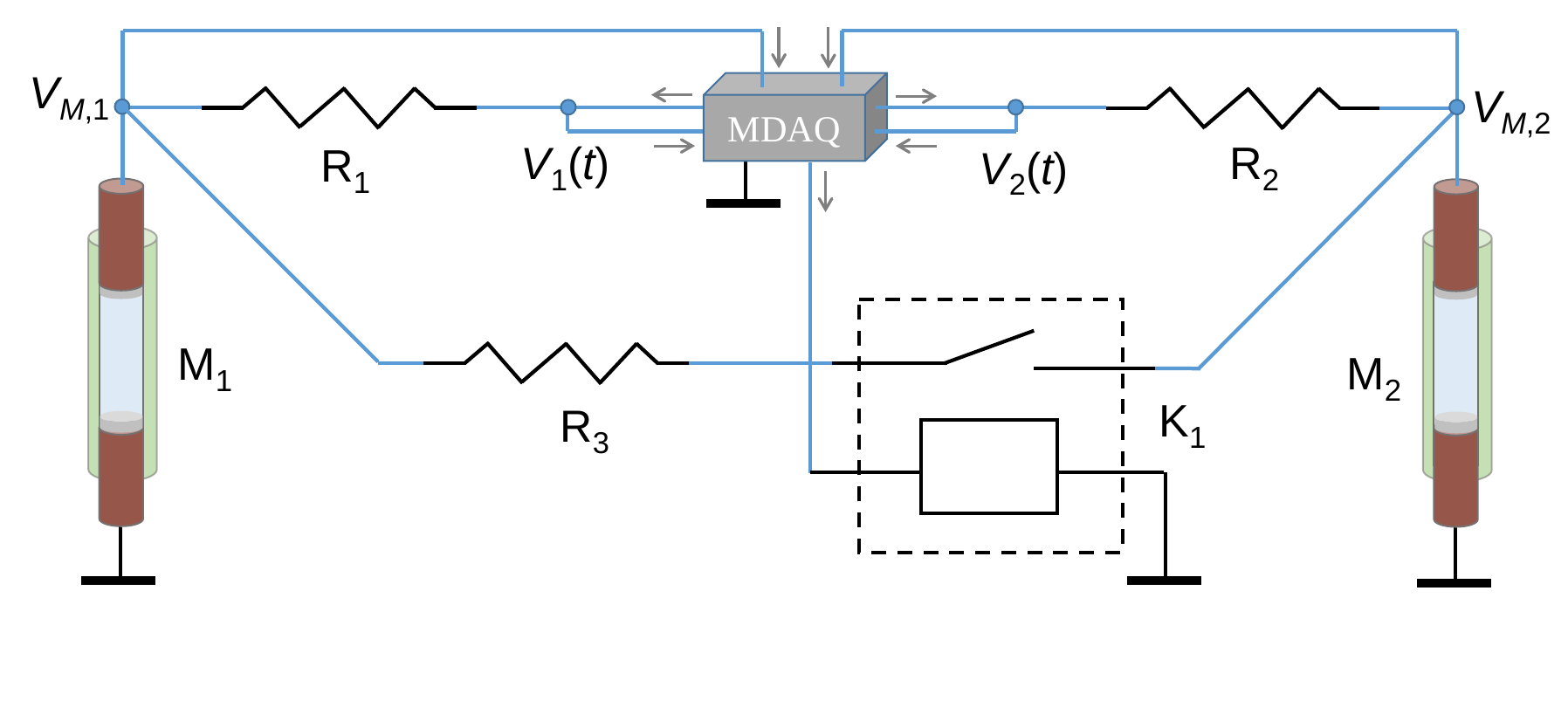} \\
(b) \includegraphics[width=0.45\textwidth]{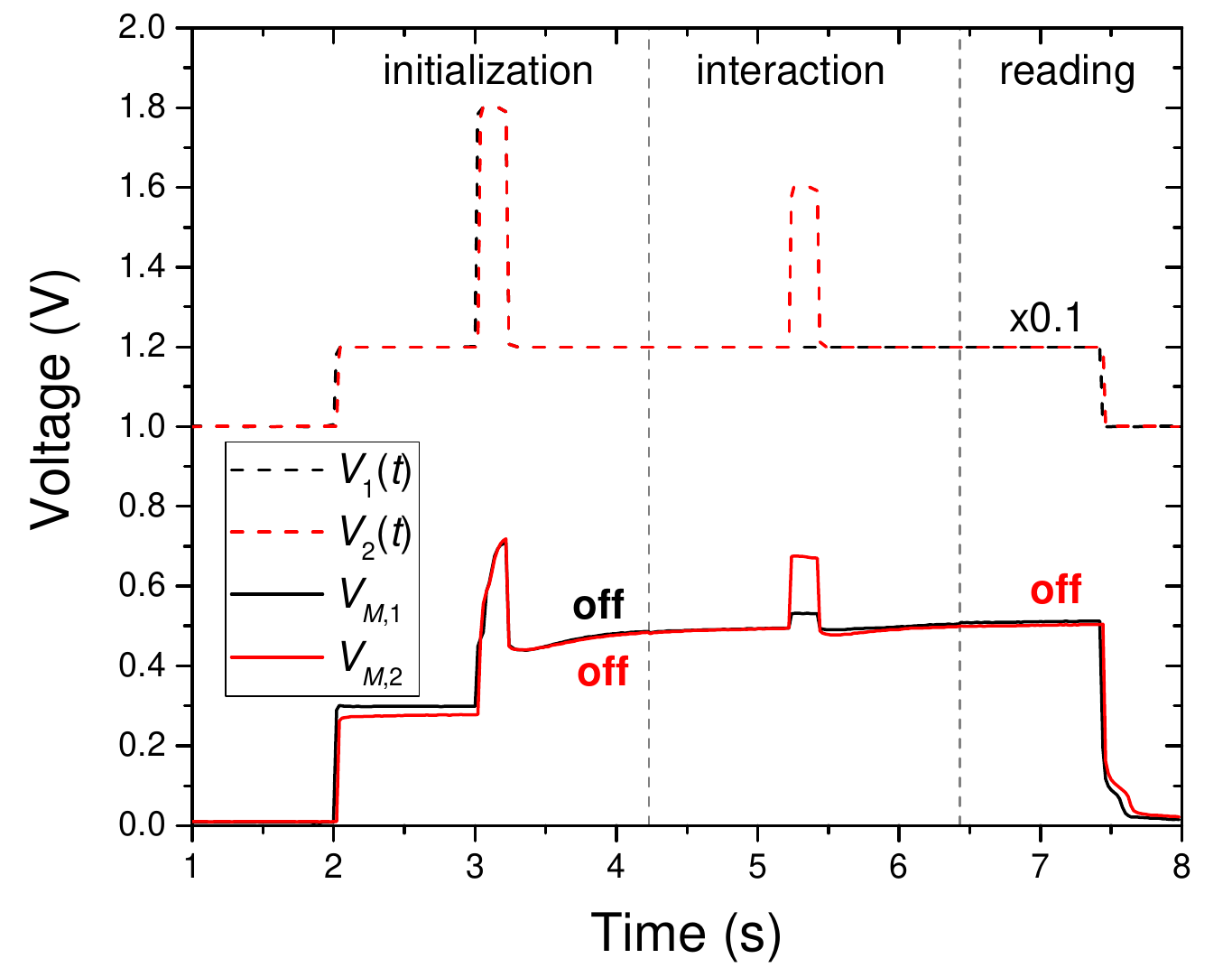} 
(c) \includegraphics[width=0.45\textwidth]{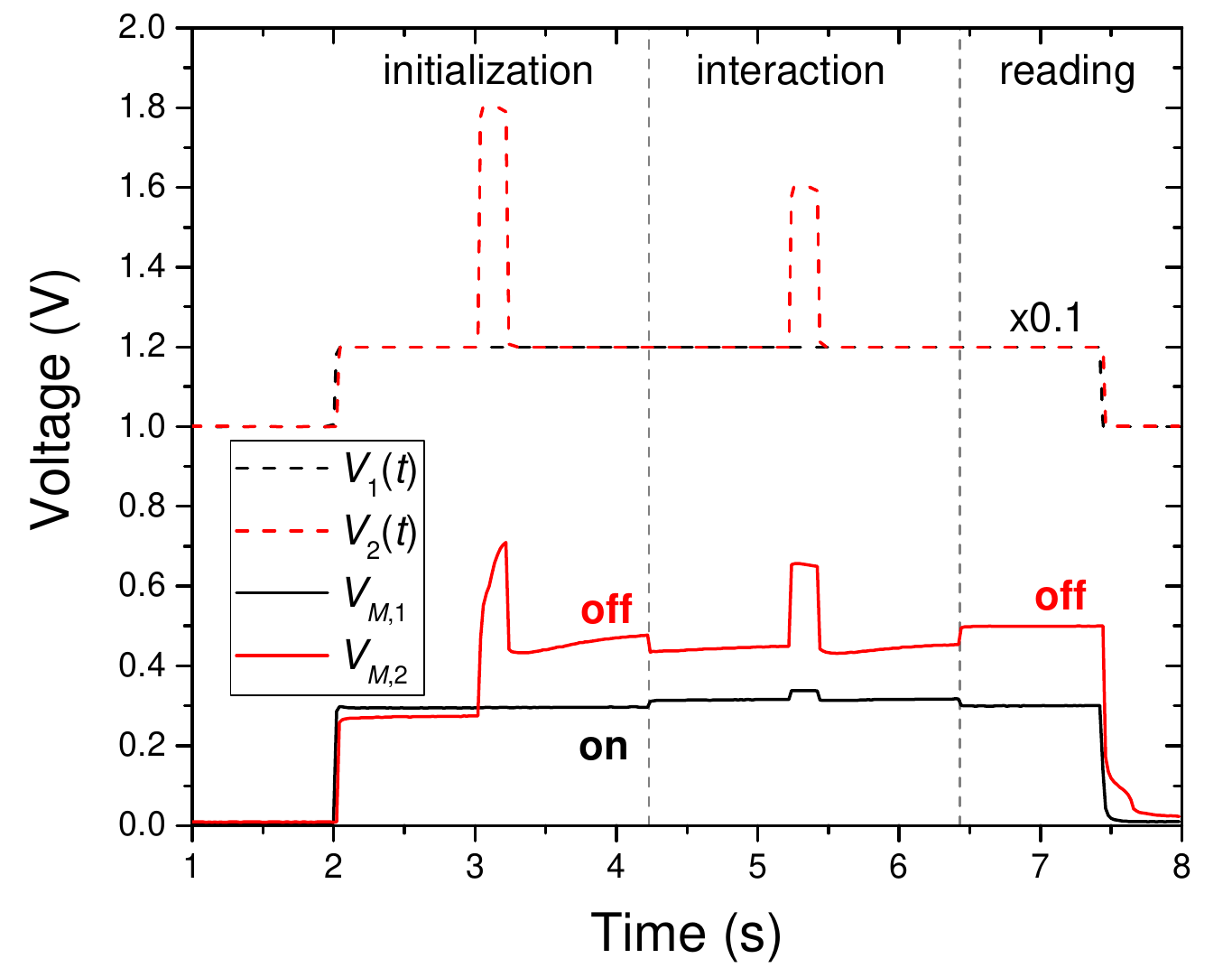}\\
(d) \includegraphics[width=0.45\textwidth]{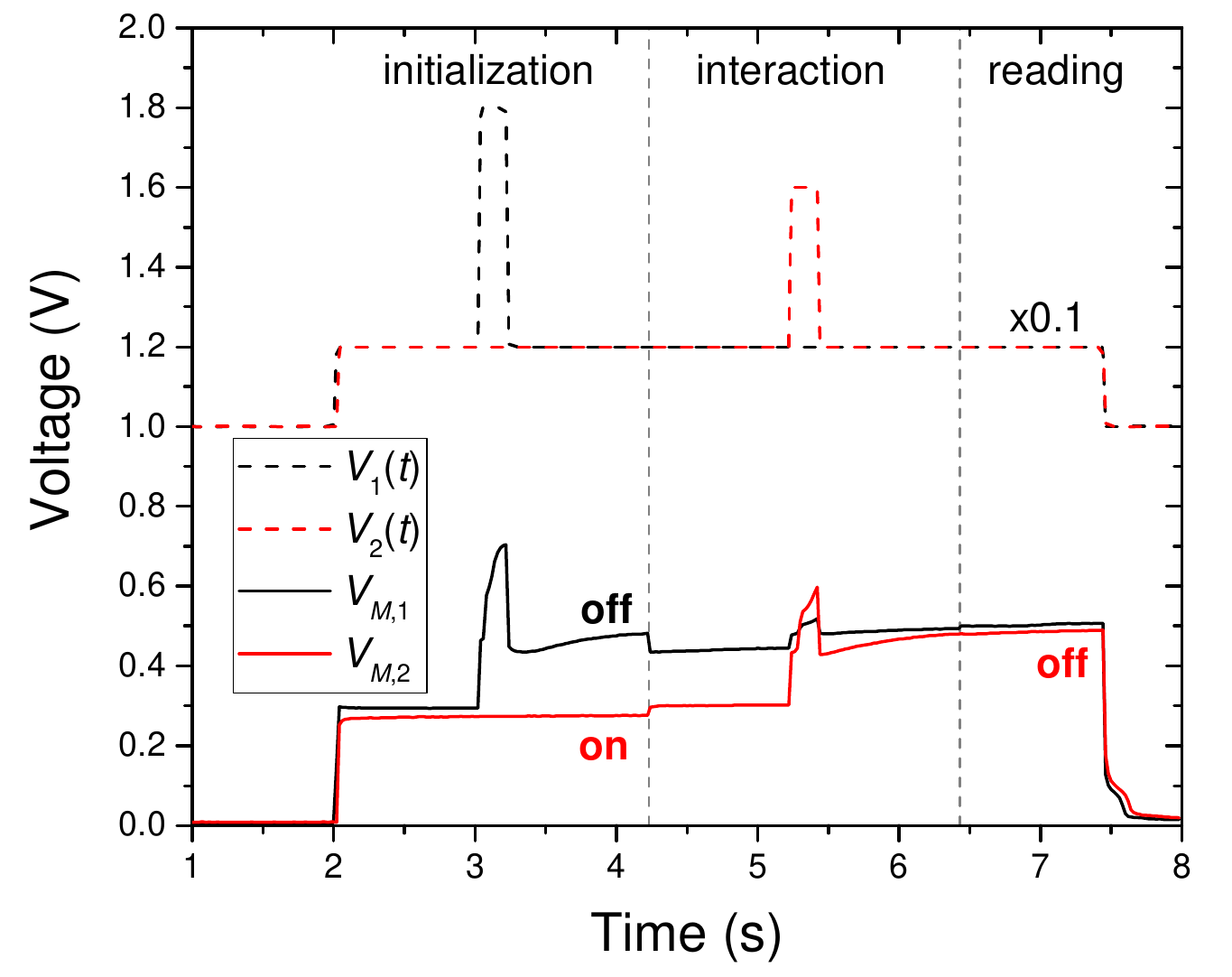} 
(e) \includegraphics[width=0.45\textwidth]{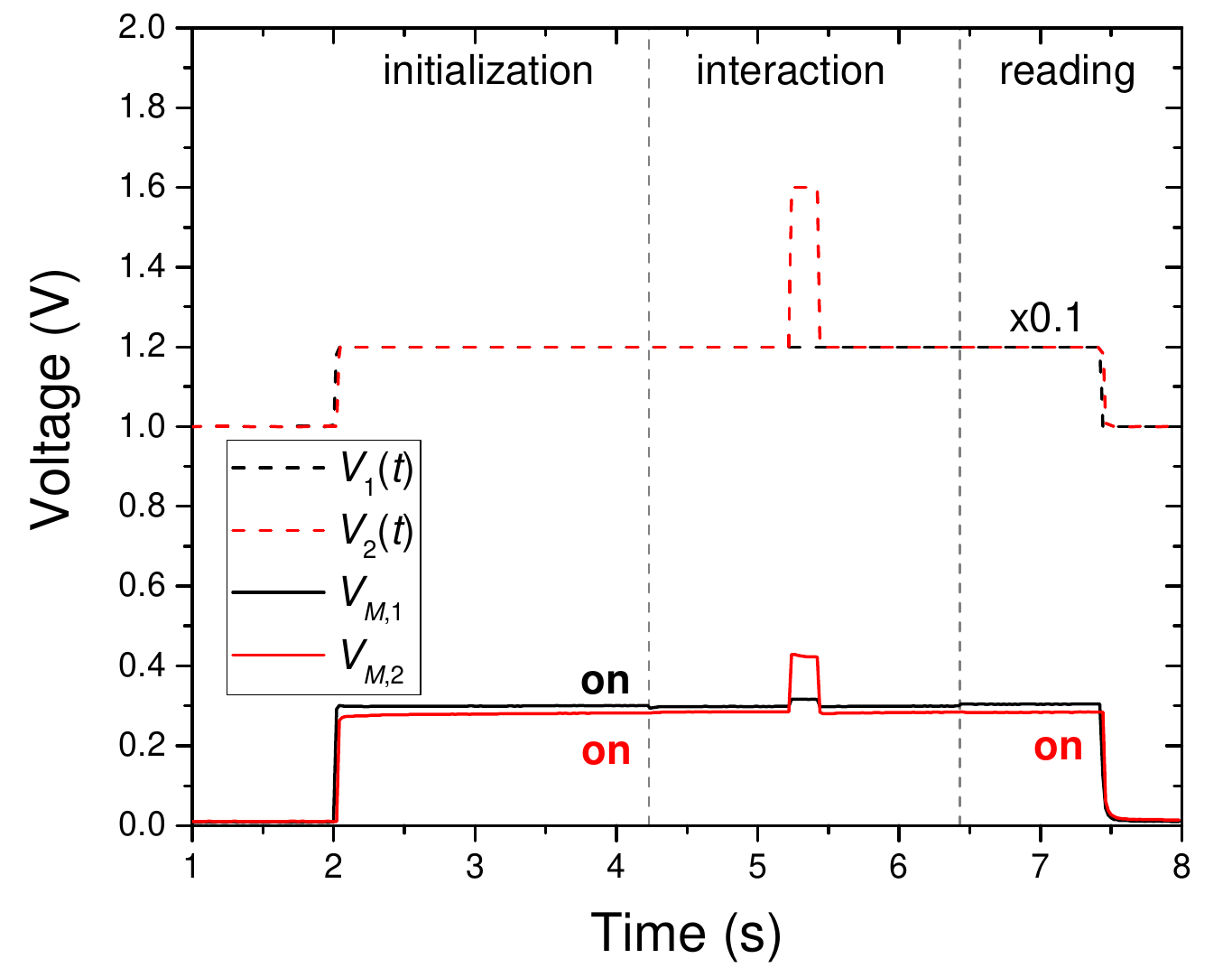}
\caption{Demonstration of AND gate.  In these measurements, we used $R_1=R_2=10$~k$\Omega$ and $R_3=3.3$~k$\Omega$. The control voltage curves in (b)-(e) were scaled and displaced by 1~V for clarity. \label{fig:4}}
\end{figure}

Next, we demonstrate that EGaIn memristive elements are suited to realize AND and OR logic gates~\footnote{In this demonstration, we associate the {\bf off} state of one memristor with Boolean 0, while the {\bf on} state of the same memristor with Boolean 1.}. For this purpose, we built the circuit shown in Fig.~\ref{fig:4}(a). A multifunctional data acquisition unit (MDAQ) was used to apply and read voltages. The circuit includes a relay to induce the interaction of two memristors, M$_1$ and M$_2$. The experiments consisted of three phases: initialization, interaction, and reading (Fig.~\ref{fig:4}(b)-(c)). In the initialization phase, the relay was open and the states of the memristors were individually programmed by applying (or not applying) a voltage pulse of 1.8~V for 200~ms at $t=3$~s and then maintaining a holding voltage of 1.2~V. The amplitude and duration of the pulse were chosen so that the memristors exposed to it switch to the {\bf off} state. Consistent with Fig.~\ref{fig:3}(c), the voltage across each memristor subjected to the programming pulse converges to a \textit{high} stable final value, indicating it is in the \textbf{off} state (Boolean 0) or, otherwise, \textit{low} stable final value, indicating the \textbf{on} state (Boolean 1).

In the interaction phase, the relay is closed and a positive voltage pulse of 1.6~V is applied to M$_2$ (through R$_2$) for 200~ms at $t= 5.2$~s, while the holding voltage is continuously applied to M$_1$ (through R$_1$). Within this setup, the pulse-induced switching of M$_2$ is controlled by the state of M$_1$. If M$_1$ is \textbf{on}, the voltage across M$_2$ is reduced so that the transition of M$_2$ from \textbf{on} to \textbf{off} is blocked (as the voltage across M$_2$ stays below the threshold $V_{off}$). In contrast, when M$_1$ is \textbf{off}, the switching of M$_2$ from \textbf{on} to \textbf{off} is allowed (as the voltage across M$_2$ exceeds the threshold $V_{off}$). Fig.~\ref{fig:4}(b)-(e) shows that M$_1$'s state remains unchanged throughout the interaction phase. In panels (b) and (c) of Fig.~\ref{fig:4}, M$_2$ ends up in the {\bf off} state, given that it started {\bf off}. When you examine Fig.~\ref{fig:4}(d), M$_2$ switches to {\bf off} due to M$_1$ being {\bf off}. Conversely, in Fig.~\ref{fig:4}(e), M$_2$ stays {\bf on} because M$_1$ is {\bf on}.

In the reading phase, the relay is opened and the states of M$_1$ and M$_2$ are evaluated in terms of memristor voltages, $V_{M,1}$ and $V_{M,2}$, which are recorded in response to the same holding voltage. Using the state of M$_2$ as the output of the logic gate, it is evident that this experiment reproduces the logic table of AND. Referring to Fig.~\ref{fig:4}(b)-(e), the truth table for the AND gate can be read by linking the {\it high} voltage level of $V_{M,2}$ to Boolean 0, and the {\it low} voltage level to Boolean 1, as previously discussed.  Thus the output of the AND gate produces a Boolean 1 only when both memristors are \textbf{on} (Boolean 1). 

To realize the OR gate, we employed the experimental setup depicted in Fig.~\ref{fig:4}(a), with the sole change being the use of a negative amplitude for $V_{M,2}$ during the interaction phase. As demonstrated in Fig.~\ref{fig:S1} (Supplementary Information), the final state for M$_1$ corresponds to the functionality of the OR gate. 

\subsection*{\bf Accelerated and decelerated switching.}
Finally, we investigated how EGaIn memristive devices behave when connected in series and parallel circuits (Fig.~\ref{fig:S3}(c) and (d) of SI). Using these circuits, we observed two notable phenomena: accelerated (simultaneous) and decelerated (time delayed) switching~\cite{Pershin13a}, both of which stem from the voltage divider effect. In the circuit of series-connected memristors subjected to a triangular voltage, the initial switching of one memristor decreases the voltage drop occurring across the other, as the resistance of the first memristor increases. This causes the second memristor to switch at a notably higher voltage, note two peaks at $V>0$ in Fig.~\ref{fig:S3}(e) of SI. In contrast, for memristors connected in parallel, the first switching event raises the voltage drop across the second memristor, leading to (almost) simultaneous switching of both memristors (a single peak at $V>0$ in Fig.~\ref{fig:S3}(e), SI).

The behaviors of the EGaIn tube memristor introduced herein demonstrate that voltage-dependent oxidation and reduction at a molecularly smooth liquid interface yields reliable resistive switching, both between devices and across activation cycles and time. These are important criteria for enabling memristive technologies in signal processing or brain-inspired computing circuitry. Unlike the localized formation of conductive filaments in solid-state memristors that results in highly stochastic device operation, EGaIn devices exhibit voltage-drive changes in resistance, and without requiring an initial \textit{forming} step, via spatially distributed electrochemical oxidation and reduction reactions occurring across the EGaIn:NaOH liquid interface. The associated reaction-diffusion kinetics that control the rates of gallium oxidation and reduction cause resistive switching to occur at similar speeds to many biological sensory and communication processes. Moreover, the recorded values of $|V_{off}|$ and $|V_{on}|$ are well below $1$ V, which makes EGaIn-based devices comparable to many solid-state devices, such as those based on transition metal oxides~\cite{Li2018}. However, the relatively low values of {\bf on} and {\bf off} resistance ($<10$ k$\Omega$, Fig.~\ref{fig:4andy}(e)) increase power consumption in EGaIn tube memristors. We expect that shrinking the characteristic device size, as governed by the EGaIn:NaOH interfacial area, will increase nominal resistance and reduce power dissipation. In addition, future works should investigate ways to achieve non-volatile memory storage, characterize the intrinsic fractional capacitance exhibited at intermediate frequencies, and explore their rich nonlinear properties in adaptive signal processing applications (e.g., physical reservoir computing).

\begin{methods}  \label{sec:el_meas}

\subsection{Device fabrication}

To make a device, we employed polyvinyl chloride (PVC) tubing approximately 22~mm long, with an internal diameter of 1/16 inch (1.5875 mm) and an external diameter of 1/8 inch (3.175 mm) (Grainger Part No. 4EGY3). First, a 10 AWG copper electrode was inserted approximately 6~mm deep into the tube. With the tube oriented vertically and the copper electrode pointing downward, a syringe filled approximately two-thirds of the empty tube volume with 3~M NaOH. A separate syringe was then used to inject EGaIn liquid metal (Indalloy\textregistered~300E (78.6Ga/21.4In),  Indium Corporation) close to the copper electrode, in an amount equating to approximately 1 to 2~mm of tube length. Subsequently, the tube was topped with NaOH and a second electrode was inserted 6~mm deep into the tube. Gentle tapping allowed a portion of EGaIn to transfer to the second electrode. We made sure that there were no trapped bubbles in the device at all. Some devices were exposed to sonication (with a Central Machinery 2.5~l ultrasonic cleaner), which contributed to lowering and achieving more consistent switching thresholds.

It was observed that small bubbles developed over time have no to little effect on the device characteristics. However, the performance of the device declines when a bubble completely obstructs the tube's cross section. In these cases, performance might be restored by eliminating the bubble through opening. To prevent or minimize bubble formation, we tried to keep the applied voltage to the device below $0.8$~V. In these circumstances, the devices were observed to operate reliably for long durations (several weeks, see Fig.~\ref{fig:andyS3}). However, testing was not conducted continuously.
Fig.~\ref{fig:S9} shows an excellent switching capability of our devices subjected to 1000 switching cycles.

Several earlier studies~\cite{khan2014giant,Tang21a,rashid2024role,hillaire2024interfacial} employed a 1~M concentration of NaOH as standard when performing experiments with EGaIn. We began with 1~M devices in our tests, but found that devices with a 3~M NaOH concentration exhibited a more consistent response. Consequently, we have opted for the 3~M concentration in our main experiments, especially since these devices demonstrated a significantly clearer transition from {\bf off} to {\bf on} in their current-voltage curves.

\subsection{Electrical Measurements}

The $I-V$ curves presented in this article were obtained using the Keithley 236 source measurement unit (SMU). In our experiments, the SMU was used to apply a voltage and record the current while limiting it to a predefined maximum (compliance) value typically $1$~mA. This value was not attained in most measurements. Additionally, we utilized the SMU to observe the device's behavior at a fixed level of voltage.

The multifunctional data acquisition unit (MDAQ), model Keysight U2542A, was used to apply and measure voltage. The DAQ was programmed using a custom code to deliver a time-dependent voltage (e.g., voltage pulses) to small memristive circuits, as described in this article. Several DAQ input channels were employed to record the voltage at the nodes of interest.

\subsection{Electrochemical measurements}
Half-cell cyclic voltammetry ($I-V$) recordings on single EGaIn/NaOH interfaces were performed in a 3-electrode electrochemical cell that was controlled by a Biologic SP300. Simulating the EGaIn volume and arrangement in the tube-memristor, a $\sim 20$~$\upmu$L volume of EGaIn droplet supported by a copper rod was submerged into NaOH solution. The copper rod was insulated such that only EGaIn made contact with the surrounding electrolyte. This represented the working electrode for the system. The reference and counter electrodes consisted of Ag/AgCl in 3 M KCl and Pt, respectively. Measurements were controlled using Biologic EC-Labs software.

Electrical impedance measurements and supplementary cyclic voltammetry measurements on whole-cell tube memristors were performed using a Biologic SP200 Potentiostat using a two-electrode configuration in which the device leads were connected directly to the copper rods supporting the two EGaIn interfaces within the tube. EIS measurements were performed using a $10$~mV sinusoidal AC voltage across the frequency range $100$~kHz and $100$~mHz. Measurements were controlled using Biologic EC-Labs software. Estimates of equivalent circuit parameters (as defined in Fig.~\ref{fig:4andy}(d)) were obtained from raw EIS data through a custom MATLAB script that employed a bounded, nonlinear least squares fitting routine.

All measurements were performed at ambient conditions.

\end{methods}




\begin{addendum}
 \item The authors acknowledge fruitful discussions with Prof. Fidel Santamaria and Prof. Christof Teuscher, as well as members of their research groups. This work was supported by the NSF grant EFRI-2318139. SAS acknowledges additional financial support from the James Conklin Faculty Fellowship at UTK.
 \item[Competing Interests] The authors declare that they have no competing financial interests.
 \item[Data availability] The authors can provide the raw and processed data that support the findings of this manuscript upon a reasonable request.
 \item[Correspondence] Correspondence and requests for materials should be addressed to Y.~V.~P.~(email: \\ \mbox{pershin@physics.sc.edu}) or S.~A.~S.~(email: \mbox{ssarles@utk.edu}).
\end{addendum}


\newpage

\setcounter{page}{1}
\setcounter{figure}{0}
\setcounter{equation}{0}
\renewcommand{\theequation}{S\arabic{equation}}
\renewcommand{\thepage}{S\arabic{page}}
\renewcommand{\thesection}{S\arabic{section}}
\renewcommand{\thetable}{S\arabic{table}}
\renewcommand{\thefigure}{S\arabic{figure}}

\begin{center}
 {\Large Supplementary Information}   
\end{center}

\section{Model}

\begin{figure}[bt]
    \centering
    (a)\includegraphics[width=0.45\linewidth]{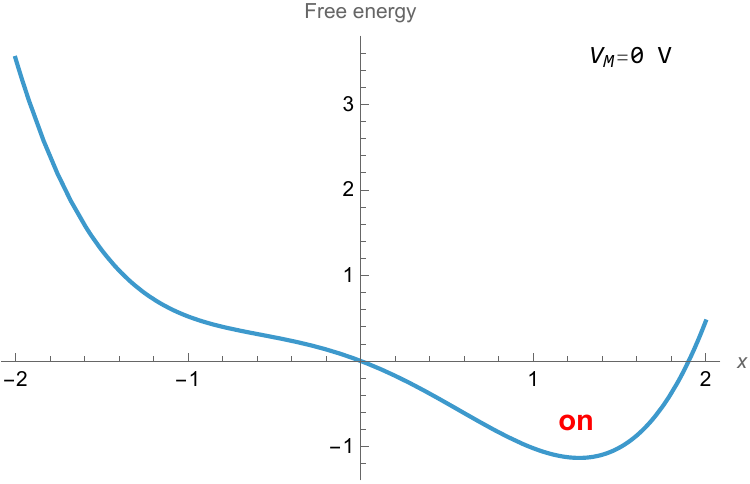}
    (b)\includegraphics[width=0.45\linewidth]{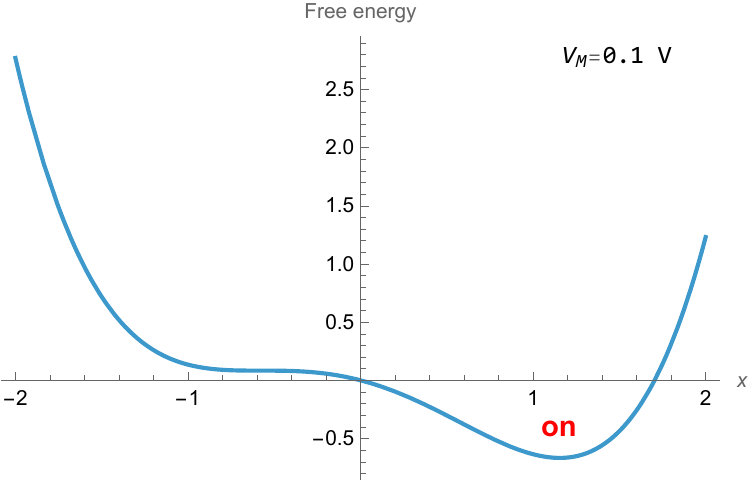} \\
    (c)\includegraphics[width=0.45\linewidth]{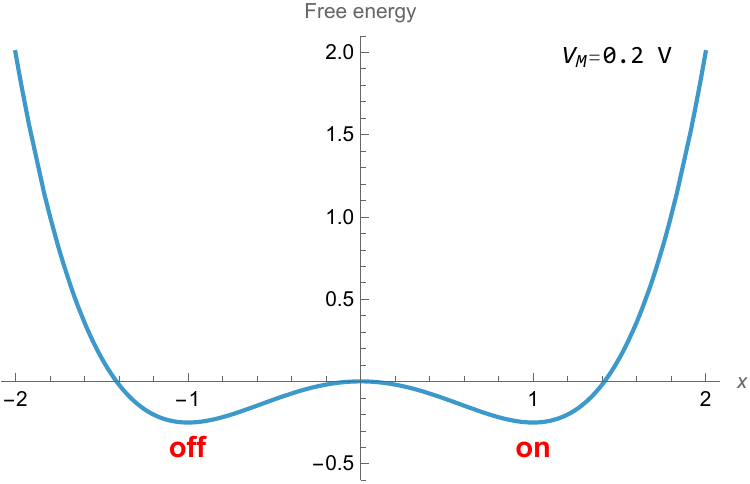}
    (d)\includegraphics[width=0.45\linewidth]{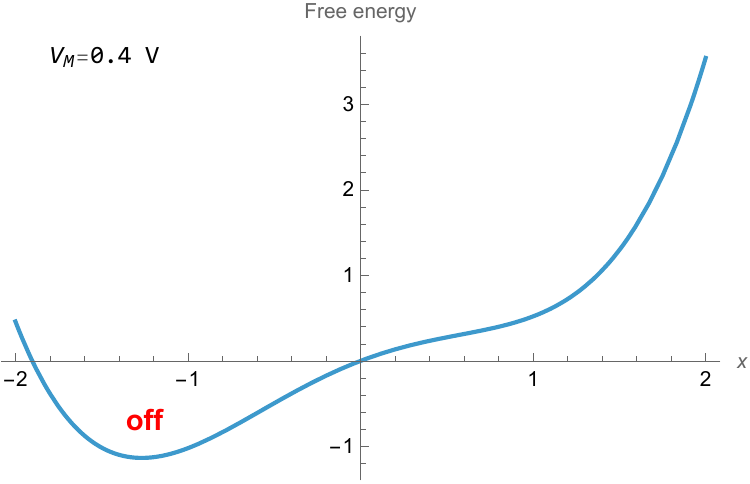}
    \caption{Free energy as a function of $x$ for selected values of $V_M$. This plot was obtained using $V_{on}=0.1$, $V_{off}=0.3$, and $b=1$. Here, the minimum at $x>0$ correspond to the {\bf on} state, while the minimum at $x<0$ correspond to the  {\bf off} state.}
    \label{fig:SFE}
\end{figure}

To find stable fixed points of Eq.~(\ref{eq:FE2}), let us explore the solutions of the algebraic equation
\begin{equation}
    -x+bx^3+h=0, \label{eq:S1}
\end{equation}
where $b>0$ and $h$ are constants. Eq.~(\ref{eq:S1}) has either one real and two complex roots or three real roots. It can be demonstrated that the equation has three real roots when
\begin{equation}
    |h| \leq \sqrt\frac{4}{27b}.
\end{equation}
Given that $h=\kappa(|V_M|-V_0)$ and the three real roots should exist only in the region of hysteresis $[V_{on},V_{off}]$, the following expressions can be derived:
\begin{eqnarray}
V_0&=&\frac{V_{on}+V_{off}}{2}\; , \\
\kappa&=&\frac{2}{V_{off}-V_{on}}\sqrt{\frac{4}{27b}}\; .
\end{eqnarray}
The remaining parameter $b$ defines the scaling along the $x$-axis and can be taken as $b=1$. In fact, at $h=0$, the minima are located at $x=\pm 1/\sqrt{b}$, see Eq.~(\ref{eq:S1}). Fig.~\ref{fig:SFE} shows examples of the free energy, Eq.~(\ref{eq:FE1}), at selected values of $V_M$.

\begin{figure}[bt]
    \centering
    \includegraphics[width=0.6\linewidth]{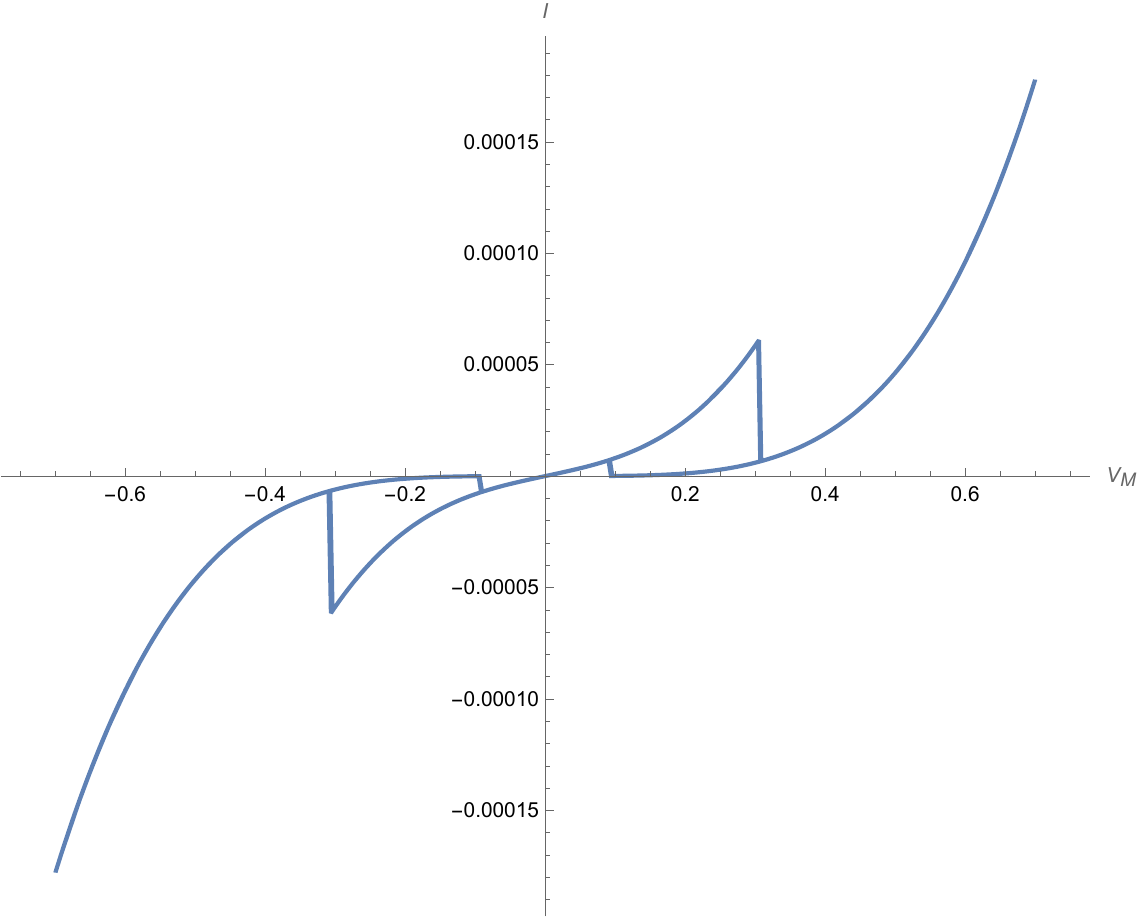}
    \caption{Model current-voltage curve. This plot was obtained using the following set of parameter values: $a_0=1/3000$~S, $a_2=1/140$~S$\cdot\textnormal{V}^{-2}$, $b_3=1/270$~S$\cdot\textnormal{V}^{-3}$, $V_{on}=0.1$~V, $V_{off}=0.3$~V, and $\Gamma=10^3$~s$^{-1}$.}
    \label{fig:SIV}
\end{figure}

In this work, the memductance (memory conductance) is approximated with
\begin{equation}
 G_M(x,V)=G_{off}(V)u(-x)+G_{on}(V)u(x) , \label{eq:Rx} \\
\end{equation}
where $u(\dots)$ is the unit step function. According to Eq.~(\ref{eq:Rx}), positive values of $x$ are associated with the {\bf on} state, while negative with {\bf off}. Based on fits of experimental data, we utilize a polynomial approximation for the voltage dependency of $G_{off}$ and $G_{on}$:
\begin{eqnarray}
G_{on}(V)&=&a_0+a_2V^2\; , \\
G_{off}(V)&=&b_3|V|^3\; .
\end{eqnarray}
Fig.~\ref{fig:SIV} presents current-voltage characteristics obtained with Eqs.~(\ref{eq:FE2})-(\ref{eq:FE3}) model.

\newpage
\section{Additional  measurements}

\begin{figure}[]
\centering
(a)\includegraphics[width=0.45\textwidth]{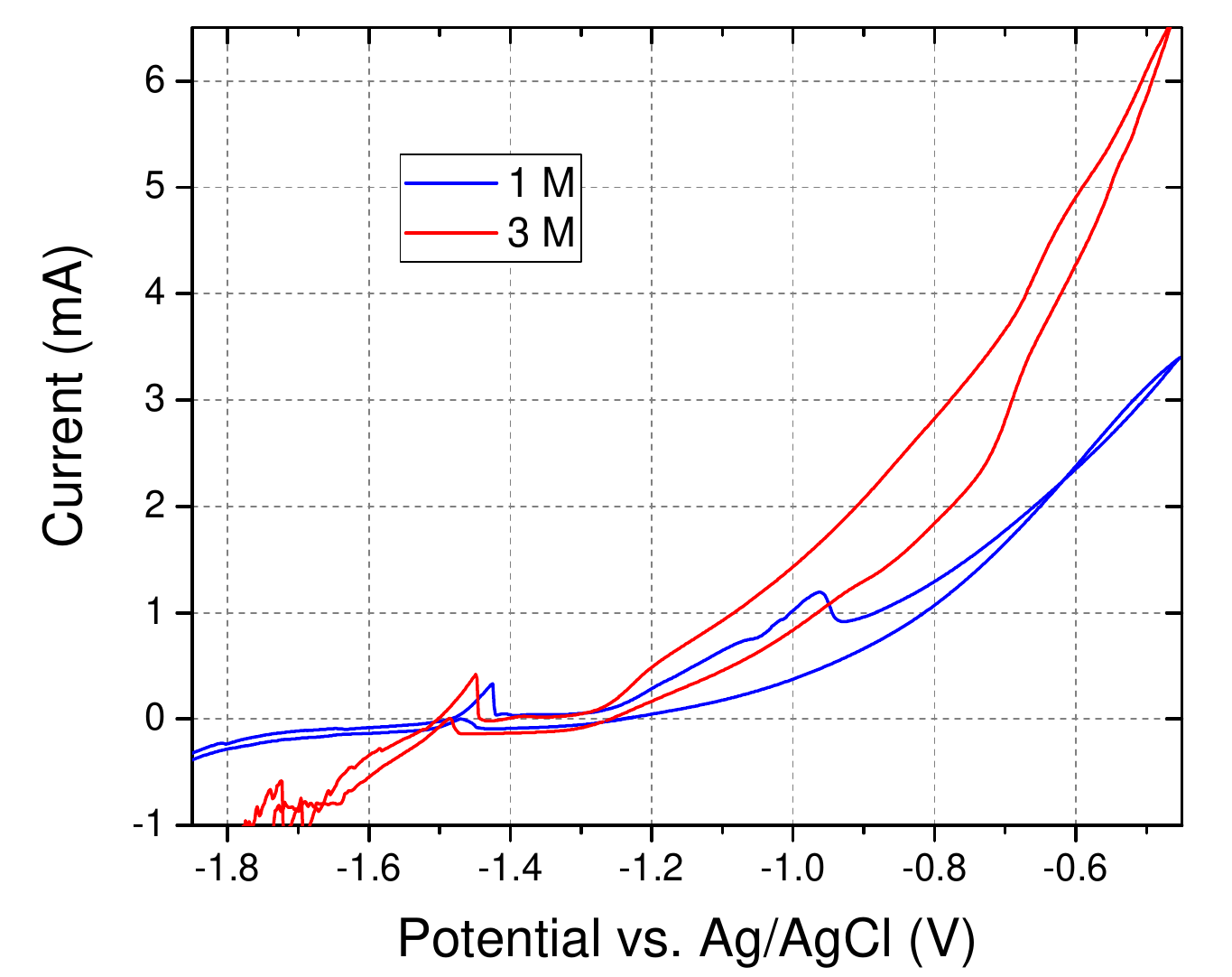}
(b) \includegraphics[width=0.45\textwidth]{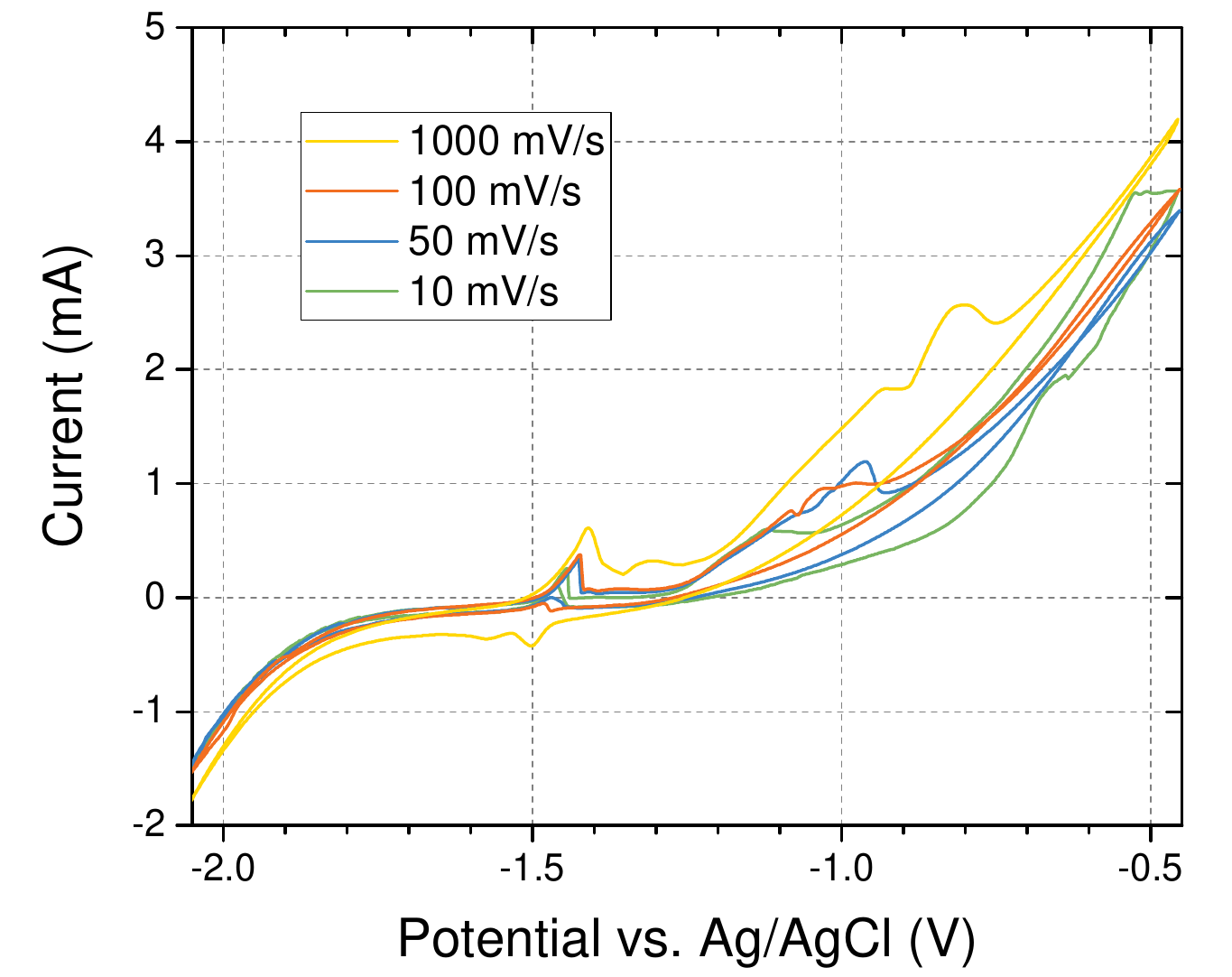}
\caption{half-cell EGaIn/NaOH current-voltage relationships.
(a) Comparison of half-cell $I-V$ responses obtained in $1$ and $3$ M NaOH. Both were measured using a scan rate of $50$ mV/s.
(b) Representative examples of half-cell $I-V$ curves in $1$ M NaOH at varying scan rates.
\label{fig:S2andy}}
\end{figure}

\begin{figure}[h]
\centering
(a) \includegraphics[width=0.45\textwidth]{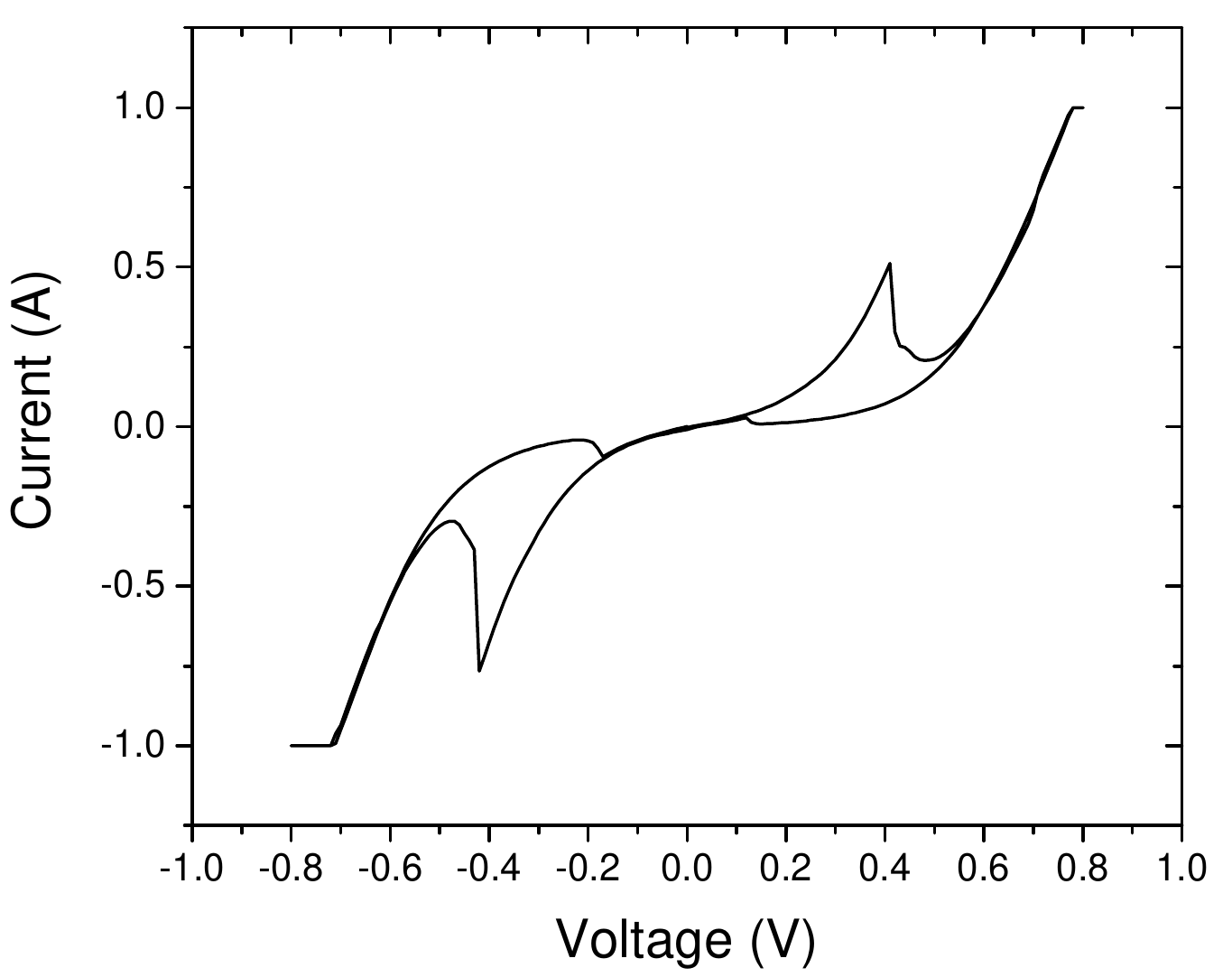} 
(b) \includegraphics[width=0.45\textwidth]{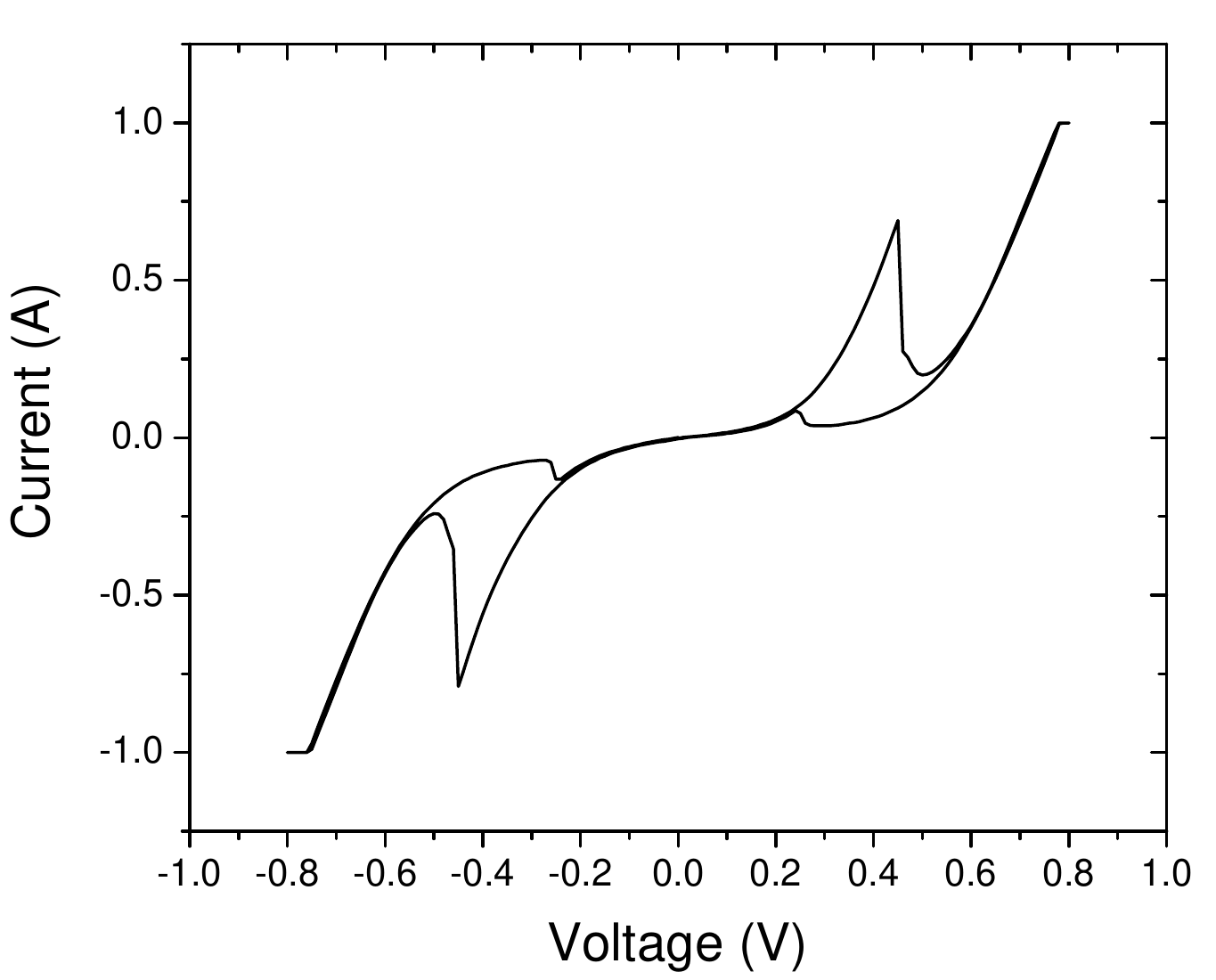}\\
(c) \includegraphics[width=0.45\textwidth]{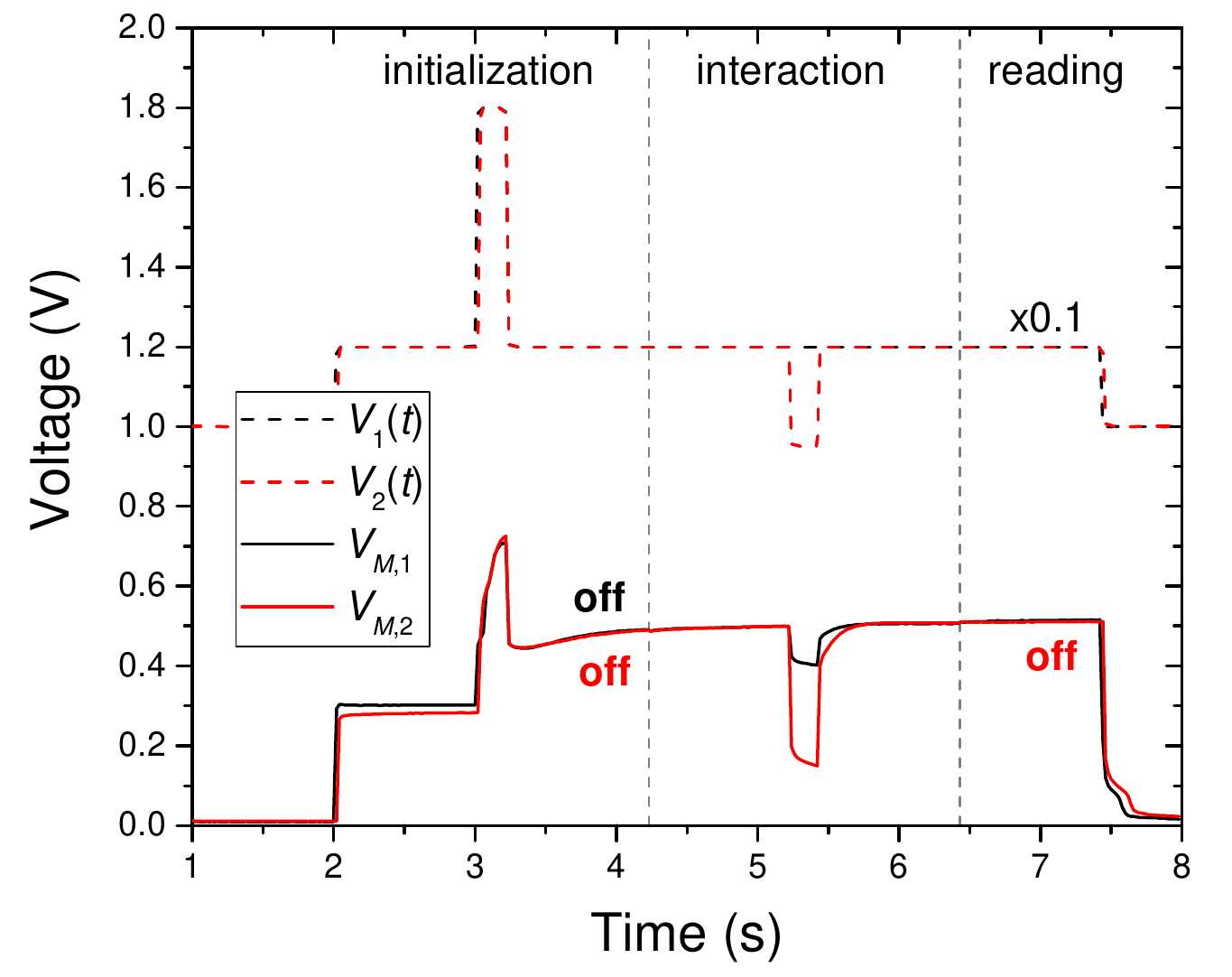} 
(d) \includegraphics[width=0.45\textwidth]{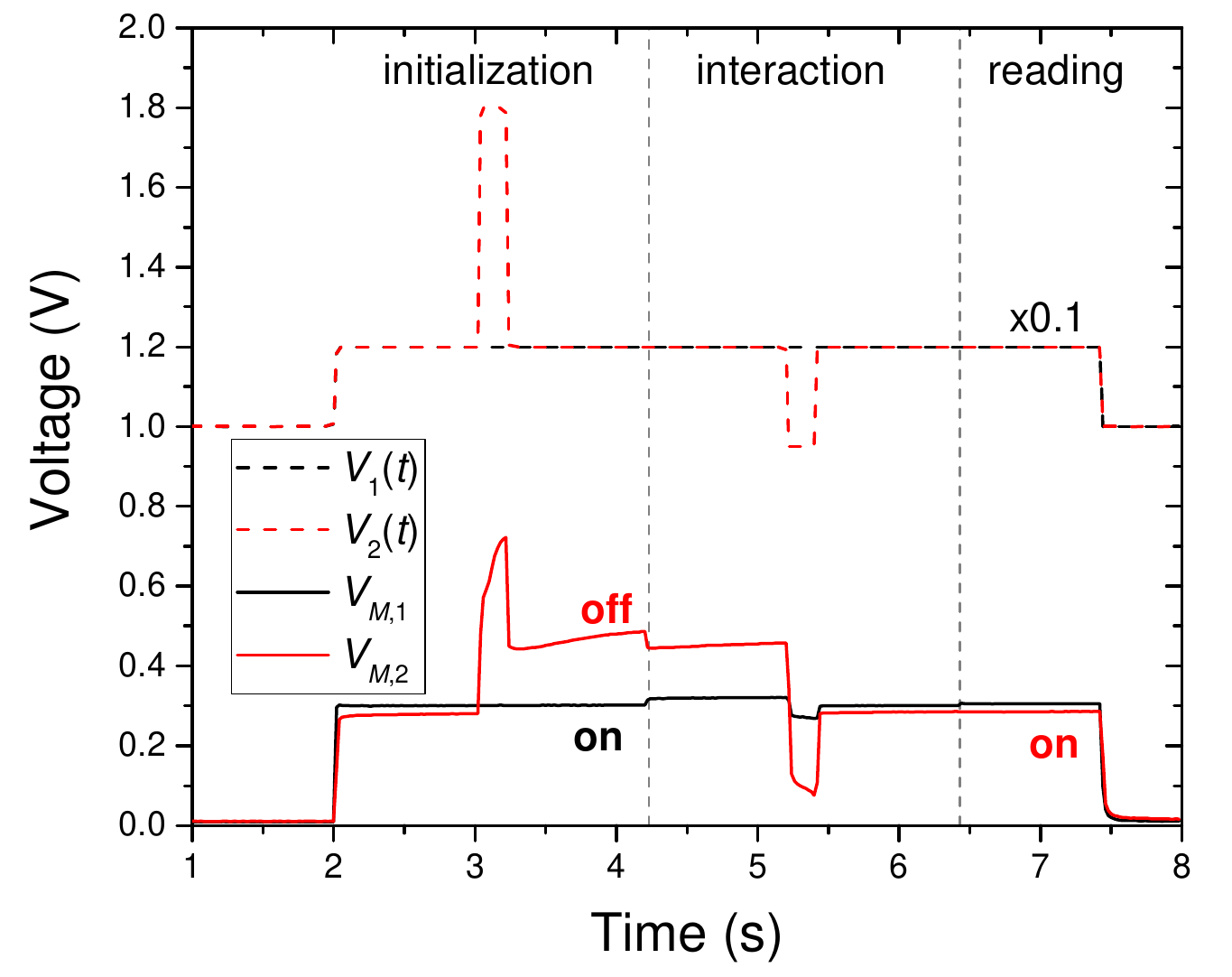}\\
(e) \includegraphics[width=0.45\textwidth]{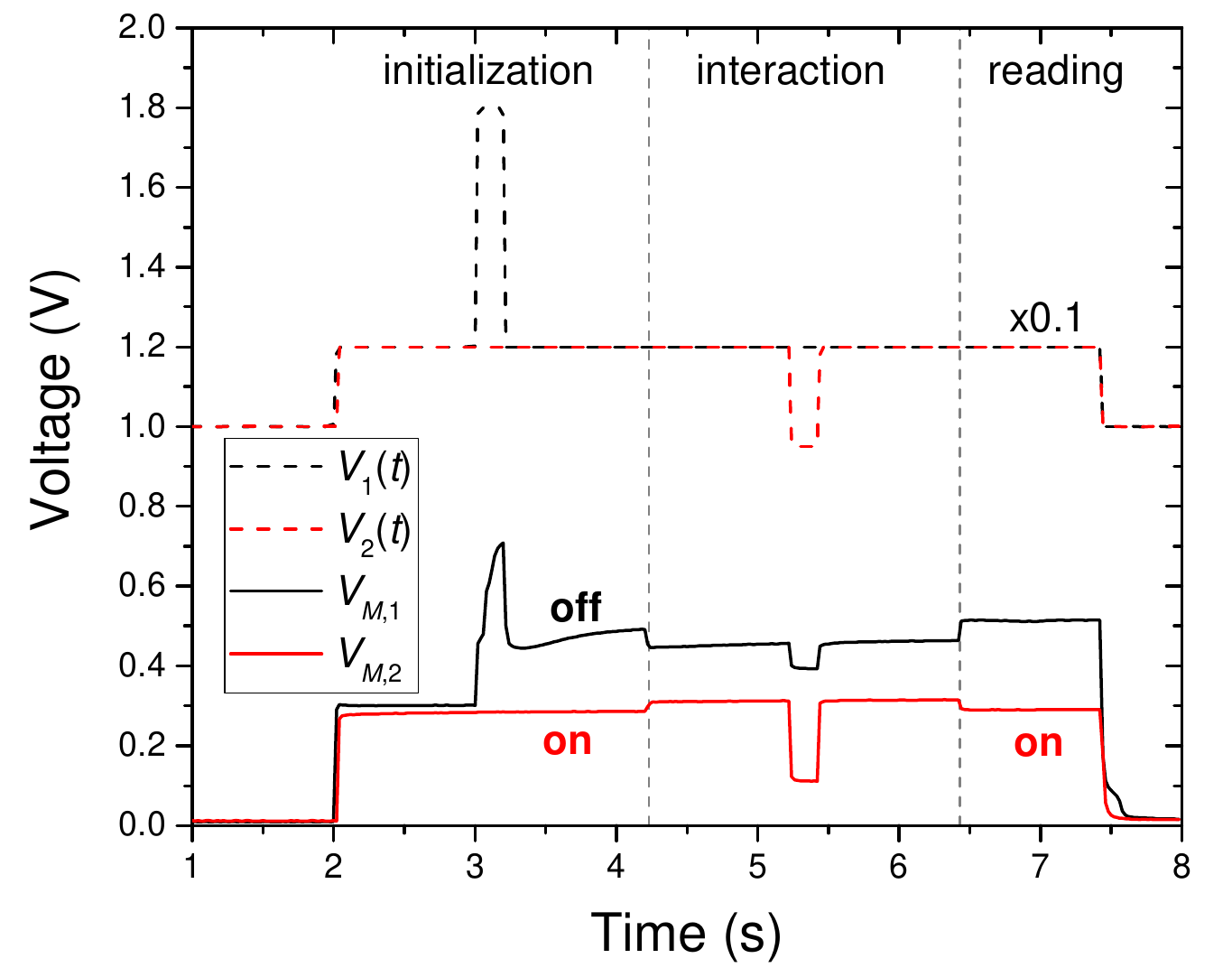} 
(f) \includegraphics[width=0.45\textwidth]{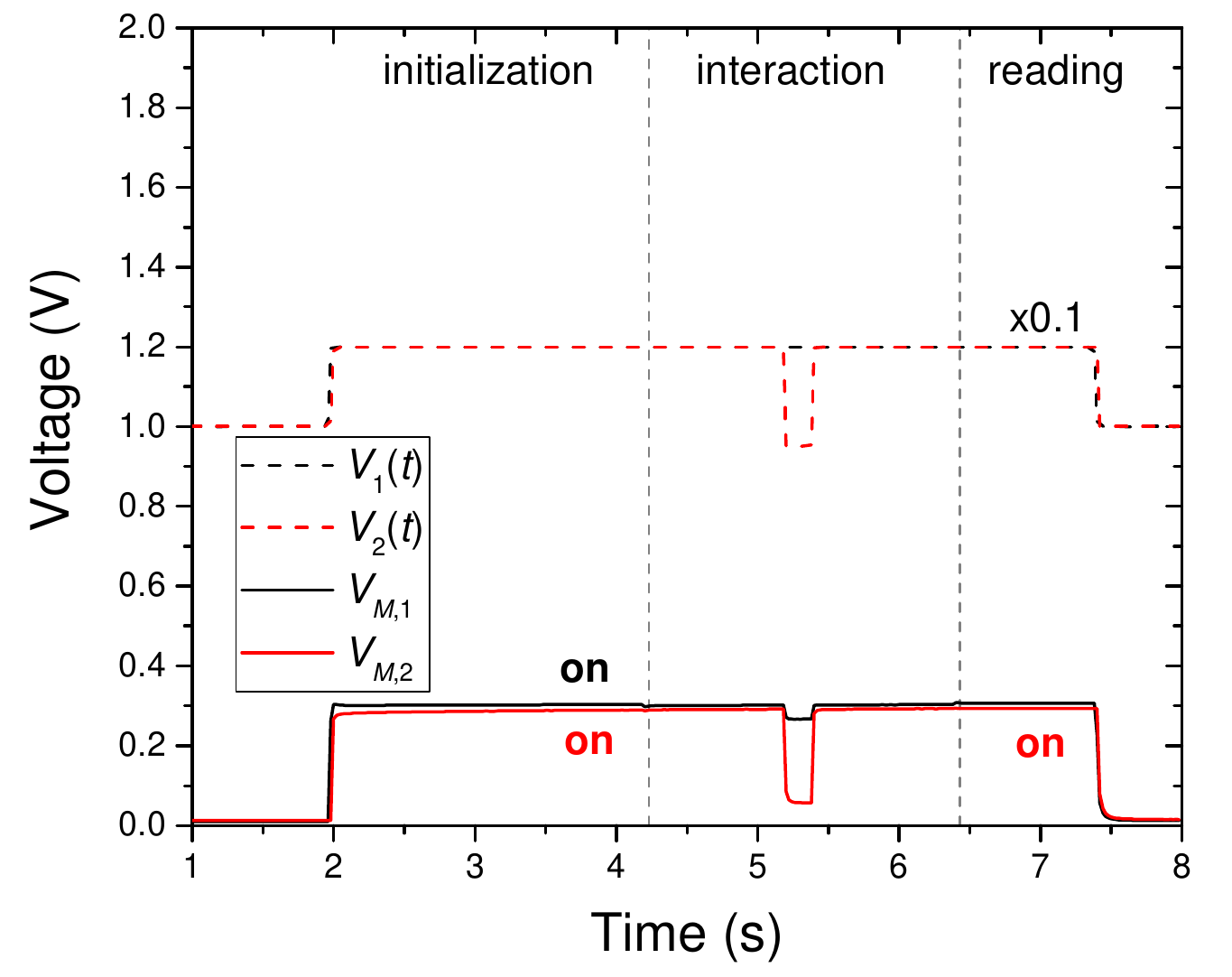}
\caption{Demonstration of OR gate. (a), (b) Current-voltage curves of devices used in the in-memory computing experiments (the voltage ramp is $0.09$~V/s). (c)-(f) Experimental demonstration of the OR gate that was obtained using the same setup as in Fig.~\ref{fig:4}(a). The control voltage curves in (c)-(f) were scaled and displaced by 1~V for clarity. \label{fig:S1}}
\end{figure}

\begin{figure}[h]
\centering
(a) \includegraphics[width=0.45\textwidth]{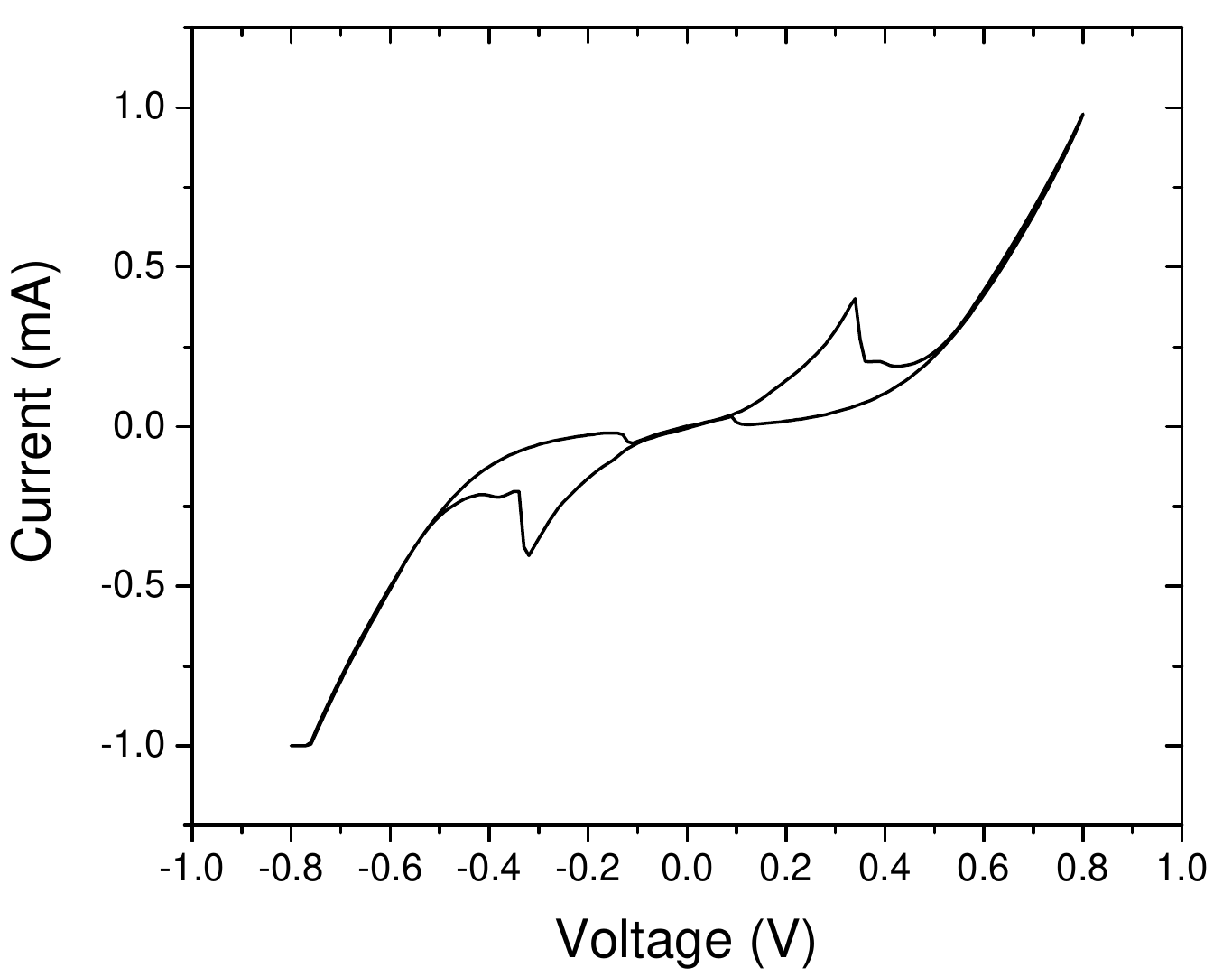} 
(b) \includegraphics[width=0.45\textwidth]{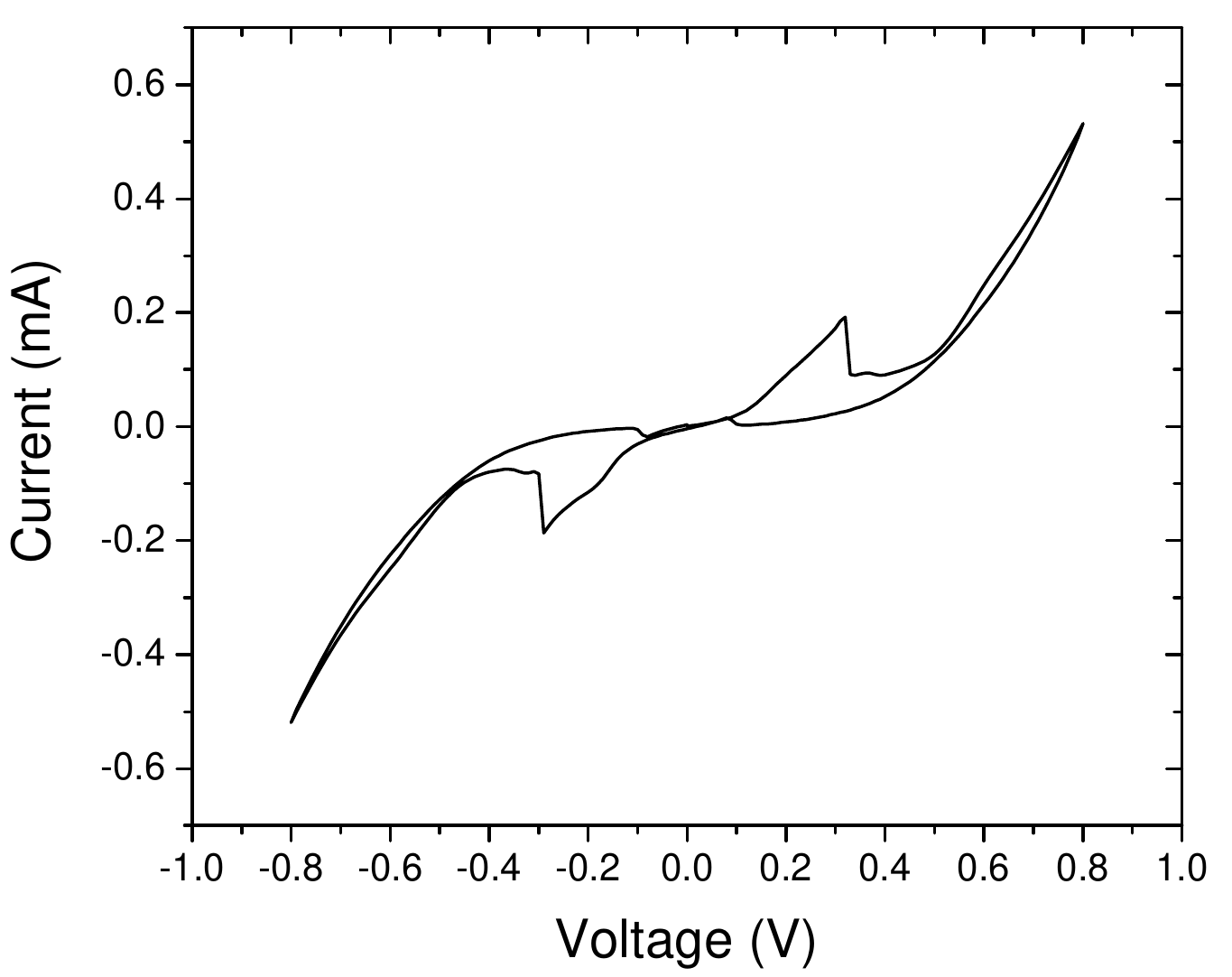} \\
(c) \includegraphics[width=0.45\textwidth]{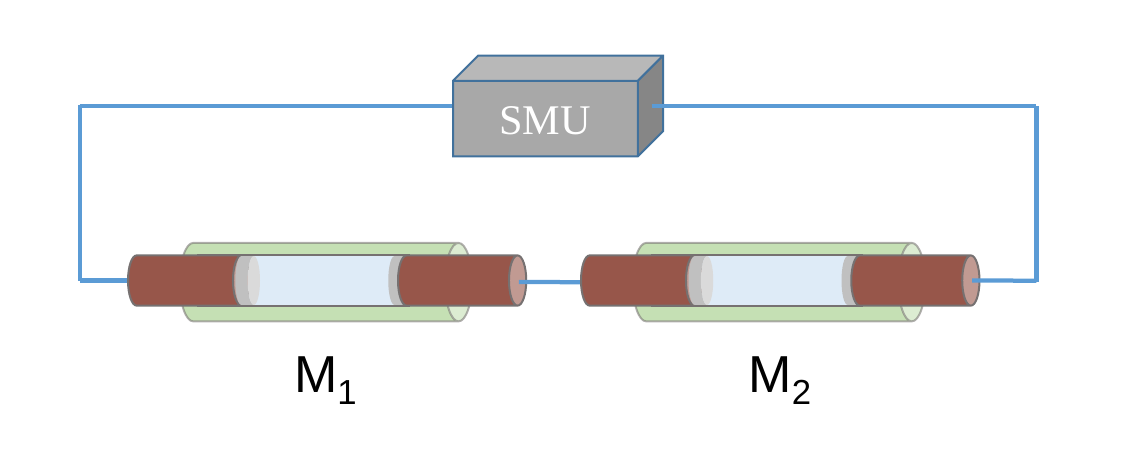}
(d) \includegraphics[width=0.45\textwidth]{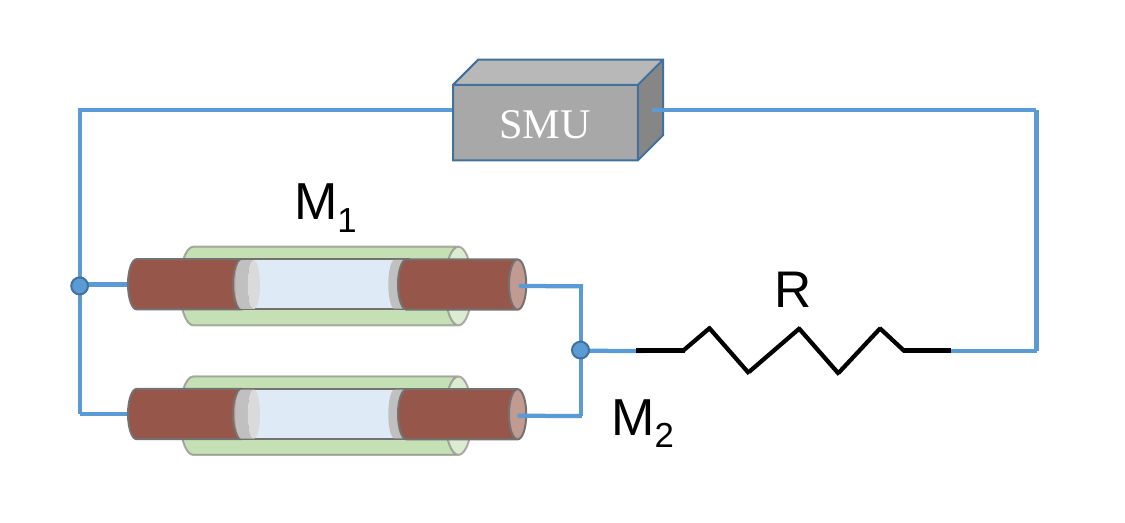}\\
(e) \includegraphics[width=0.45\textwidth]{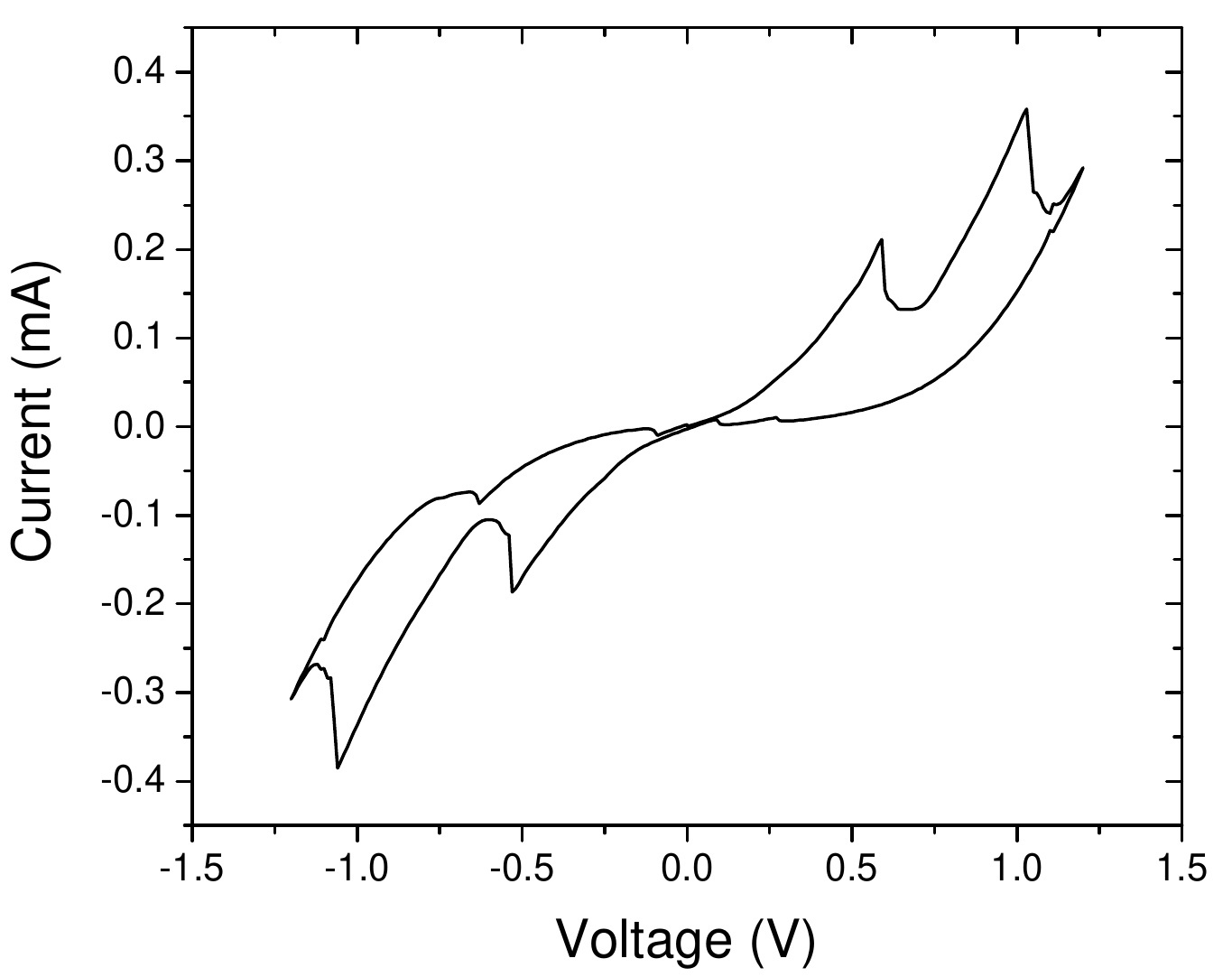}
(f) \includegraphics[width=0.45\textwidth]{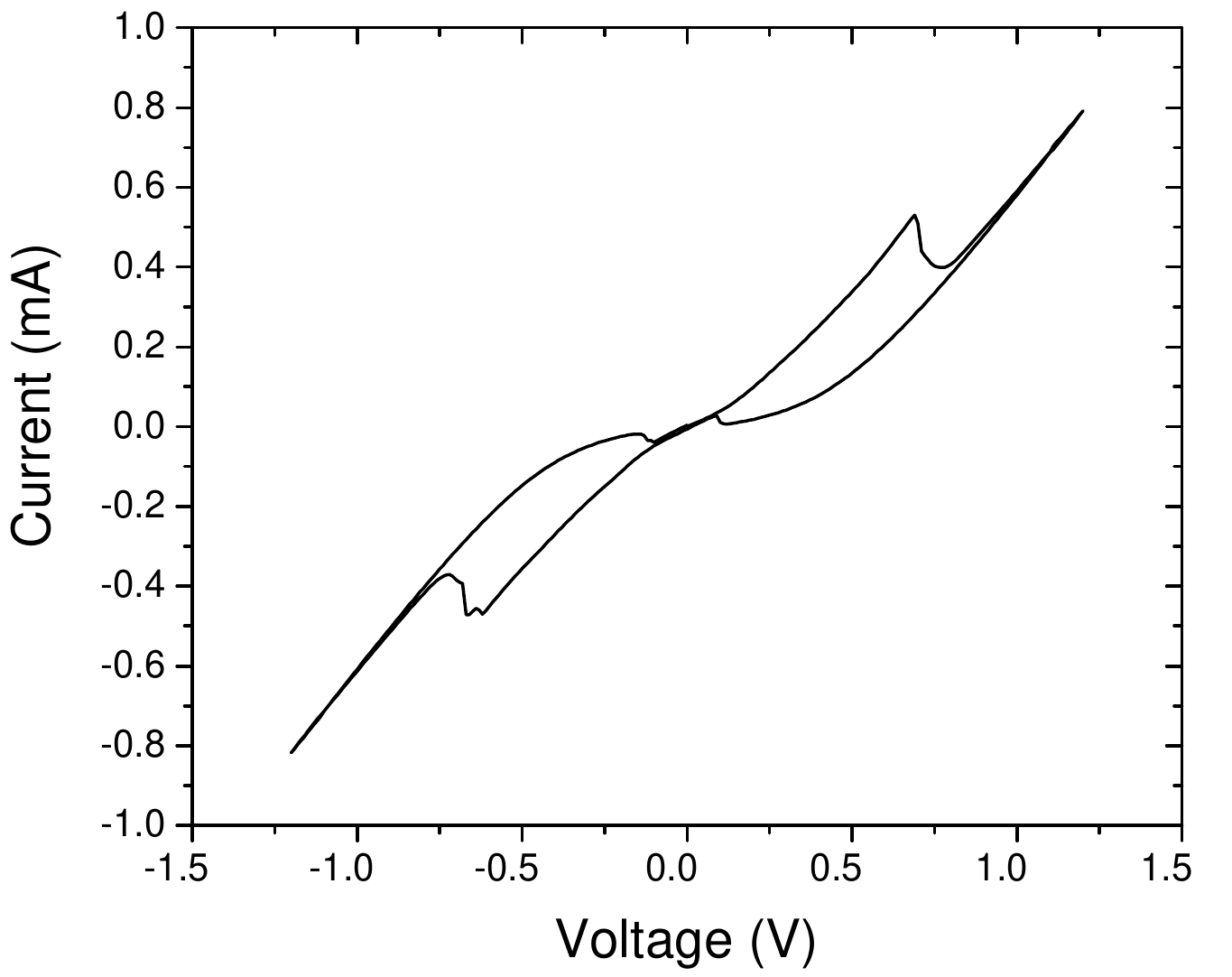}
\caption{Accelerated and decelerated switching. (a), (b): Current-voltage curves of two devices taken before the experiment. (c), (d): The measurement schematics. (e), (f): Current-voltage curves for the circuits in (c) and (d), respectively. In (d), we used $R=220$~$\Omega$.\label{fig:S3}}
\end{figure}

\begin{figure}[h]
\centering
(a) \includegraphics[width=0.45\textwidth]{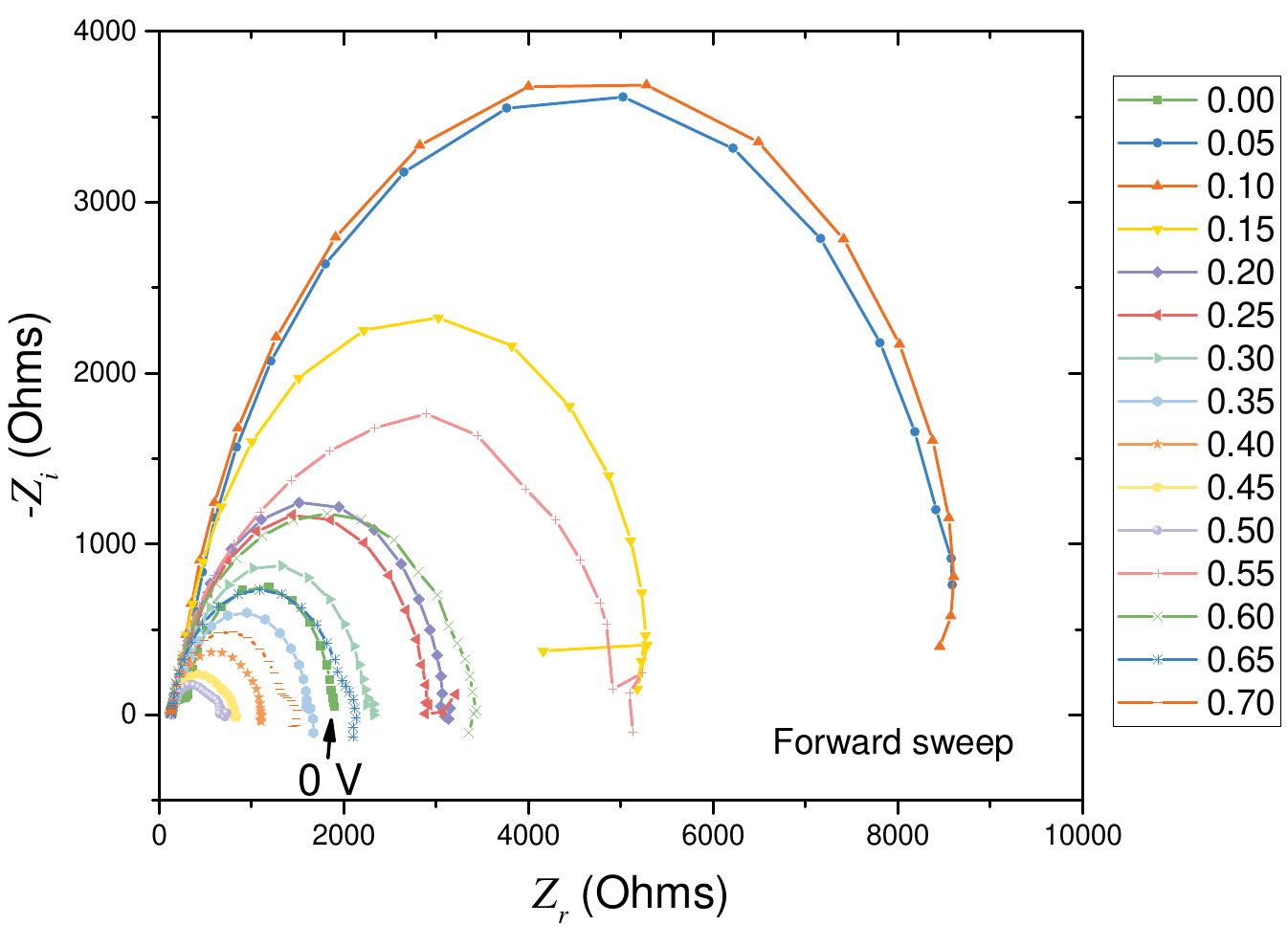} 
(b) \includegraphics[width=0.45\textwidth]{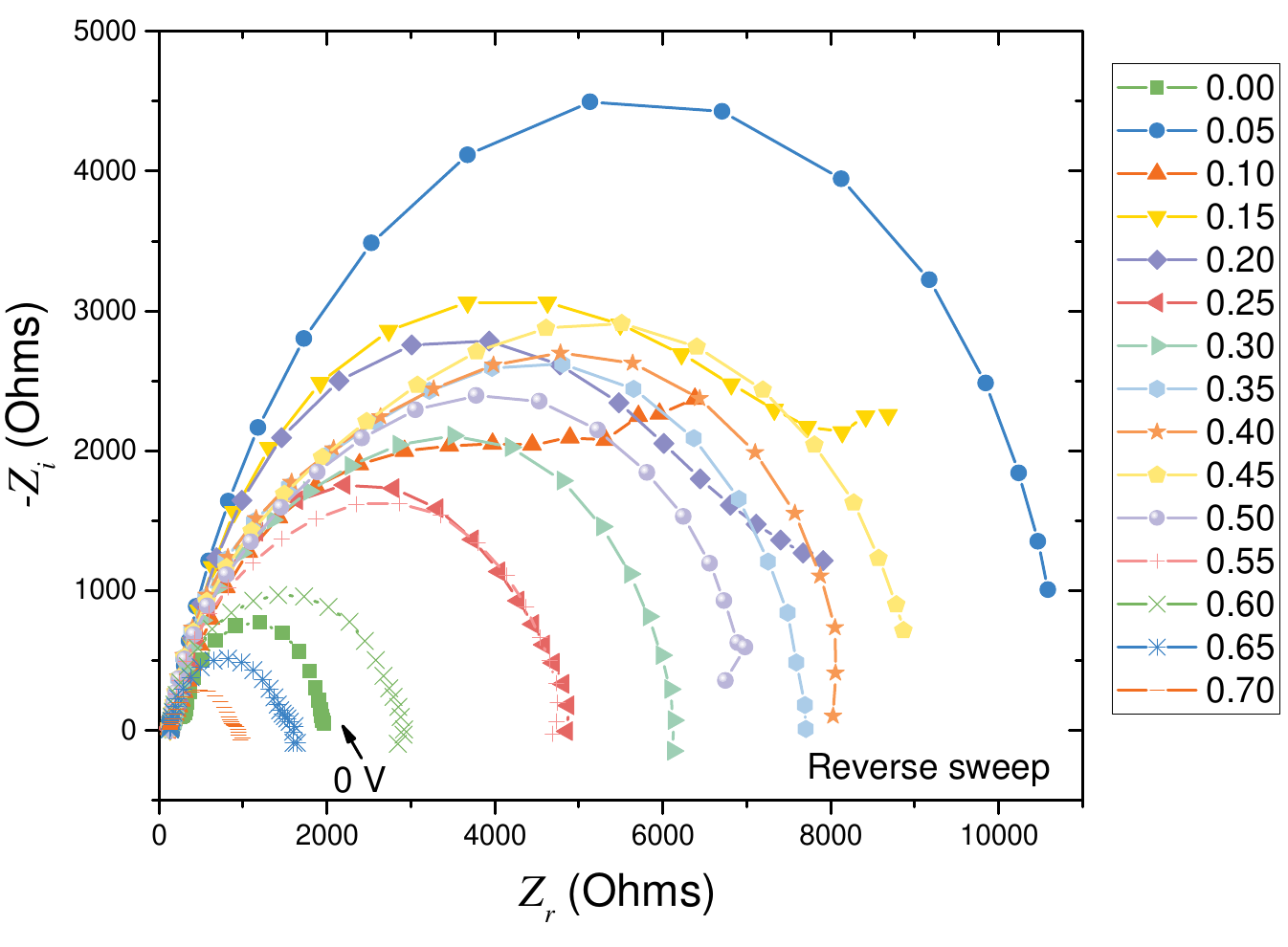}
\caption{Nyquist impedance plots during forward (a) and reverse (b) changes in dc bias corresponding to the data in Fig. ~\ref{fig:4andy}.
\label{fig:andyS1}}
\end{figure}

\begin{figure}[h]
\centering
\includegraphics[width=0.55\textwidth]{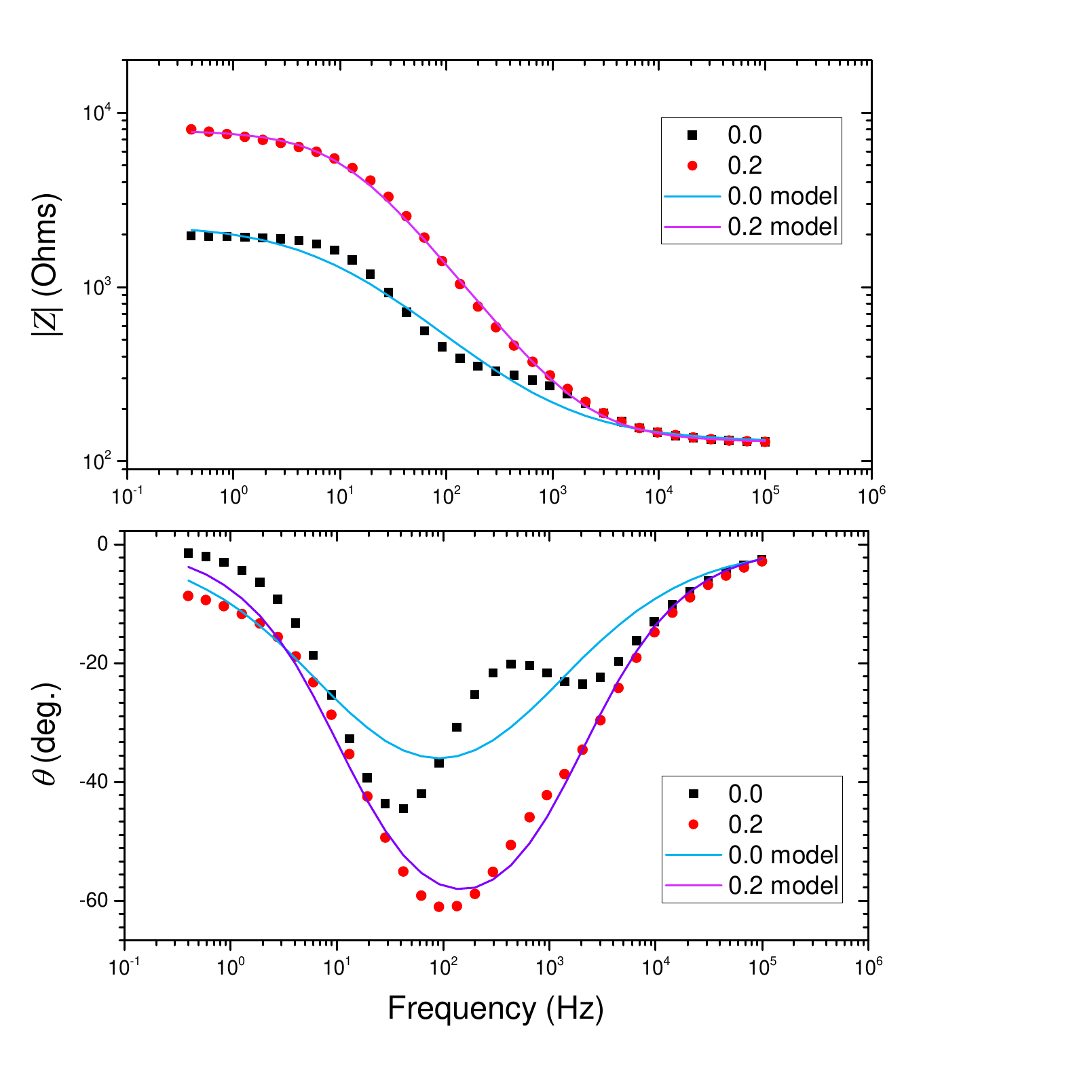} 
\caption{Bode impedance plots showing the quality of fit obtained using the equivalent circuit in Fig.~\ref{fig:4andy}(d). Black lines are the predicted impedance responses using the fitted parameters.
\label{fig:andyS2}}
\end{figure}

\begin{figure}[h]
\centering
(a)\includegraphics[width=0.48\textwidth]{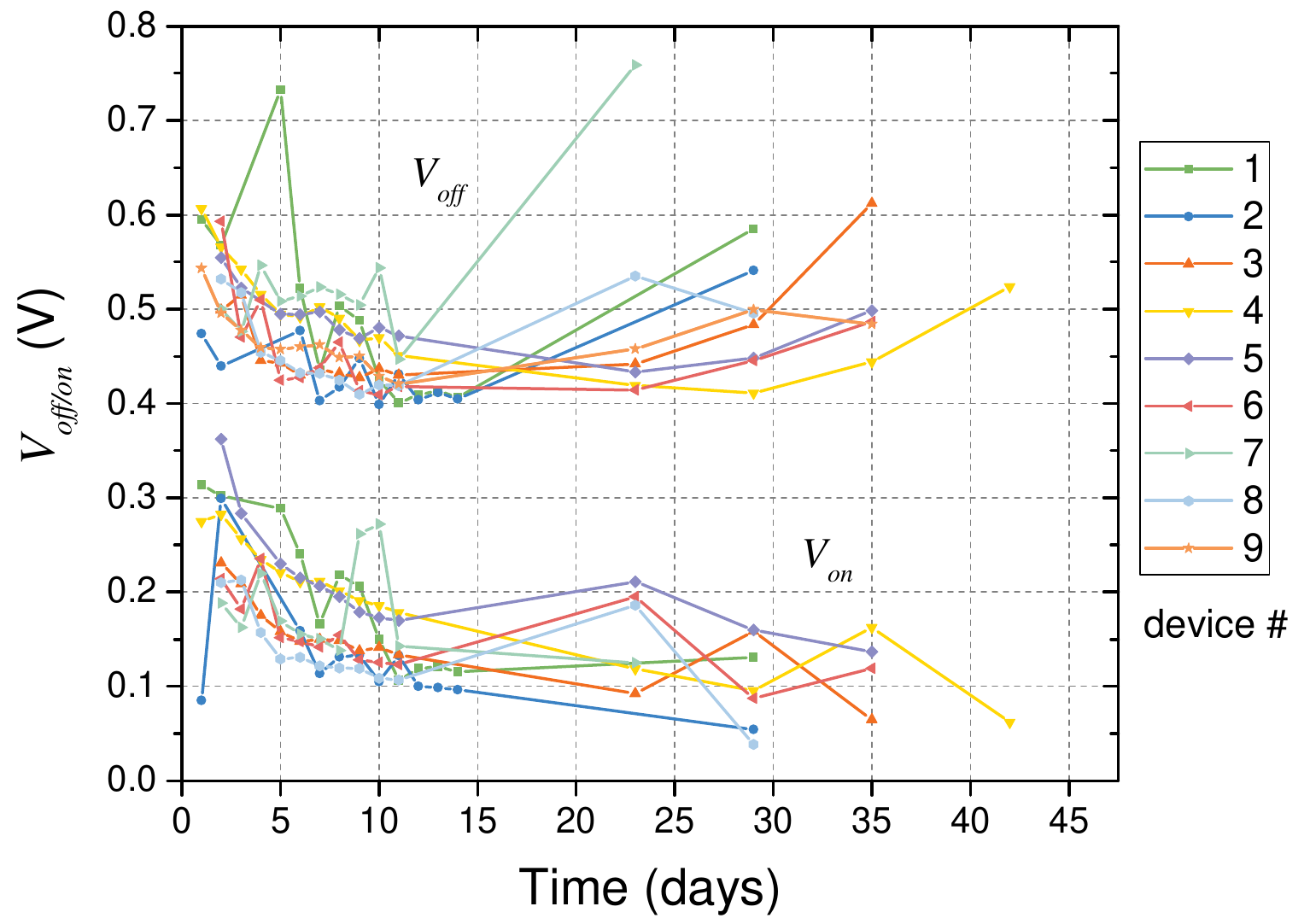} \\
(b)\includegraphics[width=0.49\textwidth]{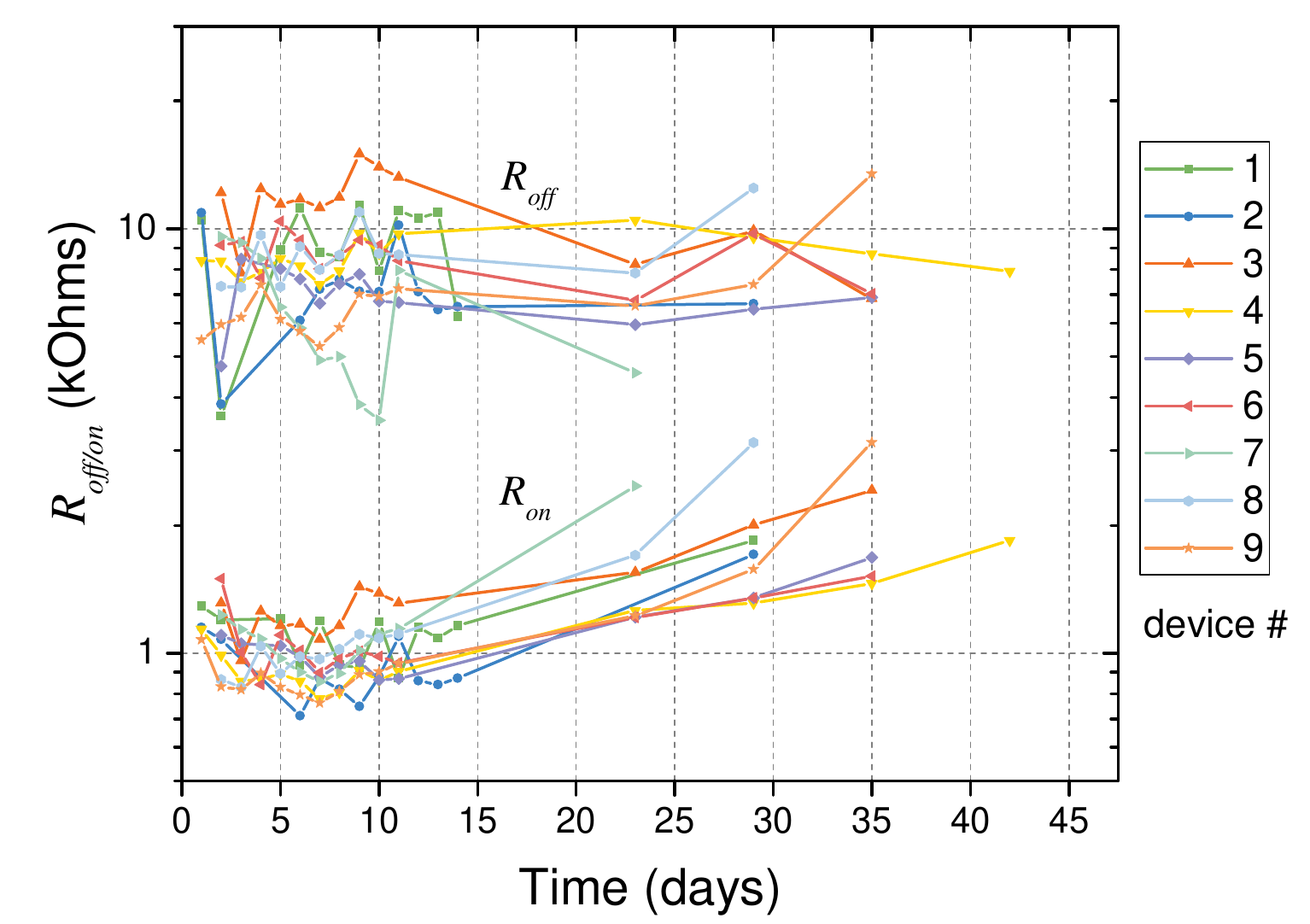}
\caption{Parameters of 9 distinct devices over time. These values were extracted from the last cycle of an $I-V$ sweep on each device on a given day.  Only the positive voltage range was considered. $V_{off}$ and $V_{on}$ in (a) are simply the locations of peak current during increasing and decreasing voltage sweeps.
$R_{on}$ and $R_{off}$  in (b) were computed as $V/I$ (i.e., the chord-wise slope from the origin) at a voltage location 50~mV below the oxidation peak determined for a given cycle. This choice is somewhat arbitrary but represents a position where the on current vs. the off current are significantly different (biggest vertical separation).}
\label{fig:andyS3}
\end{figure}

\begin{figure}[h]
\centering
(a) \includegraphics[width=0.45\textwidth]{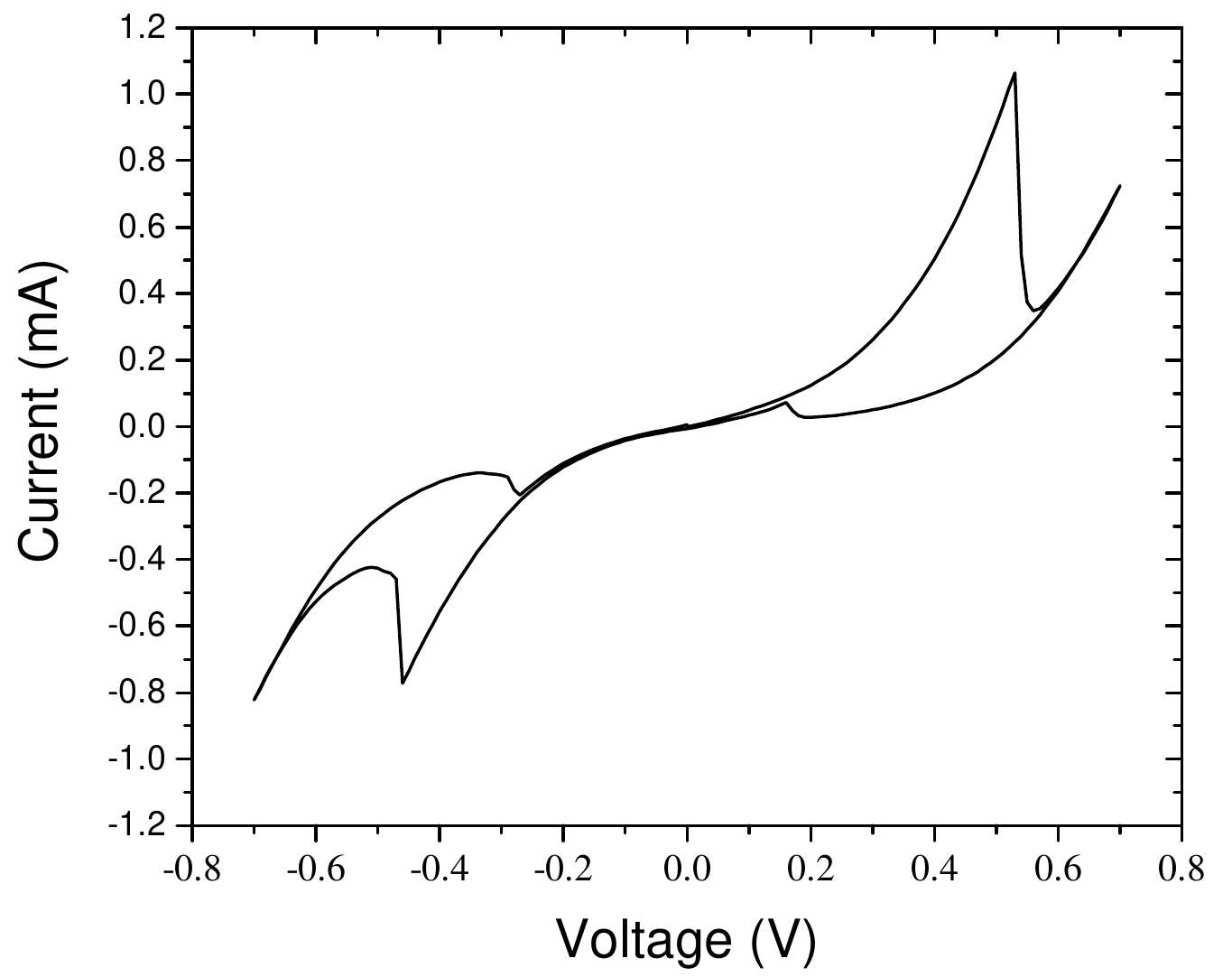} 
(b) \includegraphics[width=0.45\textwidth]{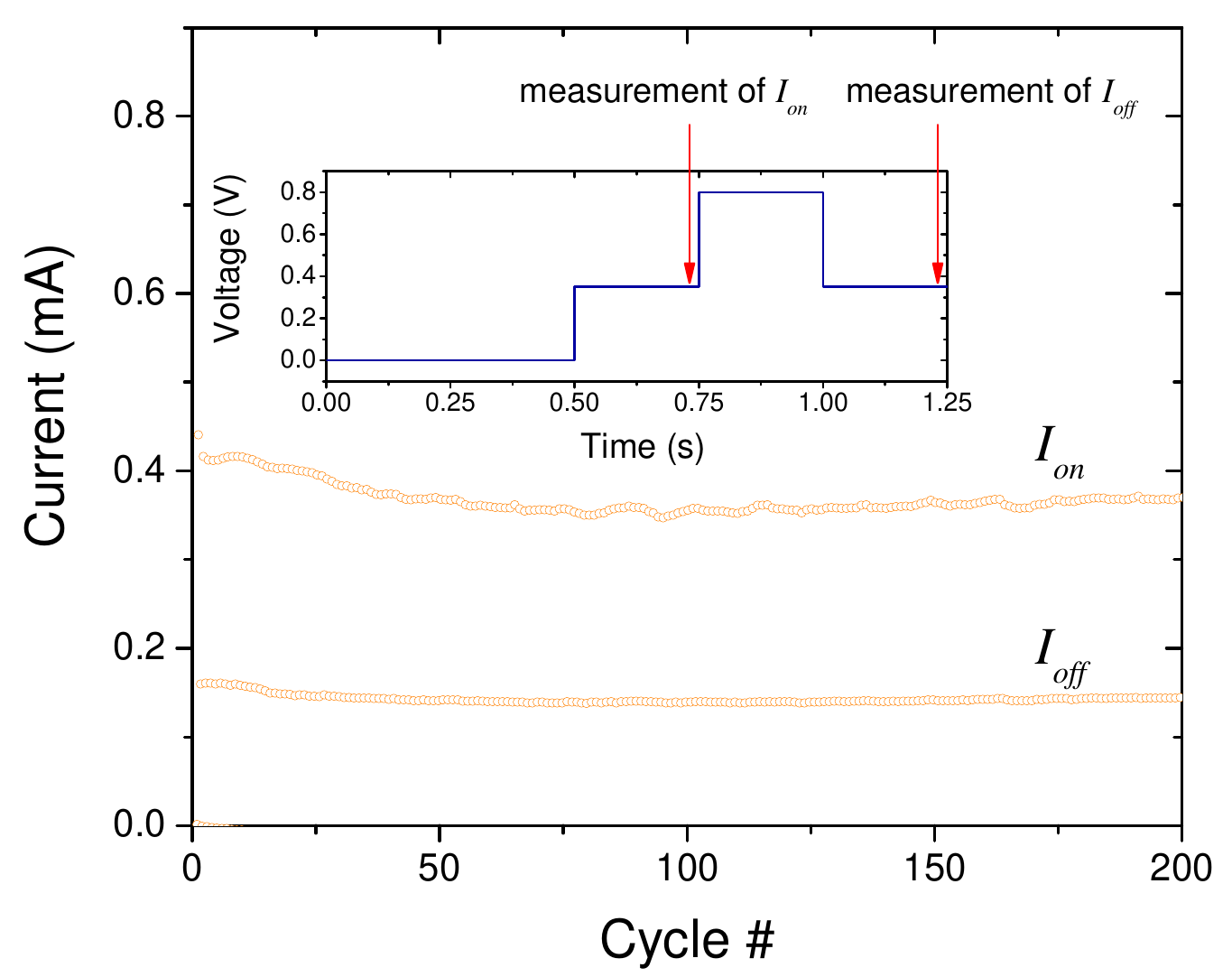}\\
(c) \includegraphics[width=0.45\textwidth]{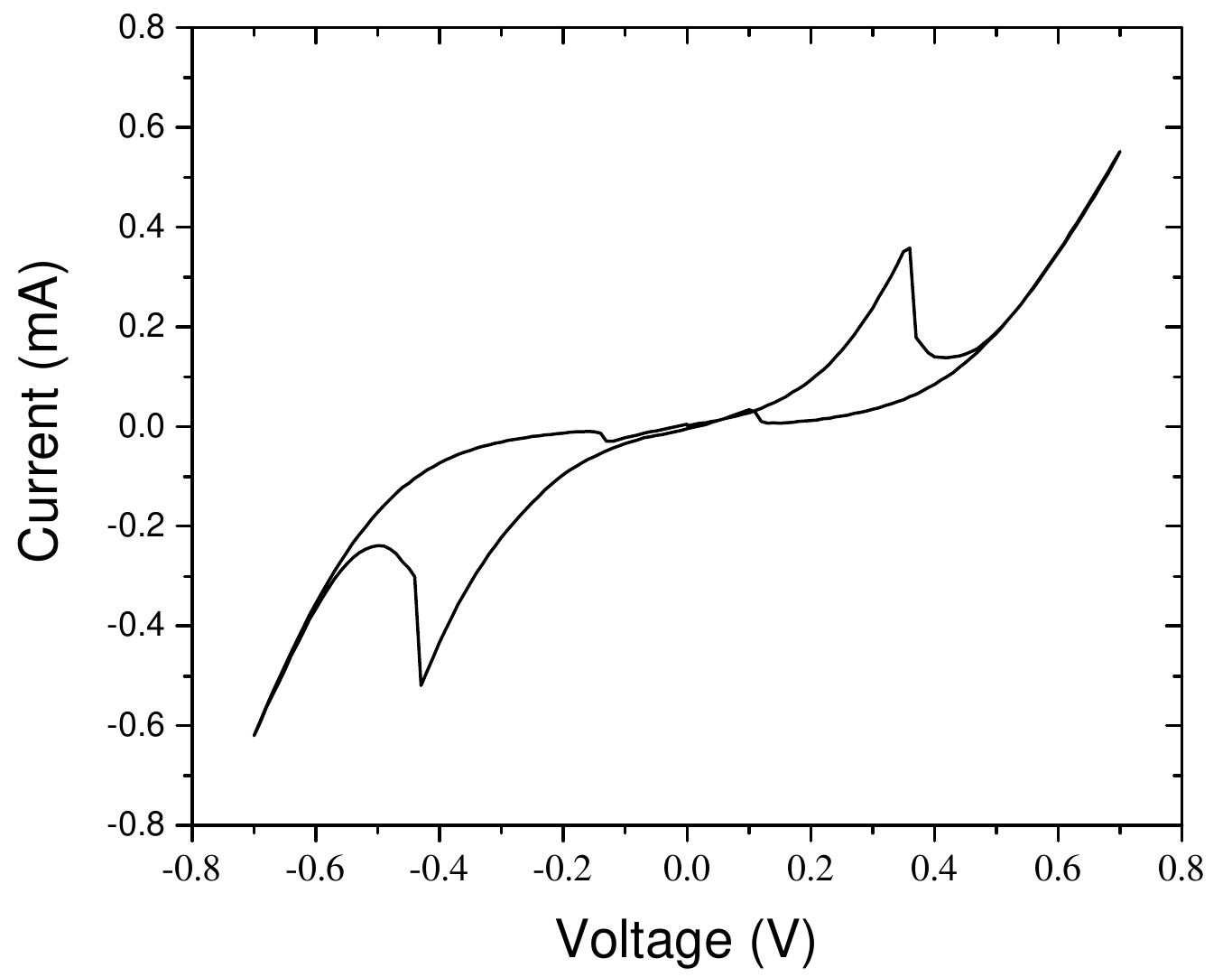} 
(d) \includegraphics[width=0.45\textwidth]{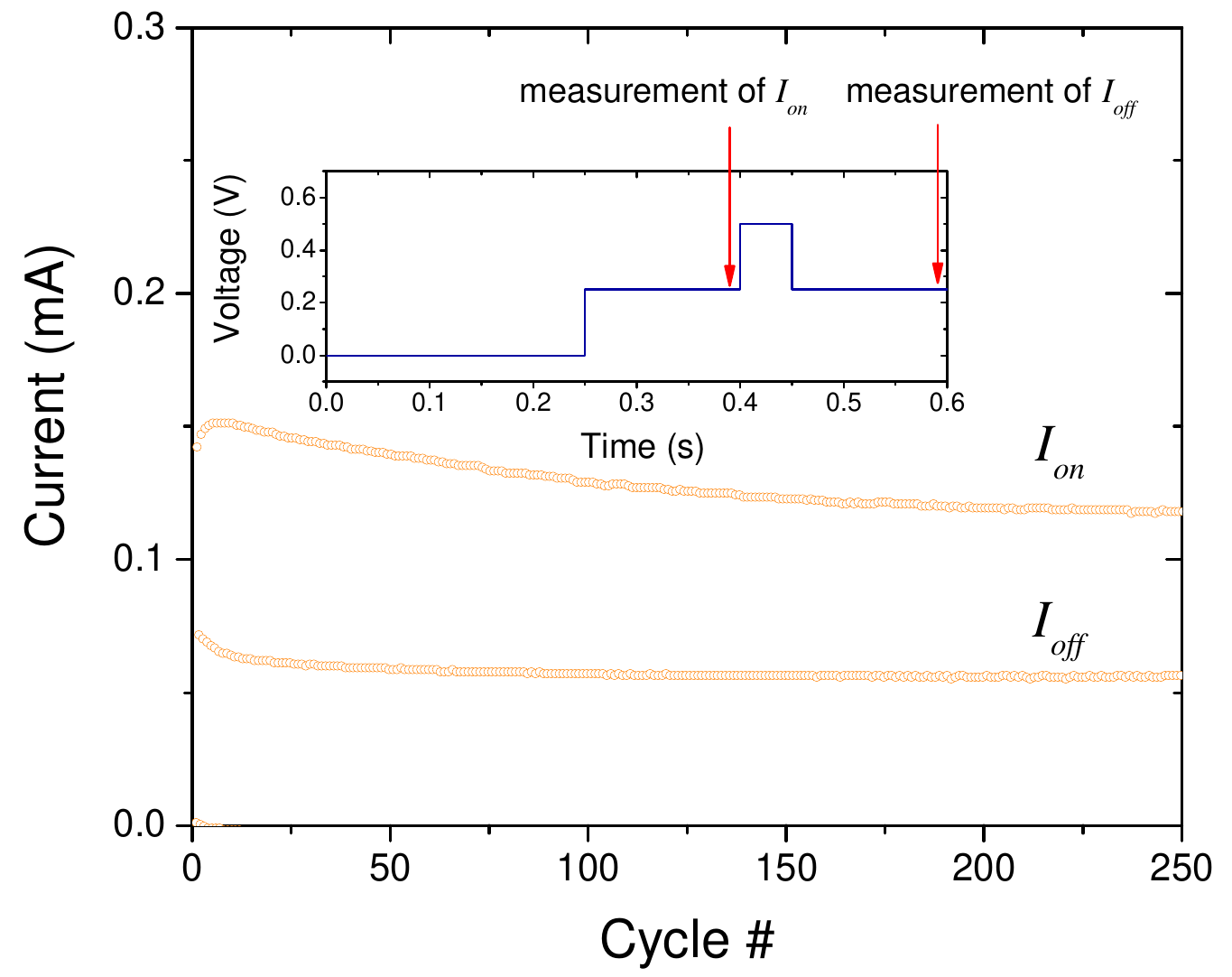}\\
(e) \includegraphics[width=0.45\textwidth]{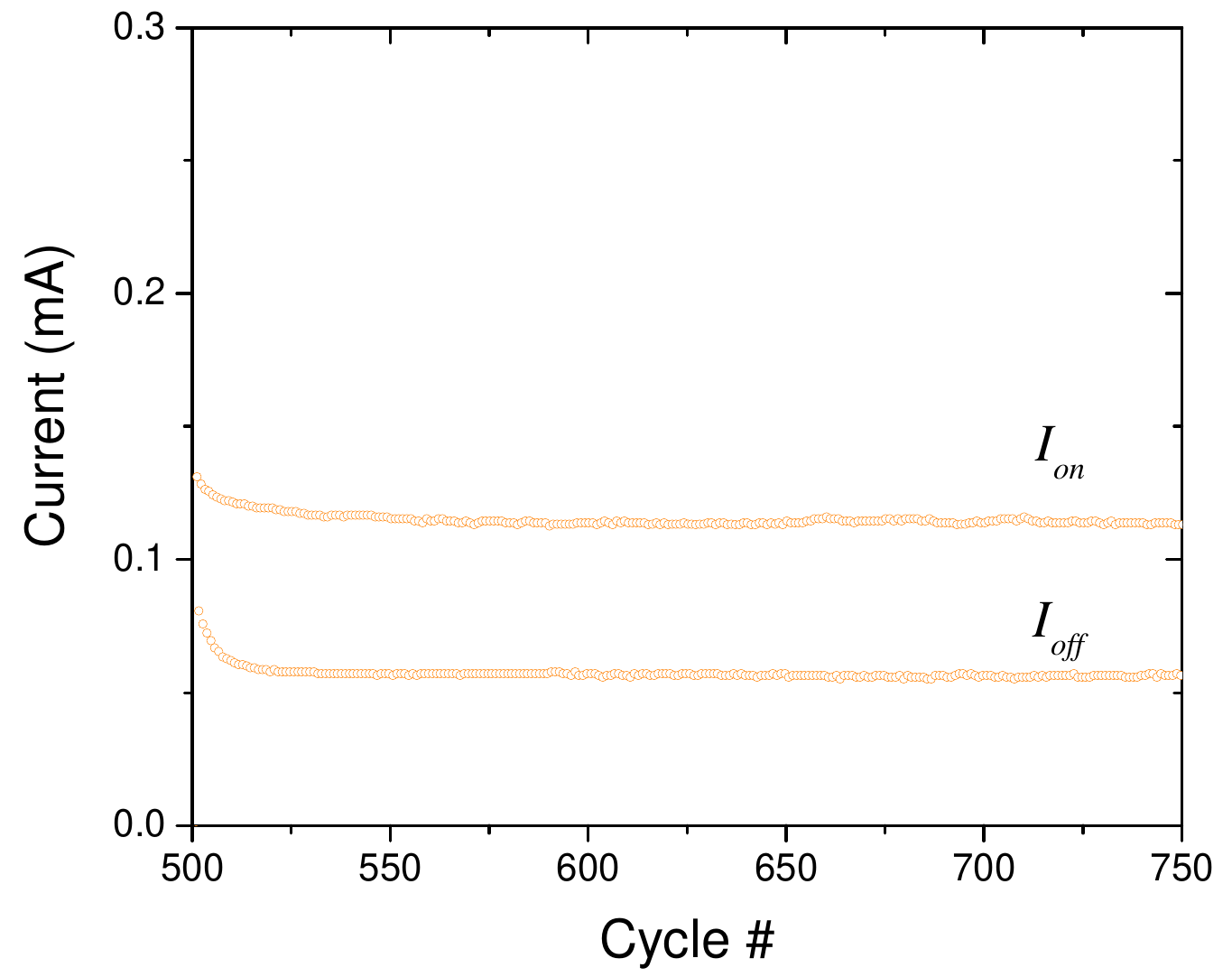} 
(f) \includegraphics[width=0.45\textwidth]{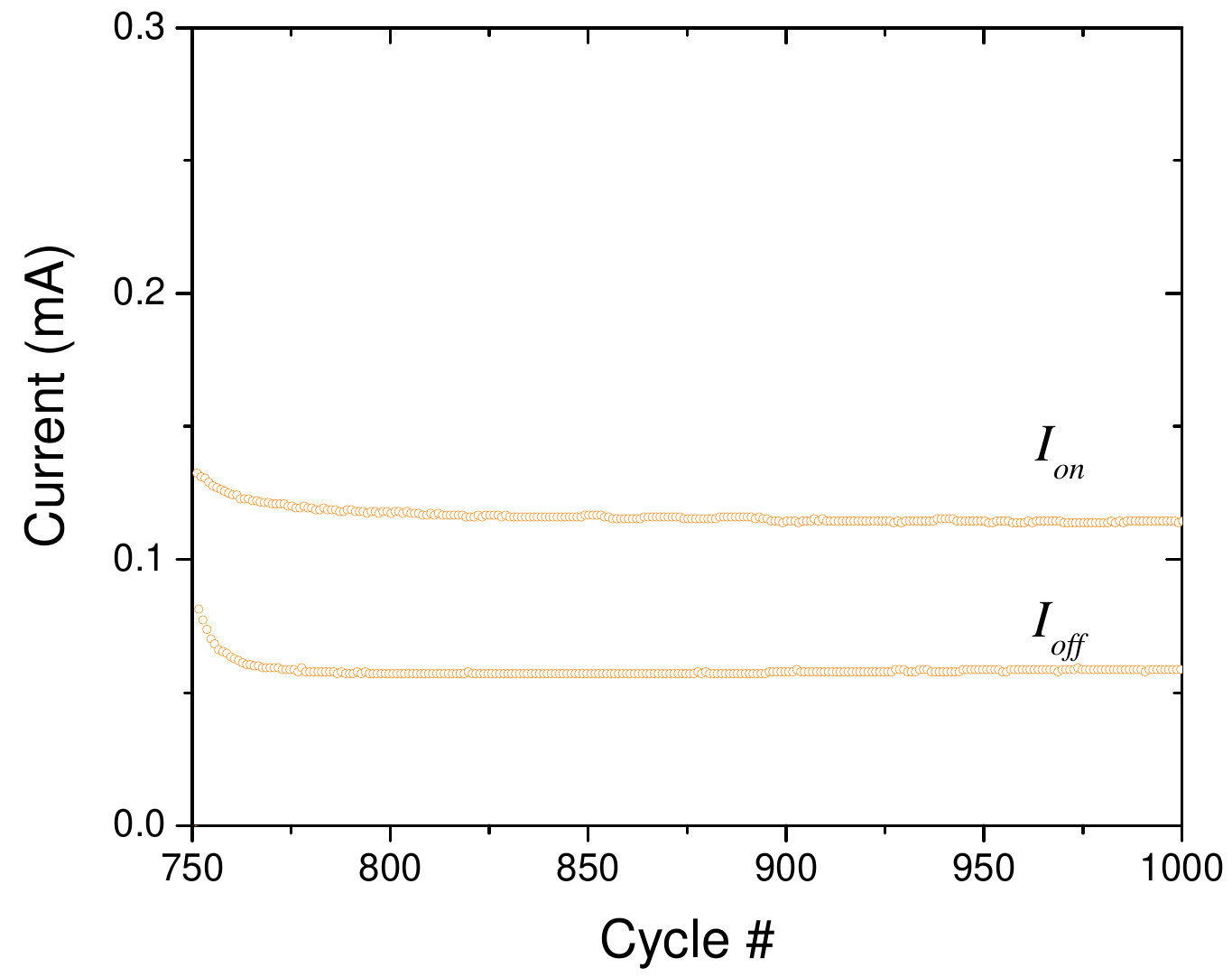} 
\caption{Performance of a periodically driven EGaIN device. (a) Current-voltage curve of the device. (b) Current in the {\bf on} and {\bf off} states. The inset of (b) illustrates the voltage waveform for a single cycle. (c) Current-voltage graphs following $10^3$~s of ultrasonic treatment. (d)-(f) Cycling behavior after sonication.
These curves were captured using the setup depicted in Fig.~\ref{fig:1}(b). \label{fig:S9}}
\end{figure}

\end{document}